\documentclass[sigconf]{acmart}

\usepackage{type1cm}     
\usepackage{graphicx}     
\usepackage{xspace}     
\usepackage{booktabs}     
\usepackage{multi row}     
\usepackage[font={bf}, tableposition=top]{caption}     
\usepackage{afterpage}
\usepackage{bold-extra}     
\usepackage{siunitx}          
\usepackage{xfrac}
\usepackage[font={small}]{subfig} 

\acmConference[WebSci\textquotesingle17]{9th International ACM Web Science Conference}{June
 26--28 2017}{Troy, NUY, USA}

\setcopyright{rightsretained}

\newcommand{\ourtitle}{{{The Effect of Collective Attention\\on Controversial Debates on Social Media}}}

\newcommand{\spara}[1]{\smallskip\noindent\textbf{#1}}

\newcommand{\rwc}{\ensuremath{\text{\sc RWC}}\xspace}

\newcommand{\core}{c\xspace}
\newcommand{\side}{s\xspace}
\newcommand{\words}{w\xspace}

\begin{document}

\copyrightyear{2017} 
\acmYear{2017} 
\setcopyright{acmlicensed}
\acmConference{WebSci '17}{June 25-28, 2017}{Troy, NY, USA}\acmPrice{15.00}\acmDOI{http://dx.doi.org/10.1145/3091478.3091486}
\acmISBN{978-1-4503-4896-6/17/06}

\title{\ourtitle}

\author{Kiran Garimella}
\affiliation{%
  \institution{Aalto University}
  \city{Helsinki} 
  \country{Finland} 
}
\email{kiran.garimella@aalto.fi}

\author{Gianmarco De~Francisci~Morales}
\affiliation{%
  \institution{Qatar Computing Research Institute}
  \city{Doha} 
  \country{Qatar} 
}
\email{gdfm@acm.org}

\author{Aristides Gionis}
\affiliation{%
  \institution{Aalto University}
  \city{Helsinki} 
  \country{Finland} 
}
\email{aristides.gionis@aalto.fi}

\author{Michael Mathioudakis}
\affiliation{%
  \institution{Aalto University}
  \city{Helsinki} 
  \country{Finland} 
}
\email{michael.mathioudakis@aalto.fi}

\renewcommand{\shortauthors}{Garimella, De Francisci Morales, Gionis, Mathioudakis}

\begin{abstract}
We study the evolution of long-lived 
controversial debates as manifested on Twitter from 2011 to 2016.
Specifically, we explore how the \emph{structure of interactions} and 
\emph{content of discussion} varies with the level of collective attention, 
as evidenced by the
number of users discussing a topic.
Spikes in the volume of users typically correspond to external events 
that increase the public attention on the topic -- as, 
for instance, discussions about `gun control' often erupt after a mass shooting.

This work is the first to study 
the dynamic evolution of polarized online debates at such scale.
By employing a wide array of network and content analysis measures,
we find consistent evidence that increased collective attention is associated
with increased network polarization and network concentration within
each side of the debate;
and overall more uniform lexicon usage across all users.
\end{abstract}

\maketitle


\section{Introduction}

Social media are a major venue of public discourse today, hosting the opinions of hundreds of millions of individuals.
Due to their prevalence they have become an invaluable instrument in the study of social phenomena and a fundamental subject of \emph{computational social science}.
In this work, we study discussions around issues that are deemed important at a societal level 
--- and in particular, ones that are controversial.
This work is a step towards understanding how the discussion about controversial topics on social media evolves, and more broadly how these topics shape the discussion at a societal and political level~\citep{abramowitz2005can,highton2011long,lindaman2002issue}.

We study how online discussions around controversial topics change as interest in them increases and decreases.
We are motivated by the observation that interest in enduring controversial 
issues is re-kindled by external events, e.g., 
when a major related story is reported.
One typical example is the gun control debate in U.S., which is revived whenever a mass shooting occurs.\footnote{See, e.g., \url{http://slate.me/1NswLLD}.}
The occurrence of such an event commonly causes an increase in collective attention, e.g., in volume of related activity in social media.

Given a controversial topic, our focus is to analyze
the interactions among users involved in the discussion, and 
quantify how certain structural properties of the interaction network vary 
with the change in volume of activity.
Our main finding is that the polarization reflected in the network structure of online interactions is correlated with the increase in the popularity of a topic.

Differently from previous studies, we study the \emph{dynamic} aspects of controversial topics on social media.
While the evolution of networks and polarization on social media have been studied in the past~\citep{conover2011political,leskovec2005graphs}, 
they have not been studied in conjunction before.
In addition, we seek to understand the \emph{response} of social media to stimuli that cause increased interest in the topics, an issue that only very recently has seen some attention~\citep{romero2016social}.

We take a longitudinal approach and collect data from Twitter 
that covers approximately five years.
This dataset gives us a very fine-grained view of the activity on social media, 
including the structure of the interactions among users, and 
the content they produced during this period.
We track four topics of discussion that are controversial in the U.S., that are recurring, and have seen considerable attention during the 2016 U.S. elections.

Our methodology relies on recent advances in quantifying controversy on social media~\citep{garimella2016quantifying}.
We build two types of networks: an \emph{endorsement} network from the retweet information on Twitter, and a \emph{communication} network from the replies.
We aggregate the data at a daily level, thus giving rise to a time series of interaction graphs.
Then, we identify the sides of a controversy via graph clustering, and find the \emph{core} of the network, i.e., the users who are consistently participating to the online discussion about the topic.
Finally, we employ a wide array of measures that characterize the discussion about a topic on social media, both from the point of view of the network structure and of the actual content of the posts.

Apart from our main result --- 
an increase in polarization linked to increased interest --- 
we also report on several other findings.
We find that most of the interactions during events of interest 
happen within the different controversy sides, and 
replies do not cross sides very often, 
in line with previous observations~\citep{smith2013role}. 
In addition, increased interest does not alter the fundamental structure of the endorsement network,
which is hierarchical, 
with a disproportionately large fraction of edges linking the periphery to the core.
This finding suggests that most casual users, who seldom participate in the discussion, 
endorse opinions from the core of the side they belong to.
When looking at the content of the posts on the two sides of a controversy, 
we find a consistent trend of \emph{convergence}, 
as the lexicons become both more uniform and more similar to each other.
This result indicates that, while the discussion is still controversial, 
both sides of the debate focus over the same fundamental issues 
brought under the spotlight by the event at hand.
Conversely, we do not find a consistent long-term trend in the polarization of discussions,
which contradicts the common narrative that our society is becoming more divided over time.
Finally, we perform similar measurements for a set of topics that are non-political and non-controversial, and highlight differences with the results for controversial discussions.\footnote{A limited subset of our results appeared in a poster at ICWSM 2017~\cite{ebbandflow2017}.}

\section{Related Work}
\label{sec:related}

A few studies exist on the topic of controversy in online news and social media. 
In one of the first papers, \citet{adamic2005political}
study linking patterns and topic coverage of political bloggers, 
focusing on blog posts on the U.S.\ presidential election of 2004.
They measure the degree of interaction between liberal and conservative blogs,
and provide evidence that  conservative blogs are linking to each other more frequently and in a denser pattern.
These findings are confirmed by a more recent study of \citet{conover2011political}, 
who focus on political communication regarding congressional midterm elections.
Using data from Twitter, 
they identify a highly segregated partisan structure 
(present in the retweet graph, but not in the mention graph), 
with limited connectivity between left- and right-leaning users.
In another recent work,  
\citet{mejova2014controversy} consider discussions of controversial and non-controversial news over a span of $7$ months.
They find a significant correlation between controversial issues and the use of negative affect and biased language. 
More recently, \citet{garimella2016quantifying} show that controversial discussions on social media have a well-defined structure, when looking at the \emph{endorsement} network.
They propose a measure based on random walks (\rwc), 
which is able to identify controversial topics, and 
\emph{quantify} the level of controversy of a given discussion 
via its network structure alone.

The aforementioned studies focus on static networks, which are a snapshot of the underlying dynamic networks.
Instead, we are interested in network dynamics and, specifically, in how it responds to increased collective attention in the controversial topic.

Several studies have looked at how networks evolve, and proposed models of network formation~\cite{leskovec2005graphs,leskovec2008microscopic}.
Densification over time is a pattern often observed~\cite{leskovec2005graphs}, i.e., social networks gain more edges as the number of nodes grows.
A change in the scaling behavior of the degree distribution has also been observed~\cite{ahn2007analysis}.
\citet{newman2011structure} offer a comprehensive review.
Most of these studies focus on social networks, and in particular, on the friendship relationship.
In our work, we are interested in studying an \emph{interaction} network, which has markedly different characteristics.

There is a large amount of literature devoted to studying the evolution of networks.
For an overview, see the book by~\citet{dorogovtsev2013evolution}.
However, none of these previous studies has devoted much attention to the evolution of interaction networks for controversial topics, 
especially when tracking topics for a long period of time.

\citet{difonzo2014network} report on a user study that shows 
how the network structure affects the formation of stereotypes 
when discussing controversial topics. 
They find that segregation and clustering lead to a stronger ``echo chamber'' effect, 
with higher polarization of opinions.
Our study examines a similar correlation between polarization and network structure, 
although in a much wider context, and focusing on the influence of external events.

\citet{garimella2017long} study polarization on Twitter over a long period of time, using content and network-based measures for polarization and find that over the past decade, polarization has increased. We find no consistent trend among the topics we study.

Perhaps the closest work to this paper is the work by \citet{smith2013role}, who study the role of social media in the discussion of controversial topics. They try to understand how positions on controversial issues are communicated via social media, mostly by looking at user level features such as retweet and reply rates, url sharing behavior, etc. They find that users spread information faster if it agrees with their position, and that Twitter debates may not play a big role in deciding the outcome of a controversial issue.

However, there are differences with our work: 
(i) they study one local topic (California ballot), over a small period of time, while we study a wide range of popular topics, spanning multiple years;
and (ii) their analysis is mostly user centric, whereas we take a global viewpoint, constructing and analyzing networks of user interaction.

\spara{The effect of external events on social networks.}
A few studies have examined the effects of events on social networks.
\citet{romero2016social} study the behavior of a hedge-fund 
company via the communication network of their instant messaging systems.
They find that in response to external shocks, i.e., when stock prices change significantly, the network ``turtles up,'' strong ties become more important, and the clustering coefficient increases.
In our case, we examine both a communication network and an endorsement network, and we focus on controversial issues.
Given the different setting, many of our findings are quite different.

Other works, such as the ones by~\citet{Lehmann2012} and~\citet{Wu2014}, examine how collective attention focuses on individual topics or items and evolves over time.
\citet{Lehmann2012} examine spikes in the frequency of hashtags and whether most frequency volume appears before or after the spike.
They find that the observed patterns point to a classification of hashtags, that agrees with whether the hashtags correspond to topics that are endogenously or exogenously driven.
\citet{Wu2014}, on the other hand, examine items posted on digg.com and how their popularity decreases over time.

\citet{morales2015measuring} study polarization over time for a single event, the death of Hugo Chavez.
Our analysis has a more broad spectrum, as we establish common trends across several topics, and find strong signals linking the volume of interest to the degree of polarization in the discussion.

\citet{andris2015rise} study the partisanship of the U.S.\ congress over a long period of time.
They find that partisanship (or non-cooperation) in the U.S.\ congress has been increasing dramatically for over 60 years.
Our study suggests that increased controversy is linked to an increase in attention on a topic,
whereas we do not see a global trend over time.




\section{Dataset}
\label{sec:dataset}

\begin{figure*}[t]
\centering
\includegraphics[width=\textwidth]{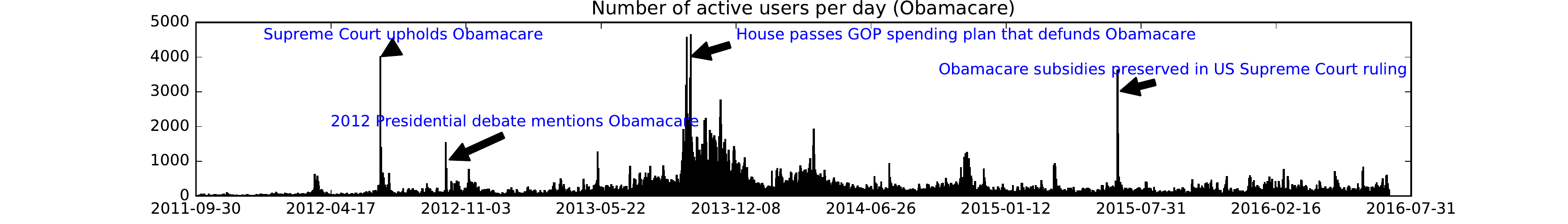}
\includegraphics[width=\textwidth]{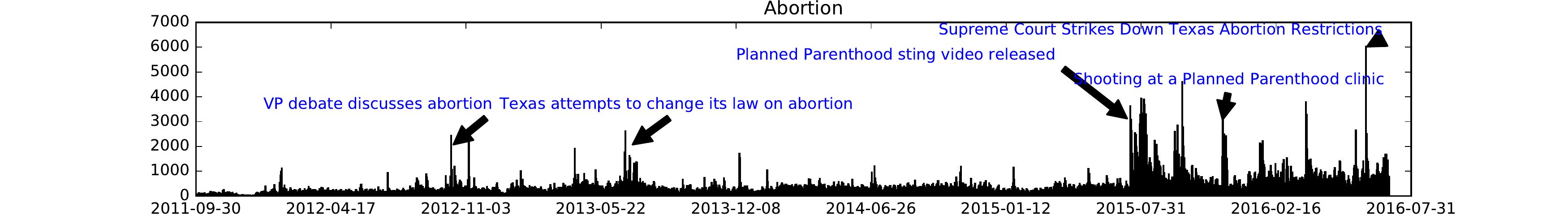}
\includegraphics[width=\textwidth]{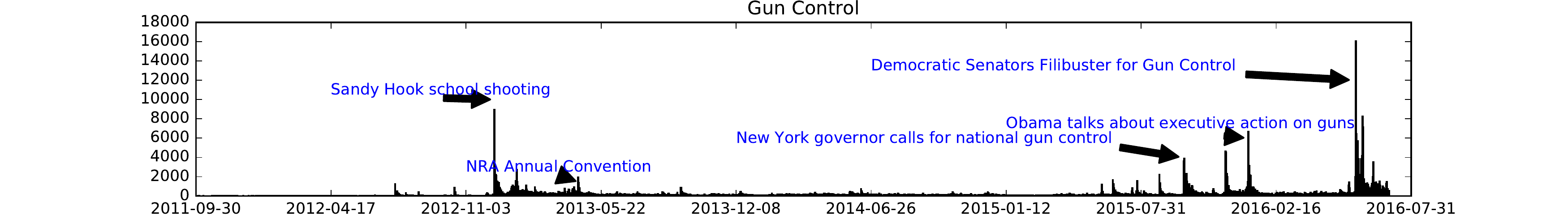}
\includegraphics[width=\textwidth]{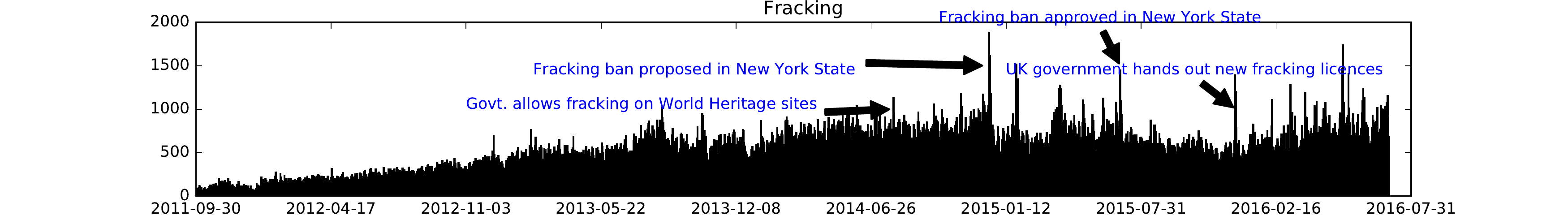}
\caption{Daily trends for number of active users for the four controversial topics under study. Clear spikes occur at several points in the timeline. Manually chosen labels describing related events reported in the news on the same day are shown in blue for some of the spikes.}
\label{fig:timeline}
\end{figure*}

Our study uses data collected from Twitter.
Using the repositories of the Internet Archive,\footnote{\small\url{https://archive.org/details/twitterstream}}
we collect a $1\%$ sample of tweets
from September 2011 to August 2016,\footnote{\small{To be precise, we have data for $57$ months from that period}}
for four topics of discussion, related to `Obamacare', `Abortion', `Gun Control', and `Fracking'.
These topics constitute long-standing controversial issues in the U.S.\footnote{\small{According to \small\url{http://2016election.procon.org}.}} and have been used in previous work~\cite{lu2015biaswatch}.
For each topic, we use a keyword list as proposed by~\citet{lu2015biaswatch} (shown in Table~\ref{tab:keywords}), and extract a base set of tweets which contain at least one topic-related keyword.
To enrich this original dataset, we use the Twitter REST API to obtain all tweets of users who have participated in the discussion at least once.\footnote{\small Up to \num{3200} due to limits imposed by the Twitter API.}
Admittedly, this dataset might suffer from sampling bias, however the topics are specific enough that the distortion should be negligible~~\citep{morstatter2013sample}.
There might also be recency bias due to the addition of the latest tweets of the users.
However, the data does not show any clear trend in this sense (see Figure~\ref{fig:timeline}).
In addition, given that we rely on detecting volume peaks, the trend does not affect our analysis.
Table~\ref{tab:keywords} shows the final statistics for the dataset.

\begin{table}[t]
\centering
\small
\caption{Keywords for the controversial topics.}
\label{tab:keywords}
\begin{tabular}{l>{\raggedright}p{10em}rr}
\toprule
Topic & Keywords & \#Tweets & \#Users \\
\midrule
Obamacare & obamacare, \#aca & \num{866484} & \num{148571} \smallskip\\
Abortion & abortion, prolife, prochoice, anti-abortion, pro-abortion, planned parenthood & \num{1571363} & \num{327702} \smallskip\\
Gun Control & gun control, gun right, pro gun, anti gun, gun free, gun law, gun safety, gun violence& \num{824364} & \num{224270} \smallskip\\
Fracking & fracking, \#frack, hydraulic fracturing, shale, horizontal drilling & \num{2117945} & \num{170835} \\ 
\bottomrule
\end{tabular}
\end{table}

We infer two types of interaction network from the dataset:
($i$) a retweet network --- a directed endorsement network of users, 
where there is an edge between two users ($u\,{\rightarrow}\,v$) if $u$ retweets $v$, and 
($ii$) a reply network --- a directed communication network of users, 
where an edge ($u\,{\rightarrow}\,v$) indicates that user 
$u$ has replied to a tweet by user~$v$.
Note that replies are characterized by a tweet starting with \texttt{`@username'} and do not include mentions and retweets.\footnote{\small See also \url{https://support.twitter.com/articles/14023} for terminology related to different types of Twitter messages.}

Polarized networks, especially the ones considered here, can be broadly characterized by two opposing \emph{sides}, which express different opinions on the topic at hand.
It is commonly understood that retweets indicate endorsement, and endorsement networks for controversial topics have been shown to have a bi-clustered structure~\cite{conover2011political,garimella2016quantifying}, i.e., they consist of two well-separated clusters that correspond to the opposing points of view on the topic.
Conversely, replies can indicate discussion, and several studies have reported that users tend to use replies to talk across the sides of a controversy~\cite{bessi2014social,liu2014twitter}.
These two types of network capture different dynamics of activity, and allow us to tease apart the processes that generate these interactions.

In this paper, we build upon the observation that the clustering structure of retweet networks reveals the opposing sides of a topic.
In particular, following an approach from previous work~\citep{garimella2016quantifying}, we collapse all retweets contained in the dataset of each topic into a single large static retweet network.
Then, we use the METIS clustering algorithm~\cite{karypis1995metis} to identify two clusters that correspond to the two opposing sides.
This process allows us to identify more consistent sides for the topic.
We evaluate the sides by manual inspection of the top retweeted users, URLs, and hashtags.
The results are consistent and accurate, and can be inspected online.\footnote{\small\url{https://mmathioudakis.github.io/polarization/}\label{footnote:website}}



Let us now consider the temporal dynamics of these interaction networks.
Given the traditional daily news reporting cycle, we build the time series of networks with the same daily granularity.
This high resolution allows us to easily discern the level of interest in the topic, and possibly identify spikes of interest linked to real world external events, as shown in Figure~\ref{fig:timeline}.
These spikes usually correspond to external newsworthy events, as shown by the annotations.
These results support the observation that Twitter is used as an \emph{agor\'{a}} to discuss the daily matters of public interest~\citep{deFrancisciMorales2012trex}.

As shown in Figure~\ref{fig:timeline}, the size of the active network for each day varies significantly.
There is, however, a \emph{hard core} set of active users who are involved in the discussion of these controversial topics most of the time.
Therefore, to understand the role of these more engaged users, we define the `core network' as the one induced by users who are active for more than $\sfrac{3}{4}$ of the observation time.
Specifically, to build a \emph{core} set of users, we first identify two subsets --- one consisting of those users who generated or received a retweet at least once per month for $45$ months; and another one defined similarly for replies.
We define the core set of users as the union of the aforementioned two sets.
Nodes of a network that do not belong to the core are said to belong to the \emph{periphery} of that network.
The size of the core ranges from around \num{600} to \num{2800} nodes for the four topics.
For any given day, the core accounts for at most around $10\%$ of the active users.

\subsection{Notation}
The set of retweets that occur within a single day $d$ gives rise to one retweet network $N^\mathit{rt}_d$.
Each user associated with a retweet is represented with one node in the network.
There is a directed edge from user $u$ to user $v$ only when user $u$ has retweeted at least one tweet authored by user $v$.
Correspondingly, the set of replies that occur within a single day give rise to a reply network $N^\mathit{re}_d$.
In addition, each node $u$ in the network is associated with a binary attribute $\core(u) \in\{\text{\tt true}, \text{\tt false}\}$ that indicates whether the node is part of the core, 
and an attribute $\side(u) \in \{1, 2\}$
that represents the side the node belongs to.
In some cases, we consider undirected versions of the networks defined above.
In such cases, we write $G^\mathit{rt}_d$, $G^\mathit{re}_d$ to denote the undirected graphs
corresponding to $N^\mathit{rt}_d$, $N^\mathit{re}_d$, respectively.

Besides these two types of network, for each day we consider the set of tweets that were generated on that day. Every tweet $m$ is associated with 
an attribute $\side(m) \in \{1, 2\}$
that indicates the side its author belongs to.
Moreover, every tweet $m$ is associated with the list of words $\words(m)$ that occur in its text.
This information gives rise to two unigram distributions $W^1_d$ and $W^2_d$, 
one for each side. Each distribution expresses the number of times each word appears in the tweets of nodes from each side.

\section{Measures}
\label{sec:measures}



For each day $d$, we employ a set of measures on the associated
networks $N^\mathit{rt}_d$, $N^\mathit{re}_d$, and unigram distributions $W^1_d$ and $W^2_d$.
We describe them below.

\spara{Polarization.}
We quantify the polarization of a network $N_d$ by using the random-walk controversy (\rwc) score introduced in previous work~\cite{garimella2016quantifying}.
Intuitively, the score captures whether the network consists of two well-separated clusters.


\spara{Clustering coefficient.}
In an undirected graph, the
clustering coefficient $\mathit{cc}(u)$ of a
node $u$ 
is defined as the fraction of closed
triangles in its immediate neighborhood.
Specifically, let $d$ be the degree of node $u$, and $T$ be the number of closed
triangles involving $u$ and two of its neighbors, then
\[
\mathit{cc}(u) = \frac{2T}{d (d - 1)}.
\]
In our case, we consider the undirected graph $G_d$ and
compute the average clustering coefficient of all nodes that
belong to each \emph{side} -- then take the mean of the two averages as the
clustering coefficient of the network.

In order to control for scale effects, i.e., correlation between the size of the network (as determined by the volume of users active on day $d$) and the clustering coefficient,
we employ a normalizer for the score.
More in detail, we use an Erd\H{o}s-R\'{e}nyi graph as null model (with edges drawn at random among pair of nodes), and normalize the score by the expected value for a null-model graph of the same size.
Unless otherwise specified, we apply the same type of normalization for all the methods defined below.

\spara{Tie strength.}
For each node $u$ in a graph $G_d$, we consider all nodes $v$ it is connected
to across all days, and order them decreasingly by the number of occurrences $|\{ d : (u, v) \in G_{d} \}|$.
That is, the node $v$ at the top of the list for $u$ is the node to which $v$ connects the most consistently throughout the time span of the dataset.
Then, we define the \emph{strong ties} of a node $u$ as the top $10\%$
of the nodes ordered as described above.
For a given day $d$, we define the tie strength of a node as the number
of strong ties it is connected to in the corresponding graph $G_d$.
The tie-strength measure for the day is defined as the average tie strength for 
all nodes on either side.
As described for the previous measure above, 
we normalize the reported measure by the expected value for a random
graph with the same number of nodes and edges.

\spara{Cross--side openness.}
This measure reports the number of edges that connect nodes from opposing sides,
and captures the inter-side interaction happening in the network on a given day.
Formally, it is defined as
\[
\mathit{CSO} = | \{ (u,v) \in G_d : \side(u) \neq \side(v) \} |.
\]
We apply the same normalization as described above.

\spara{Sides edge composition.}
For a given network, we distinguish two types of edges: \emph{within-sides}, where both adjacent nodes belong to the same side,
and \emph{across-sides}, where the adjacent nodes belong to different sides.
For each day and network, we track the fraction of the two types of edges.

\spara{Core--periphery openness.}
This measure is defined as the number of edges that connect a node from the core to the periphery.
It captures the amount of interaction between the hard core users and the casual ones.
Formally, 
\[
\mathit{CPO} = | \{ (u,v) \in G_d : \core(u) \wedge \neg \core(v) \} |.
\]

\spara{Bimotif.}
For a network $N_d$, we define the bimotif measure as the number of directed edges $(u, v) \in N_d$ for which the opposite edge $(v, u)$ also appears in the network
\[
\mathit{Bimotif} = | \{ (u,v) \in N_d : (v,u) \in N_d \} |.
\]
This measure captures the mutual interactions happening within the network. It is also known as `reciprocity' in the literature. 

\spara{Core Density.}
This measure captures the number of edges that connect exclusively members of the core
\[
\mathit{CoreDens} = | \{ (u,v) \in N_d : \core(u) \wedge \core(v) \} |.
\]

\spara{Core--periphery edge composition.} 
For a given network, we distinguish three types of edges: \emph{core--core}, where both adjacent nodes belong to the core we have identified, \emph{core--periphery}, where one node belongs to the core and one to the periphery, and \emph{periphery--periphery}, where both nodes belong to the periphery.
For each day and network, we track the fraction of each type of edges. 

\spara{Cross--side content divergence.}
This measure captures the difference between the word distributions $W^1_d$ and $W^2_d$, 
and is based on the Jensen-Shannon divergence~\cite{lin1991divergence}.
The Jensen-Shannon divergence is undefined when one of the two distributions is zero at a point where
the other is not.
Thus, we smooth the distributions by adding Laplace counts $\beta = 10^{-5}$ to avoid zero entries in either distribution.

The traffic volume on a given day can increase the vocabulary size, and thus induce an unwanted bias in the measure.
In order to counter this bias, we employ a sampling procedure similar to bootstrapping from the two distributions.
For each smoothed distribution $W^1_d$ and $W^2_d$, we sample with replacement
$k = \num{10000}$ words at random, and compute the Jensen-Shannon divergence of these equal-sized samples.
We repeat the process $100$ times and report the average sample Jensen-Shannon divergence
as the `cross-side content divergence' for day $d$.
Intuitively, the higher its value, the more different the word distributions across the two sides.

\spara{Within-side entropy.}
This measure captures how `concentrated' each of the two distributions $W^1_d$ and $W^2_d$ is.
For each side, we compute the entropy for each distribution.
The higher its value, the more widely spread is the corresponding distribution.
We use the same bootstrap sampling method described above to avoid bias due to activity volume.

\spara{Topic variance.}
This measure captures, to some extent, $what$ is being talked on the two sides of the discussion. 
We extract a large number of topics by using Latent Dirichlet Allocation (k=100) on the complete tweet corpus.
We then compute the distribution of topics in each bucket.
This distribution gives an estimate on which of the 100 topics are being talked about in the bucket.
We report the variance of this distribution.
If the distribution is focused on a small number of topics, the variance is high.
Conversely, a low variance indicates a uniform distribution of topics.

\spara{Sentiment variance.}
This measure captures the variance of sentiment valence (positive versus negative) in all the tweets of one day $d$~\cite{garimella2016quantifying}.

\spara{Psychometric analysis.}
To understand the if there are behavioral changes in terms of content generated and shared by users with increasing activity, we use the Linguistic Inquiry and Word Count (LIWC) dictionary,\footnote{\small\url{http://liwc.net}} which identifies emotions in words~\cite{kramer2014experimental}.
We measure the fraction of tweets containing the LIWC categories: anger, sadness, posemo, negemo, and anxiety. 

\subsection{Analysis}
We explore how the aforementioned measures vary with the number of active users in the networks, which is a proxy for the amount of collective attention the topic attracts.
We sort the time series of networks by volume of active users, and partition it into ten quantiles (each having an equal number of days),
so that days of bucket $i$ are associated with smaller volume than those of bucket $j$, 
for $i < j$.
For each bucket, we report the mean and standard deviation of the values for each measure, and observe the trend from lower to higher volume.

Note that the measures presented in this section are carefully defined so that their expected value does not depend on the volume of underlying activity (i.e., number of network nodes and edges or vocabulary size).

\section{Findings}
\label{sec:findings}

In what follows, we report our findings on the measures defined in Section~\ref{sec:measures} --- starting from the ones related to the retweet and reply networks (Section~\ref{sec:network_findings}), then proceeding to the ones related to content (Section~\ref{sec:content_findings}) and network cores (Section~\ref{sec:core_findings}). We provide additional analysis for the periods around the spikes in interest (Section~\ref{sec:local_findings}), as well as for the evolution of measures over time (Section~\ref{sec:time_findings}).

\subsection{Network}
\label{sec:network_findings}
\begin{figure}[tb]
\centering
\begin{minipage}{.25\linewidth}
\centering
\subfloat{\label{}\includegraphics[width=\textwidth]{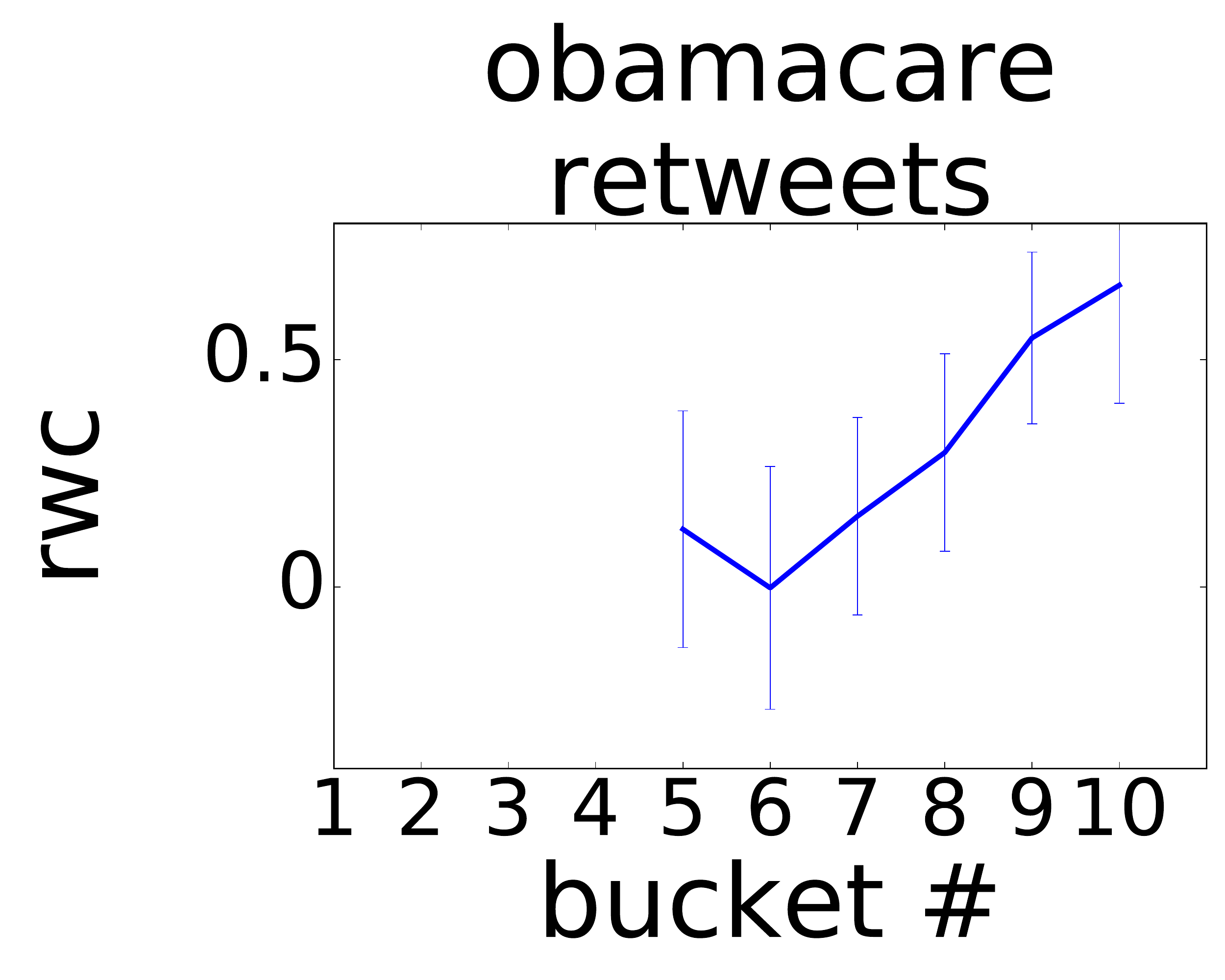}}
\end{minipage}%
\begin{minipage}{.25\linewidth}
\centering
\subfloat{\label{}\includegraphics[width=\textwidth]{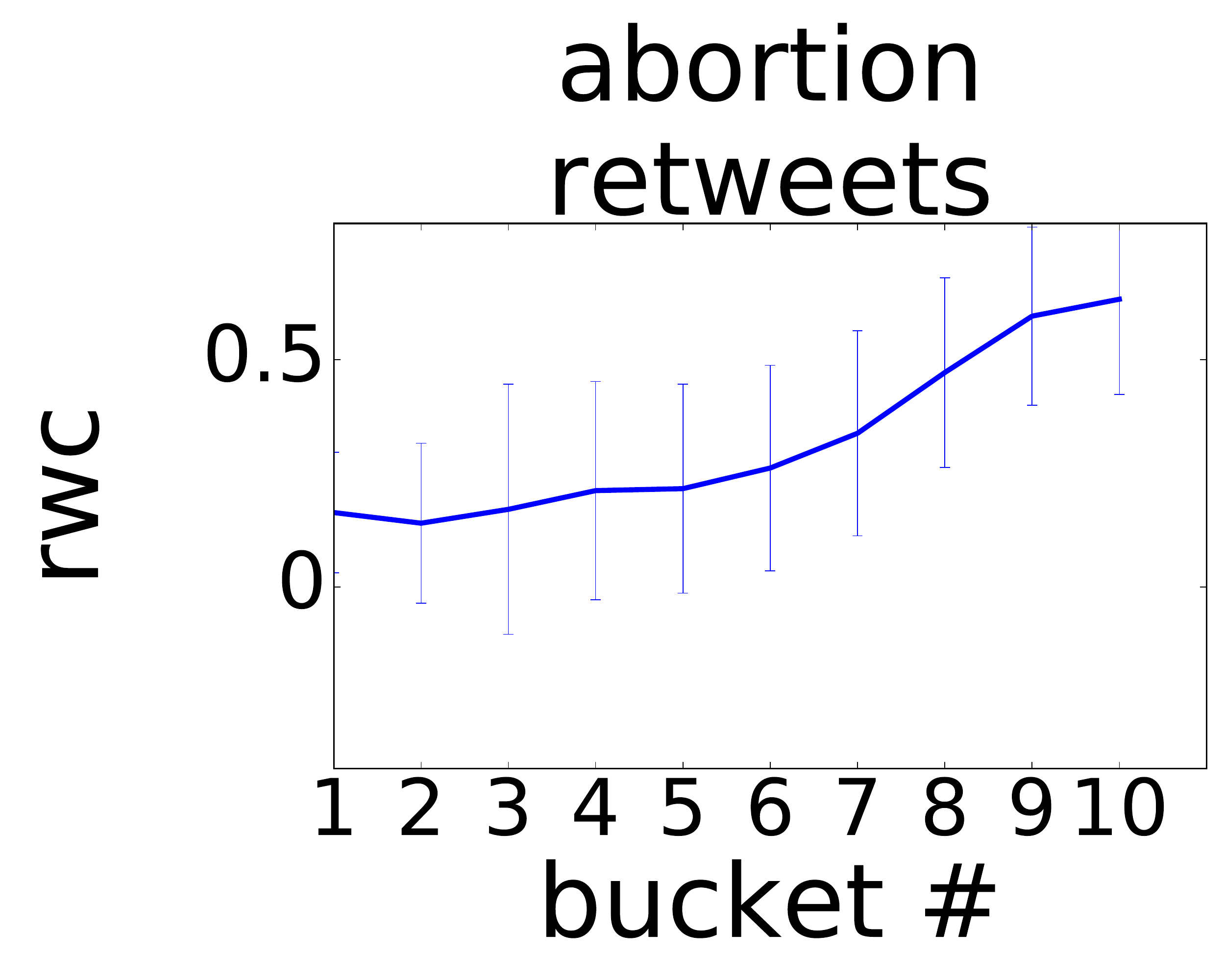}}
\end{minipage}%
\begin{minipage}{.25\linewidth}
\centering
\subfloat{\label{}\includegraphics[width=\textwidth]{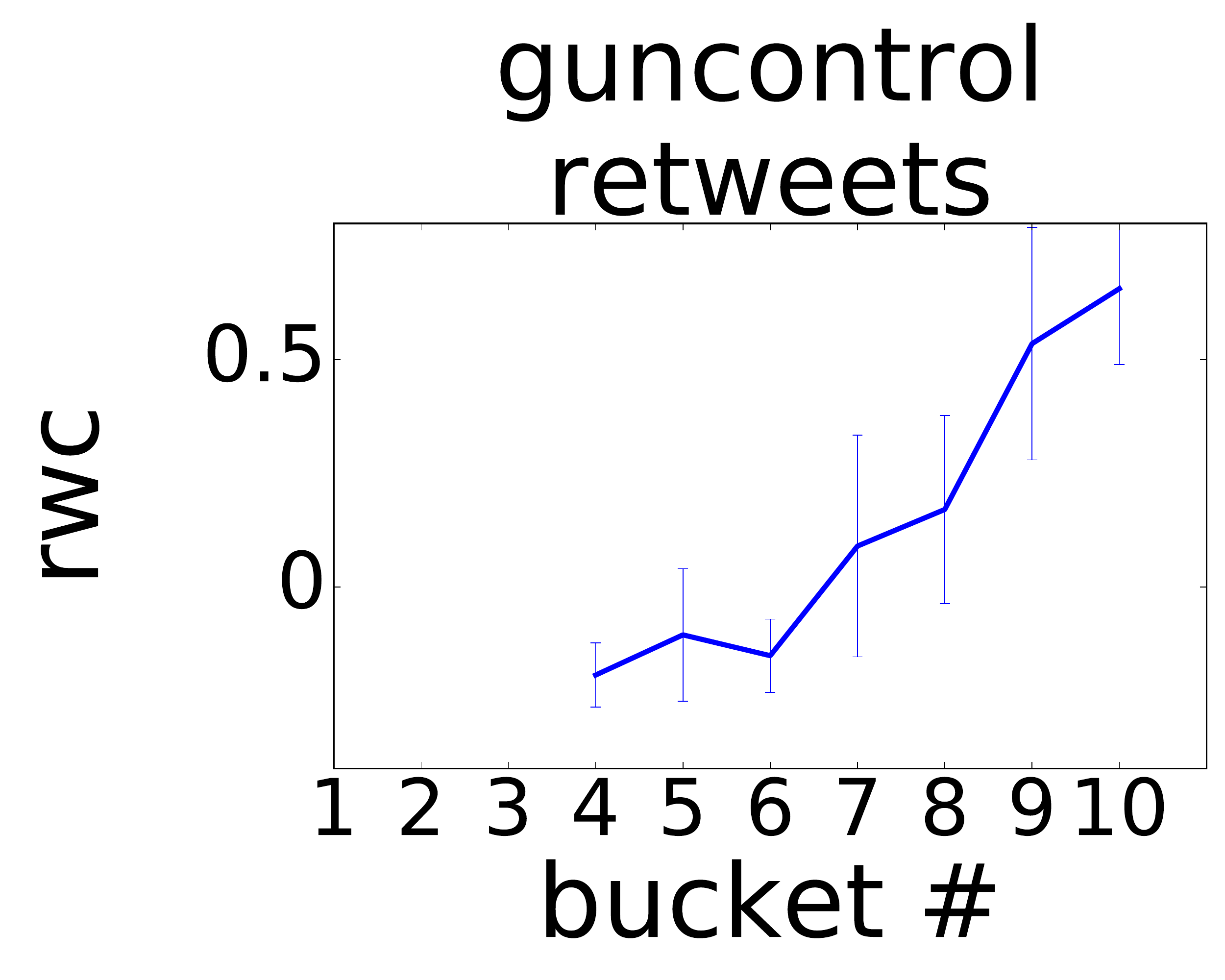}}
\end{minipage}%
\begin{minipage}{.25\linewidth}
\centering
\subfloat{\label{}\includegraphics[width=\textwidth]{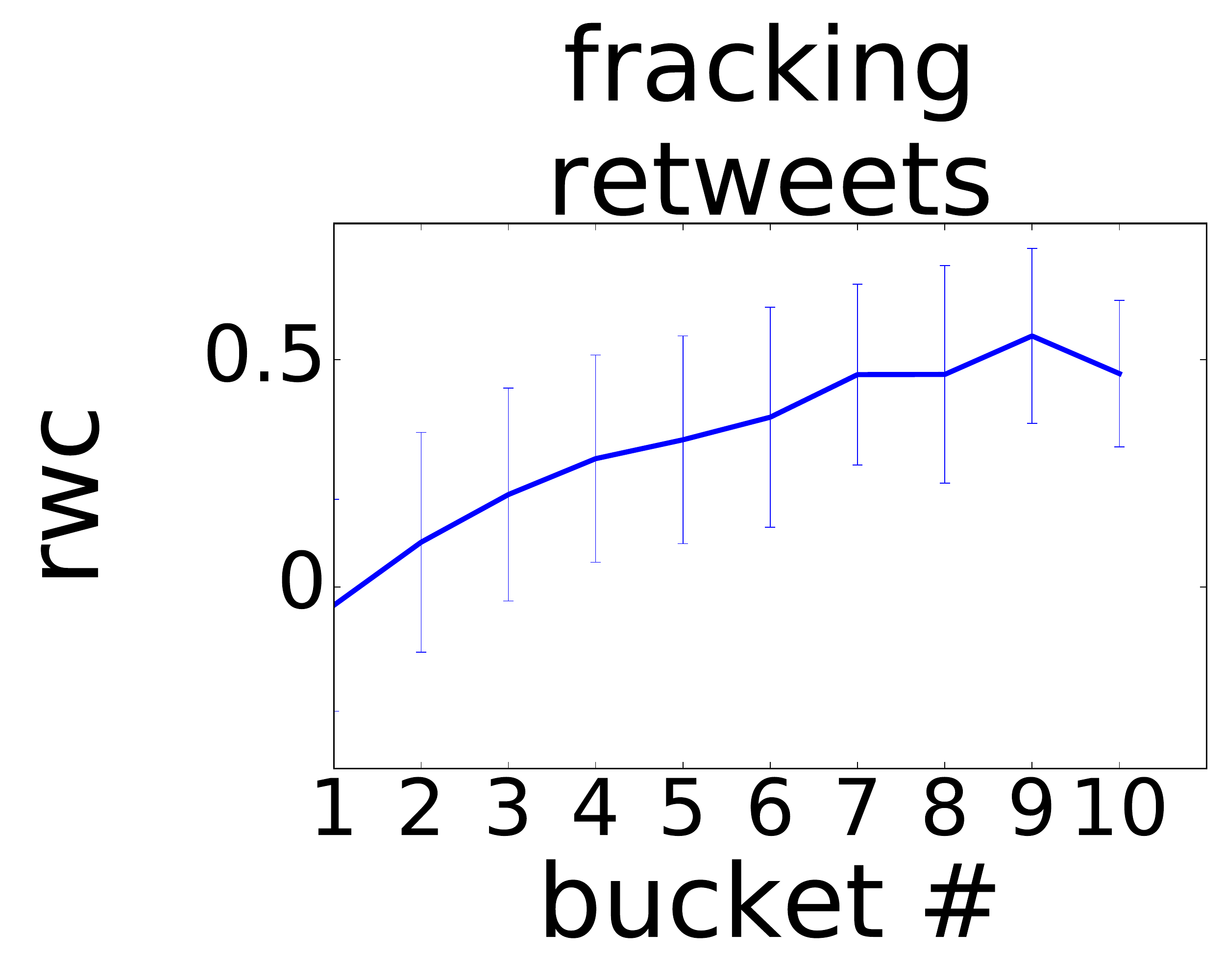}} 
\end{minipage}%
\caption{RWC score as a function of the activity in the retweet network. An increase in interest in the controversial topic corresponds to an increase in the controversy score of the retweet network.}
\label{fig:rwc}
\end{figure}

We observe a significant correlation between RWC score and interest in the topic.
Figure~\ref{fig:rwc} shows the RWC score as a function of the quantiles of the network by retweet volume (as explained in the previous section).
There is a clear increasing trend, which is consistent across topics.
This trend suggests that increased interest in the topic is correlated with an increase in controversy of the debate, 
and increased polarization of the retweet networks for the two sides.
Conversely, reply networks are sparser and more disconnected, thus, 
the RWC score is not meaningful in this case (not shown due to space constraints).
This difference is expected, and was already observed in the work that  introduced RWC~\citep{garimella2016quantifying}.

\begin{figure}[tb]
\centering
\begin{minipage}{.25\linewidth}
\centering
\subfloat{\label{}\includegraphics[width=\textwidth]{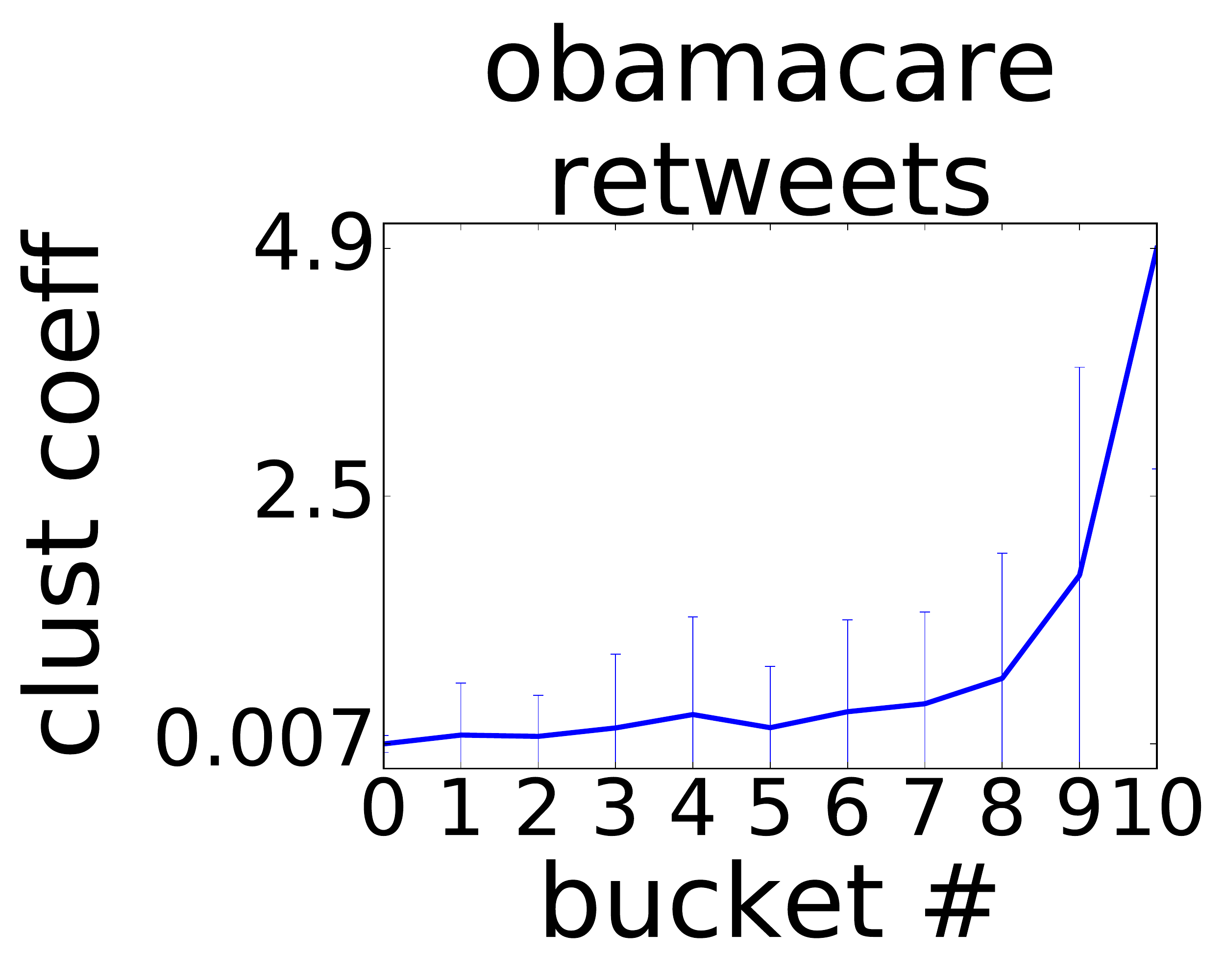}}
\end{minipage}%
\begin{minipage}{.25\linewidth}
\centering
\subfloat{\label{}\includegraphics[width=\textwidth]{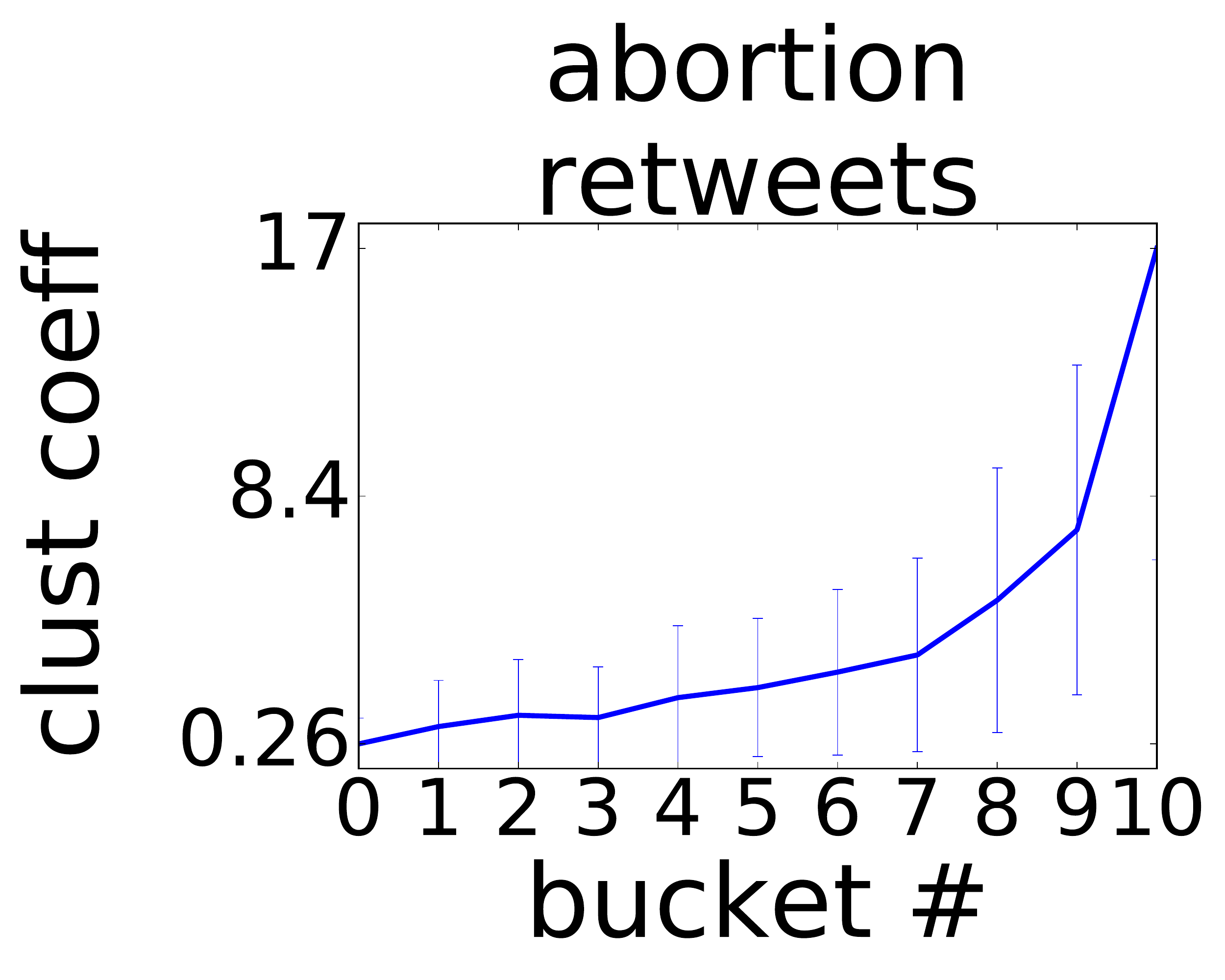}}
\end{minipage}%
\begin{minipage}{.25\linewidth}
\centering
\subfloat{\label{}\includegraphics[width=\textwidth]{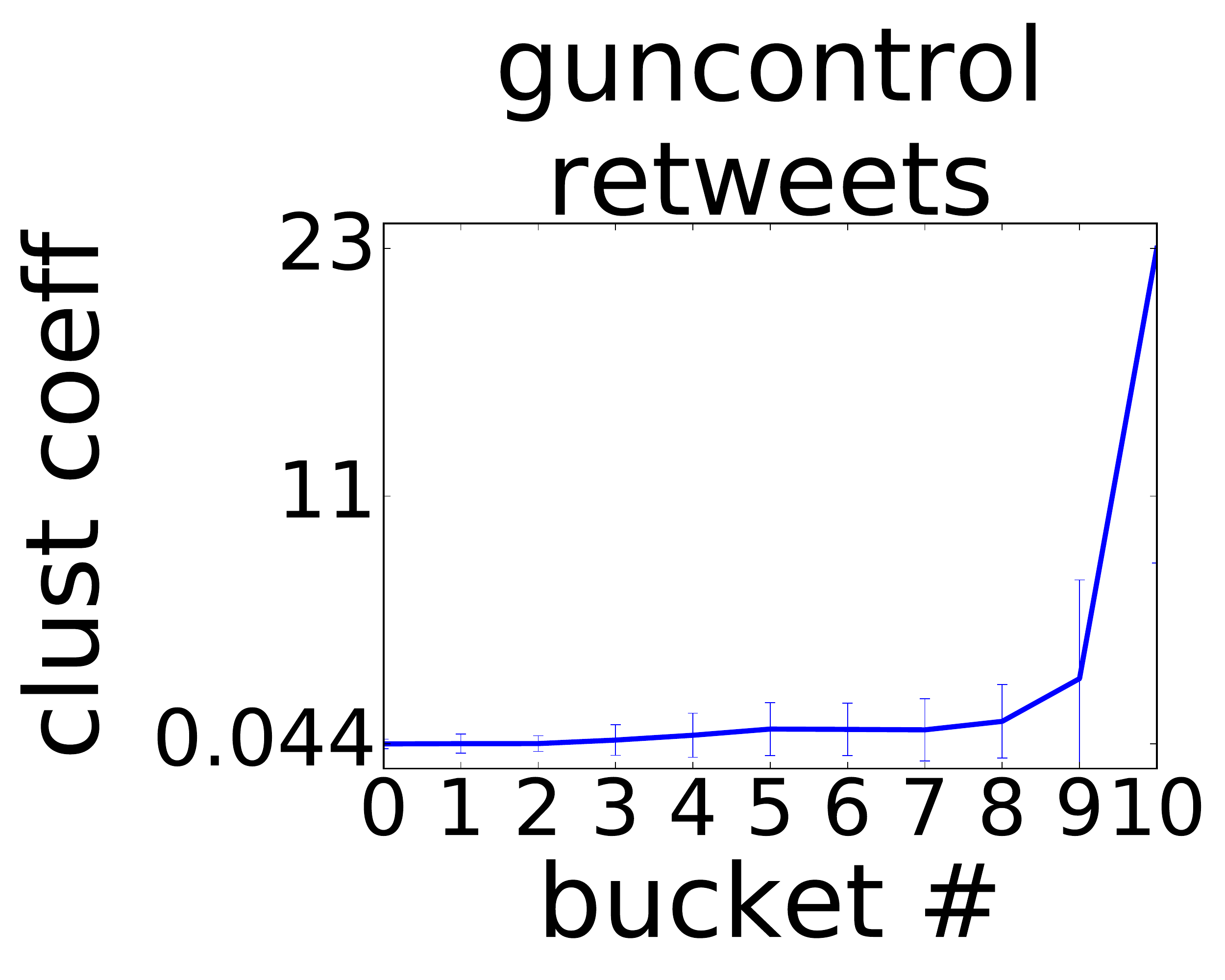}}
\end{minipage}%
\begin{minipage}{.25\linewidth}
\centering
\subfloat{\label{}\includegraphics[width=\textwidth]{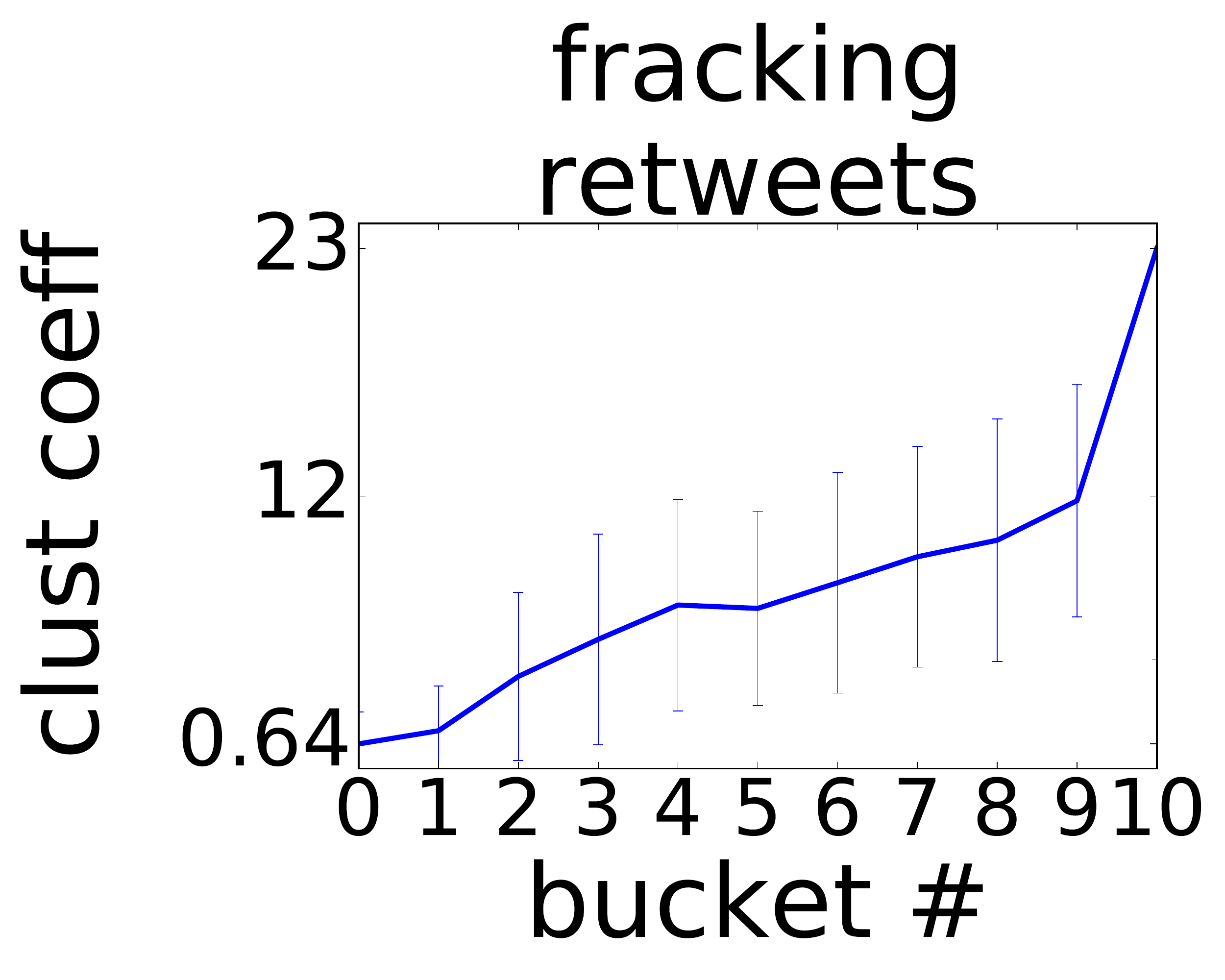}}
\end{minipage}%
\caption{Average clustering coefficient of as a function of the activity in the retweet network. Spikes in interest correspond to an increase in the clustering coefficient on both sides of the discussion, which indicates the retweet networks tend to close up.}
\label{fig:cc-rt}
\end{figure}

A similar result can be observed for the clustering coefficient, as shown in Figure~\ref{fig:cc-rt}.
As the interest in the topic increases, the two sides tend to \emph{turtle up}, and form a more close-knit retweet network.
This result suggests that the \emph{echo chamber} phenomenon gets stronger when the discussion sparks.
Our finding is also consistent with results by~\citet{romero2016social}.
As for the previous measure, the clustering coefficient does not show a significant pattern for the reply networks.
Replies are often linked to dyadic interactions, while the clustering coefficient measures triadic ones, so we expect such a difference between the two types of network.

\begin{figure}[tb]
\centering
\begin{minipage}{.25\linewidth}
\centering
\subfloat{\label{}\includegraphics[width=\textwidth]{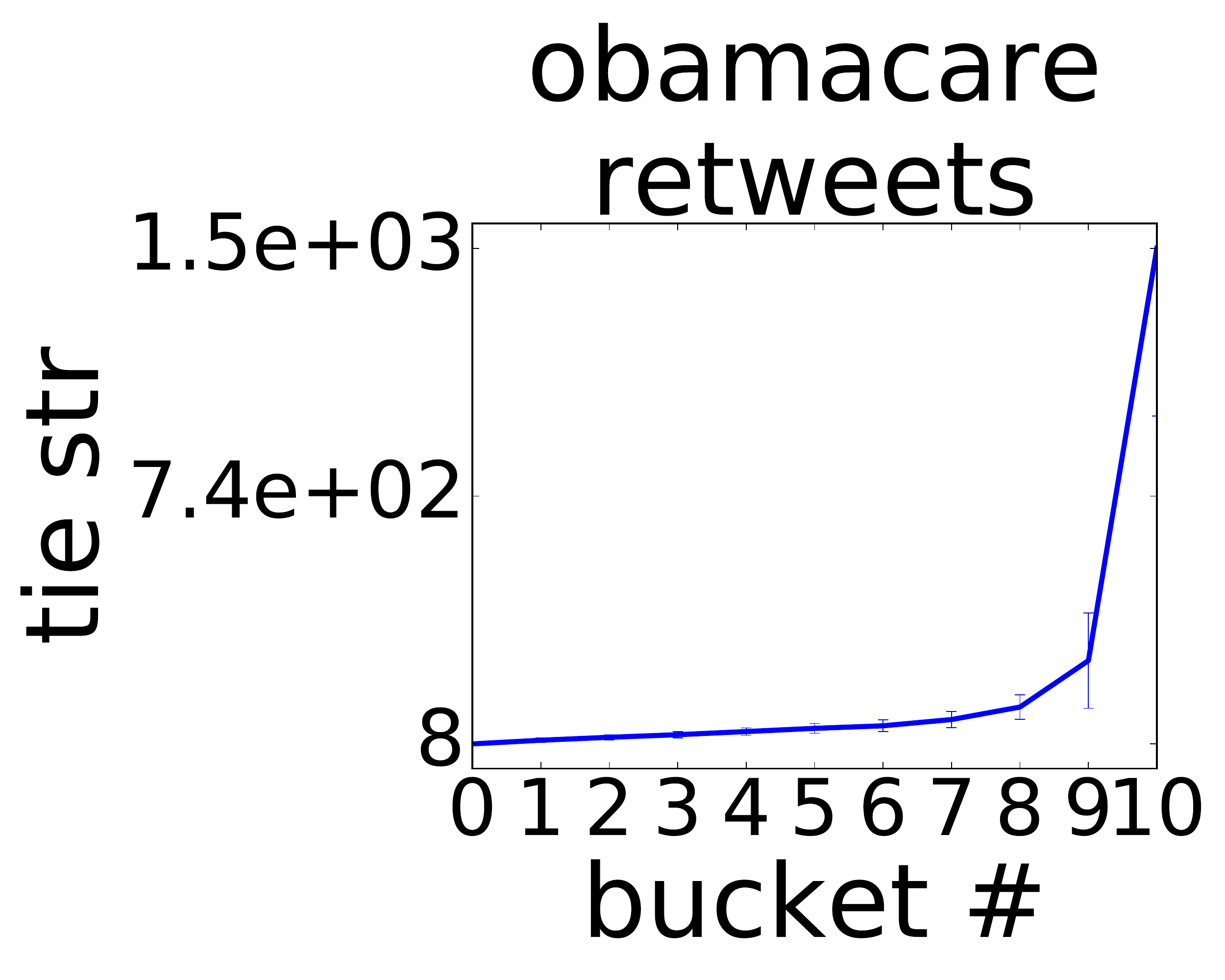}}
\end{minipage}%
\begin{minipage}{.25\linewidth}
\centering
\subfloat{\label{}\includegraphics[width=\textwidth]{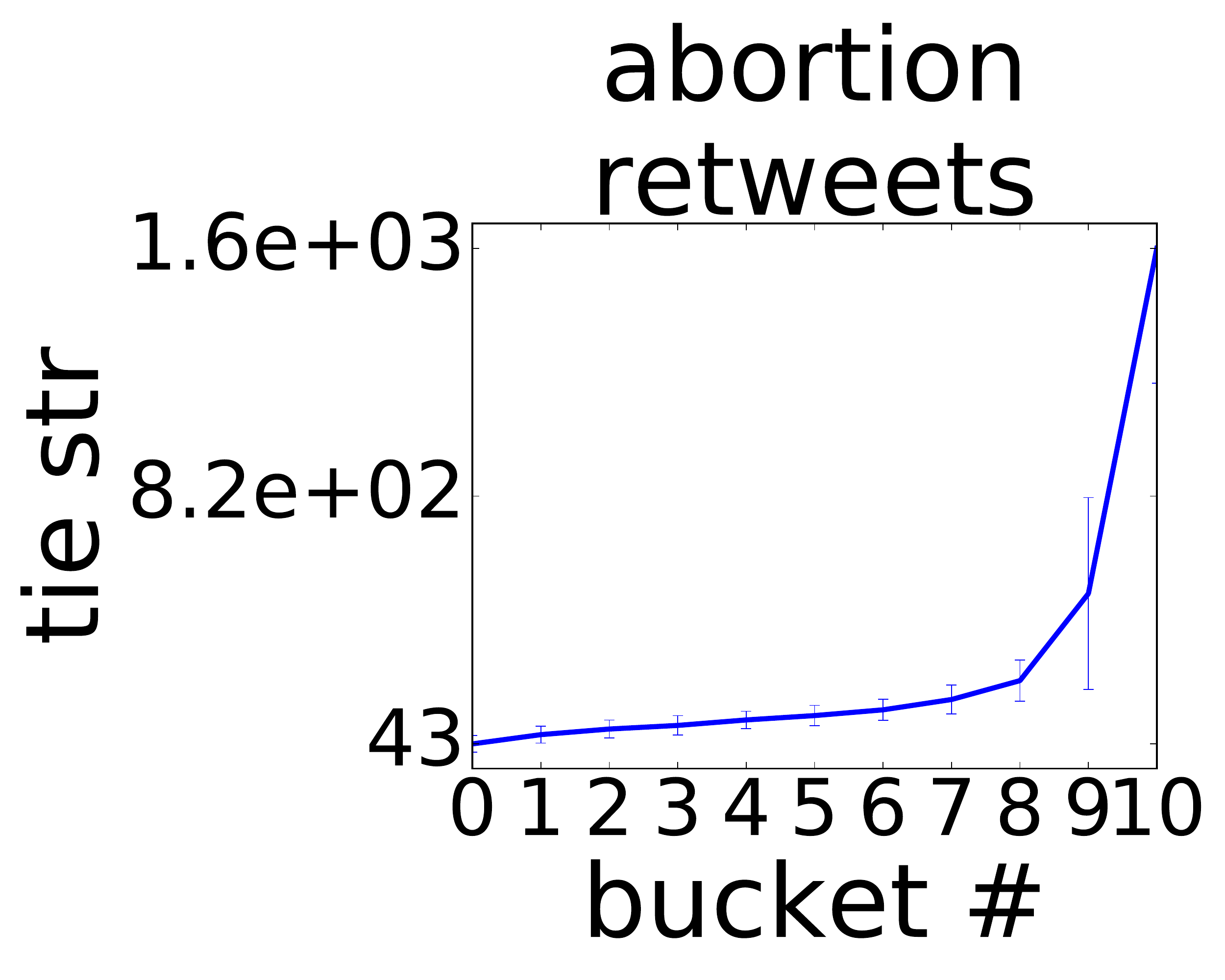}}
\end{minipage}%
\begin{minipage}{.25\linewidth}
\centering
\subfloat{\label{}\includegraphics[width=\textwidth]{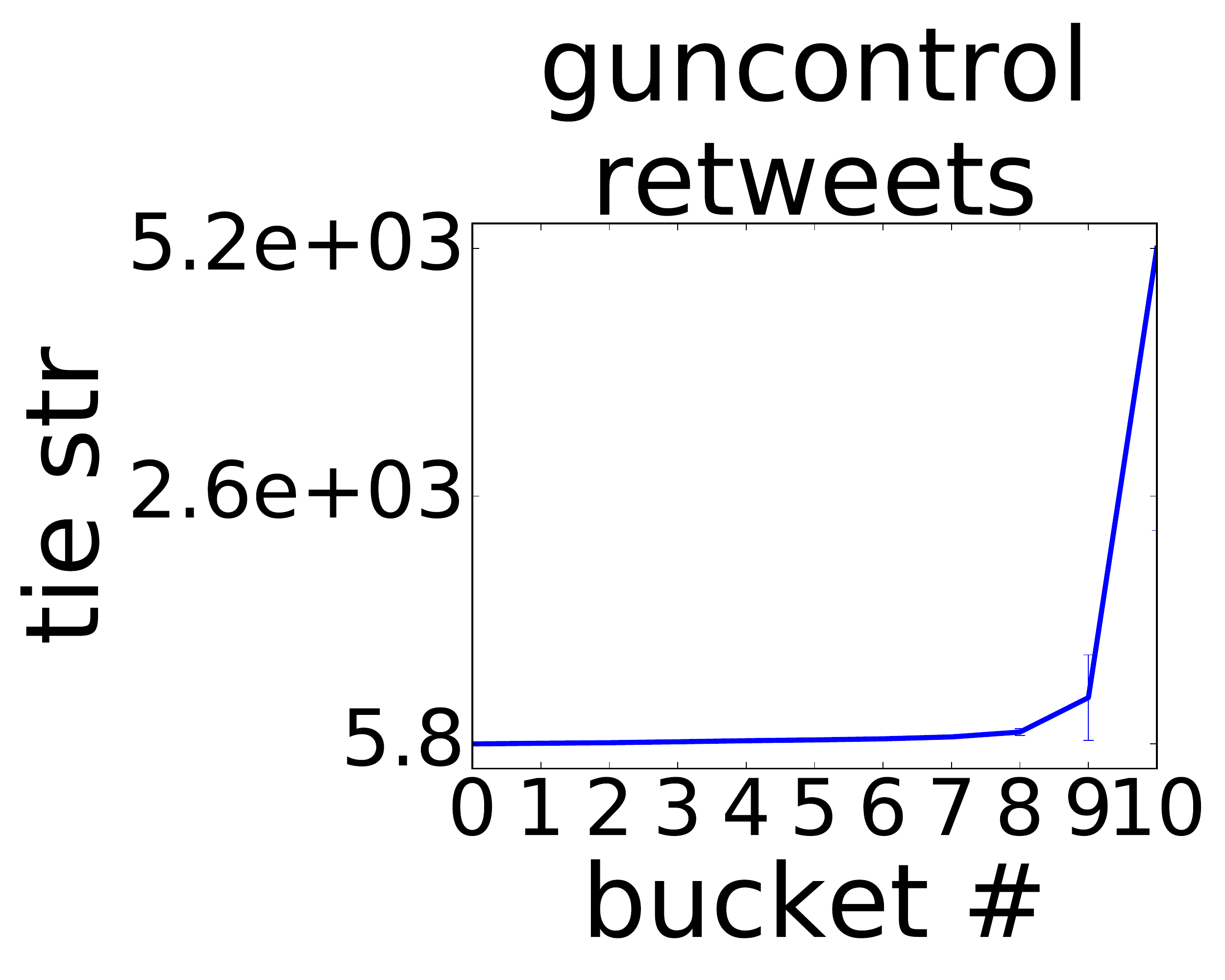}}
\end{minipage}%
\begin{minipage}{.25\linewidth}
\centering
\subfloat{\label{}\includegraphics[width=\textwidth]{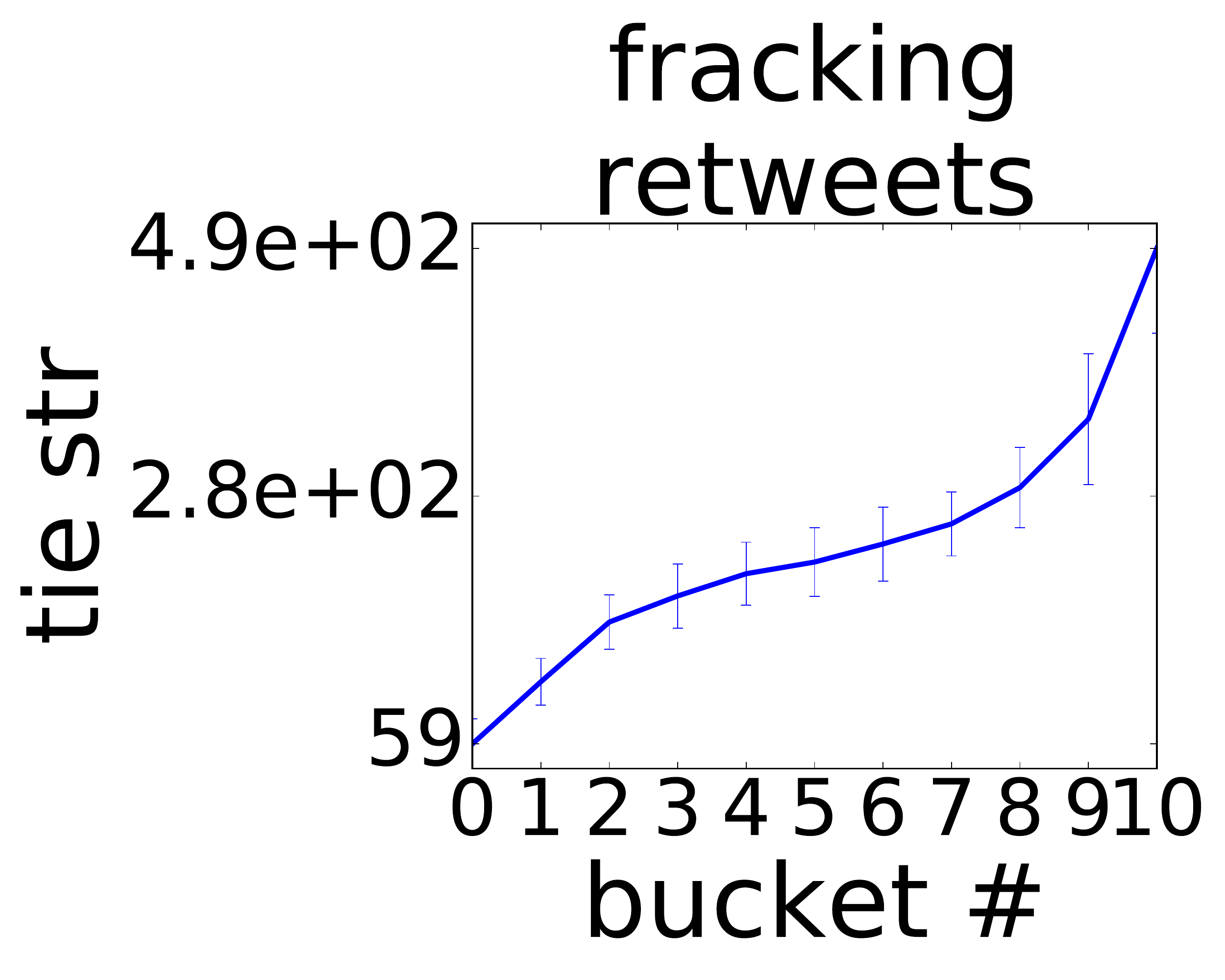}}
\end{minipage}%
\caption{Tie strength as a function of the activity in the retweet network. Spikes in activity correspond to more interaction with stronger ties, which indicates a closing up of the retweet network.}
\label{fig:ts-rt}
\end{figure}

\begin{figure}[tb]
\centering
\begin{minipage}{.25\linewidth}
\centering
\subfloat{\label{}\includegraphics[width=\textwidth]{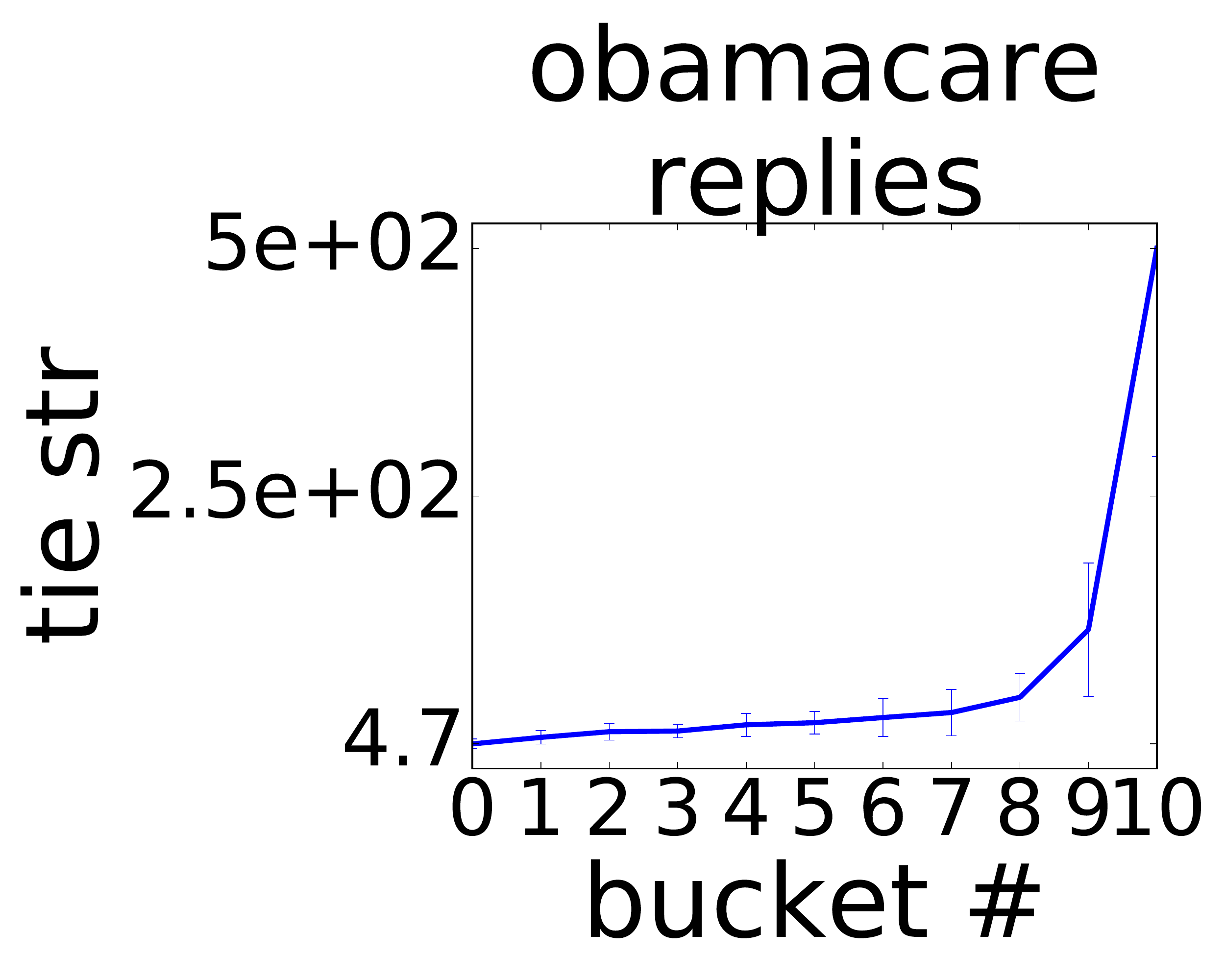}}
\end{minipage}%
\begin{minipage}{.25\linewidth}
\centering
\subfloat{\label{}\includegraphics[width=\textwidth]{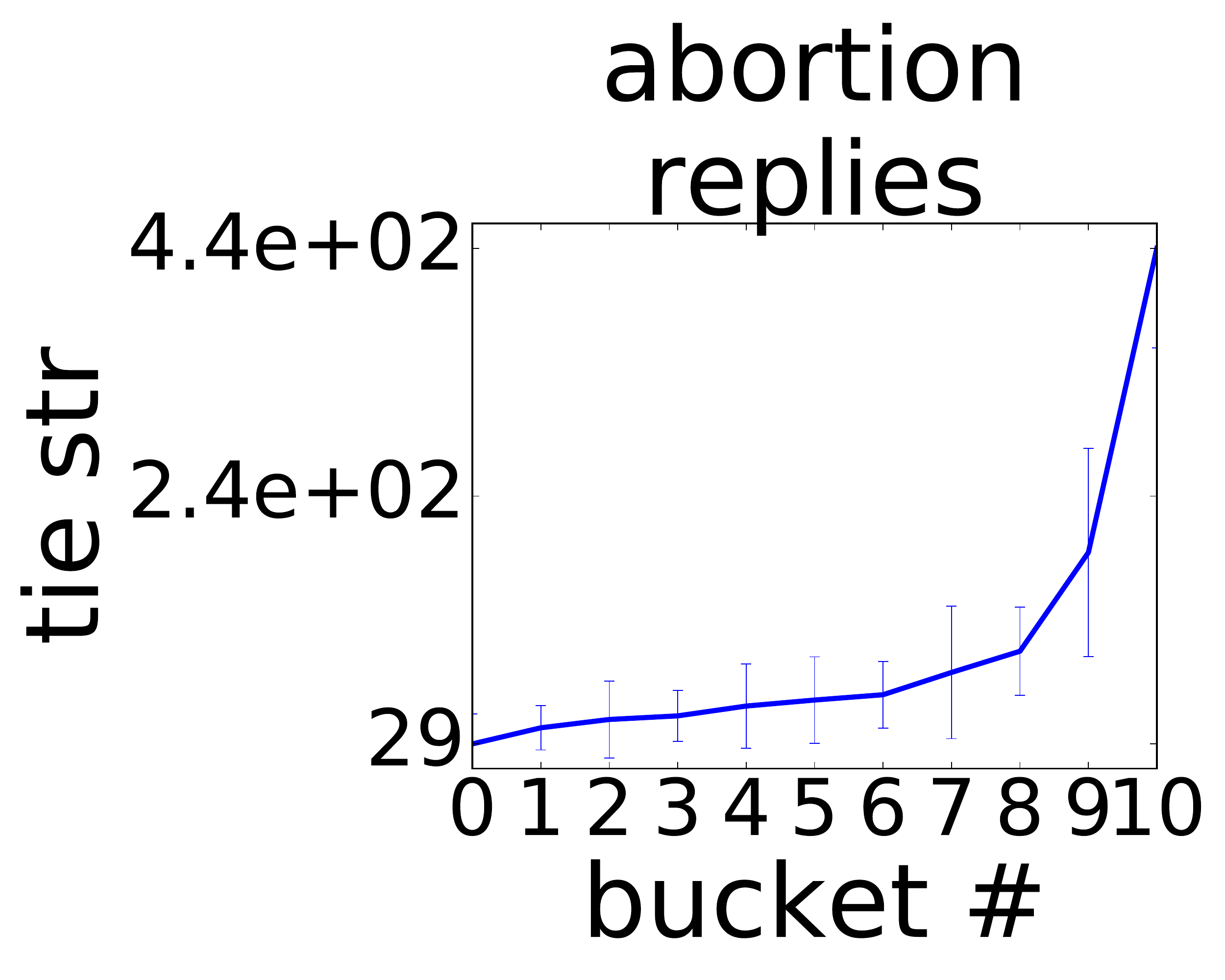}}
\end{minipage}%
\begin{minipage}{.25\linewidth}
\centering
\subfloat{\label{}\includegraphics[width=\textwidth]{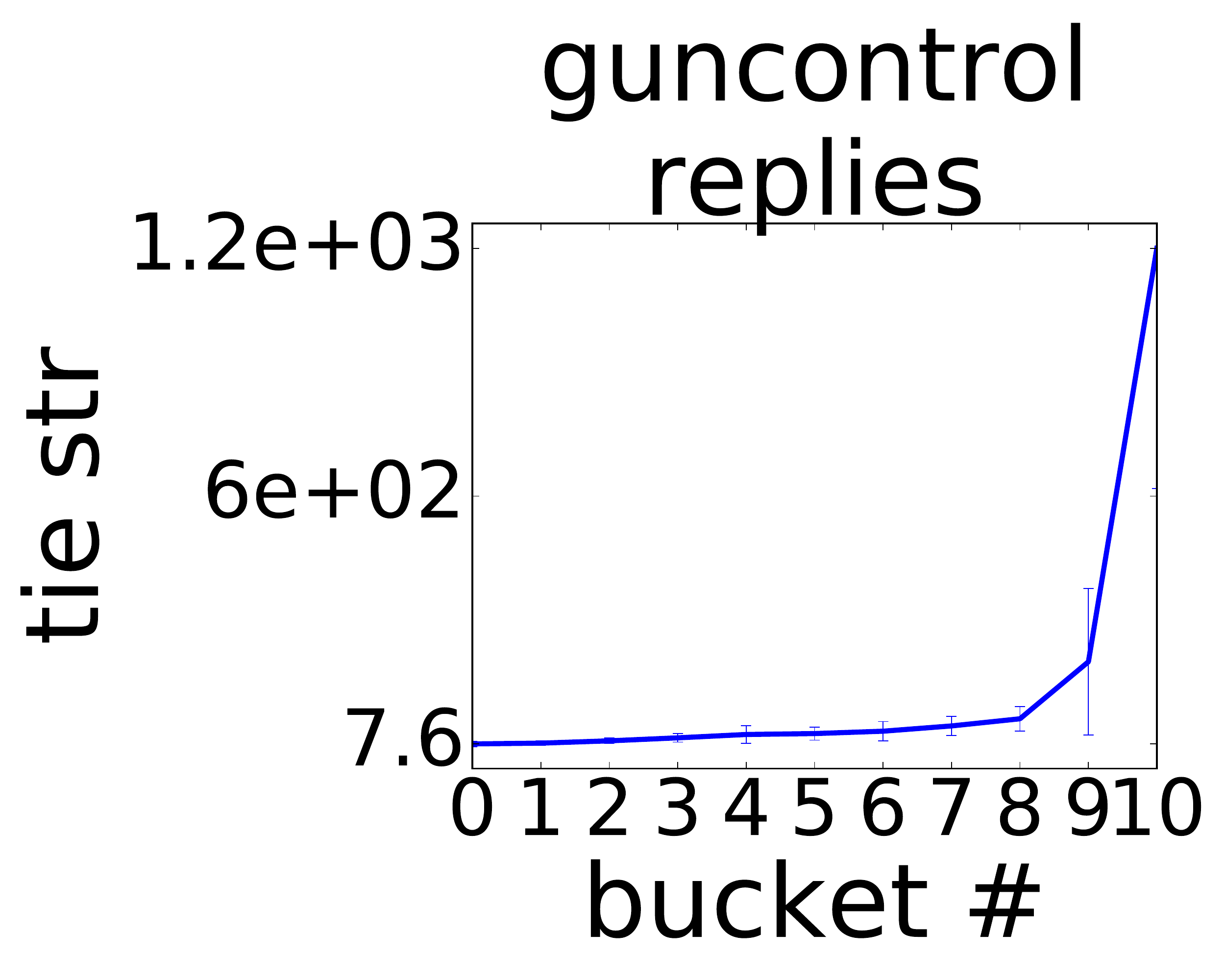}}
\end{minipage}%
\begin{minipage}{.25\linewidth}
\centering
\subfloat{\label{}\includegraphics[width=\textwidth]{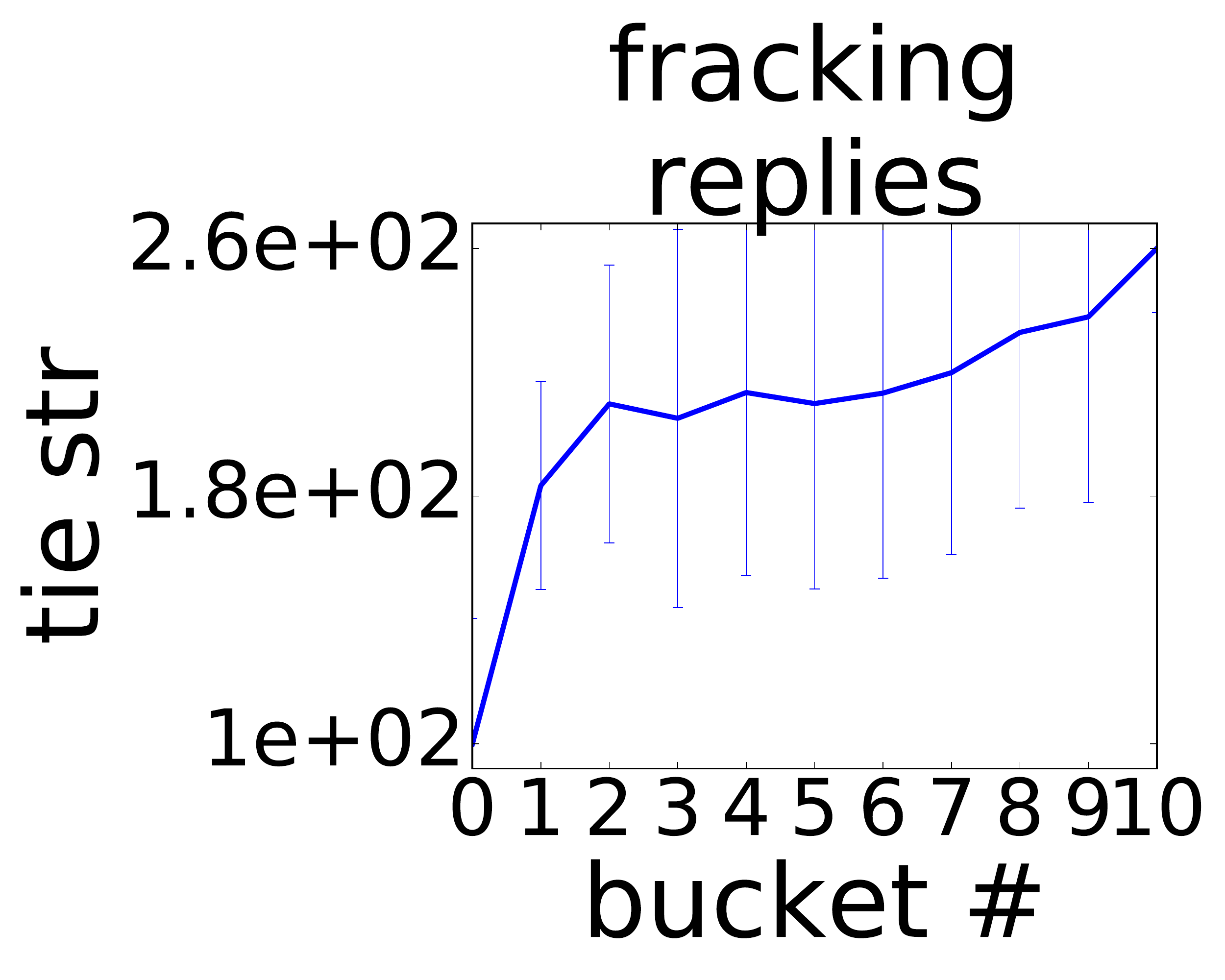}}
\end{minipage}%
\caption{Tie strength as a function of the activity in the reply network. Users tend to communicate proportionally more with closer ties when interest spikes, which reveals a further closing up of the network.}
\label{fig:ts-re}
\end{figure}

\begin{figure}[tb]
\centering
\begin{minipage}{.25\linewidth}
\centering
\subfloat{\label{}\includegraphics[width=\textwidth]{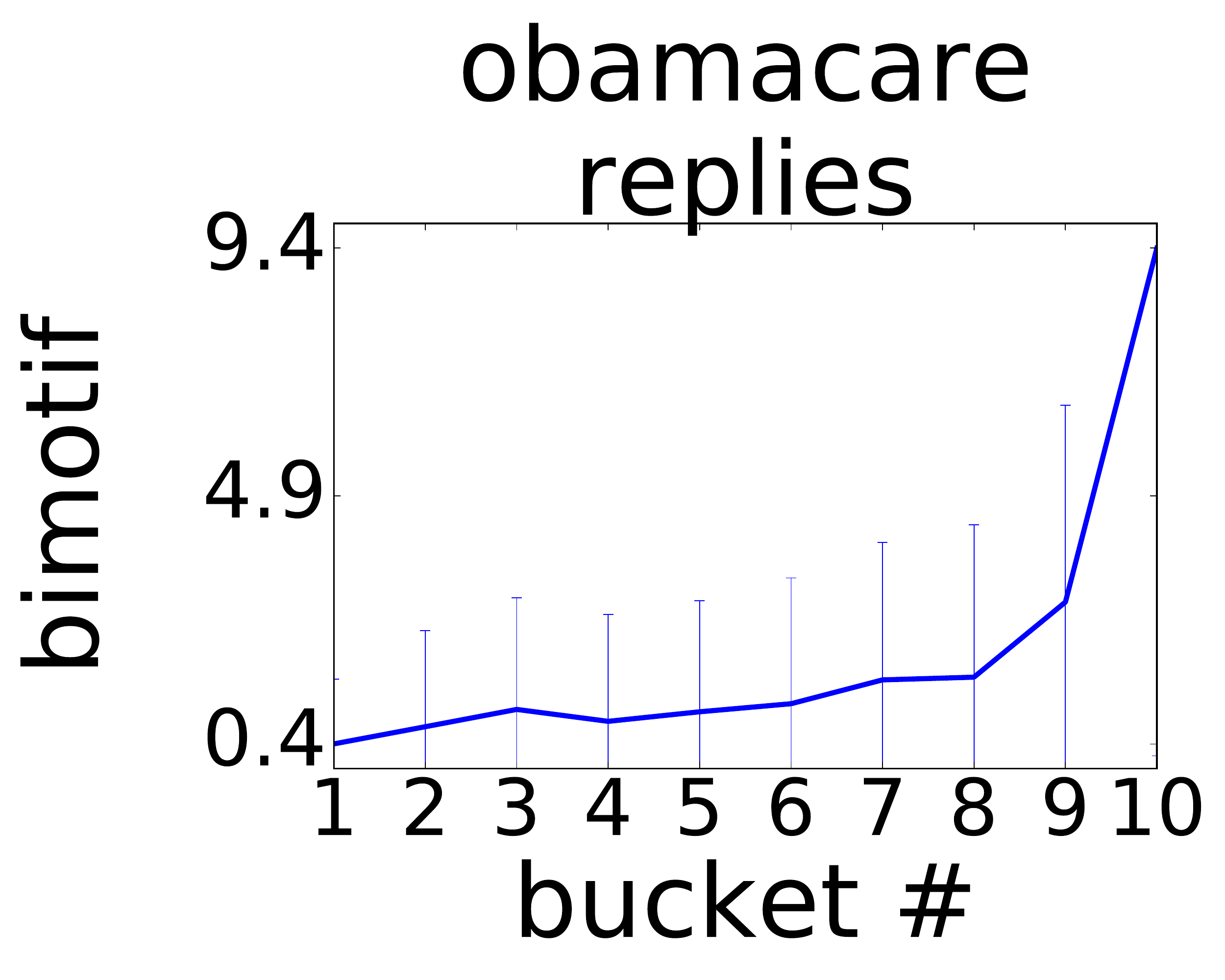}}
\end{minipage}%
\begin{minipage}{.25\linewidth}
\centering
\subfloat{\label{}\includegraphics[width=\textwidth]{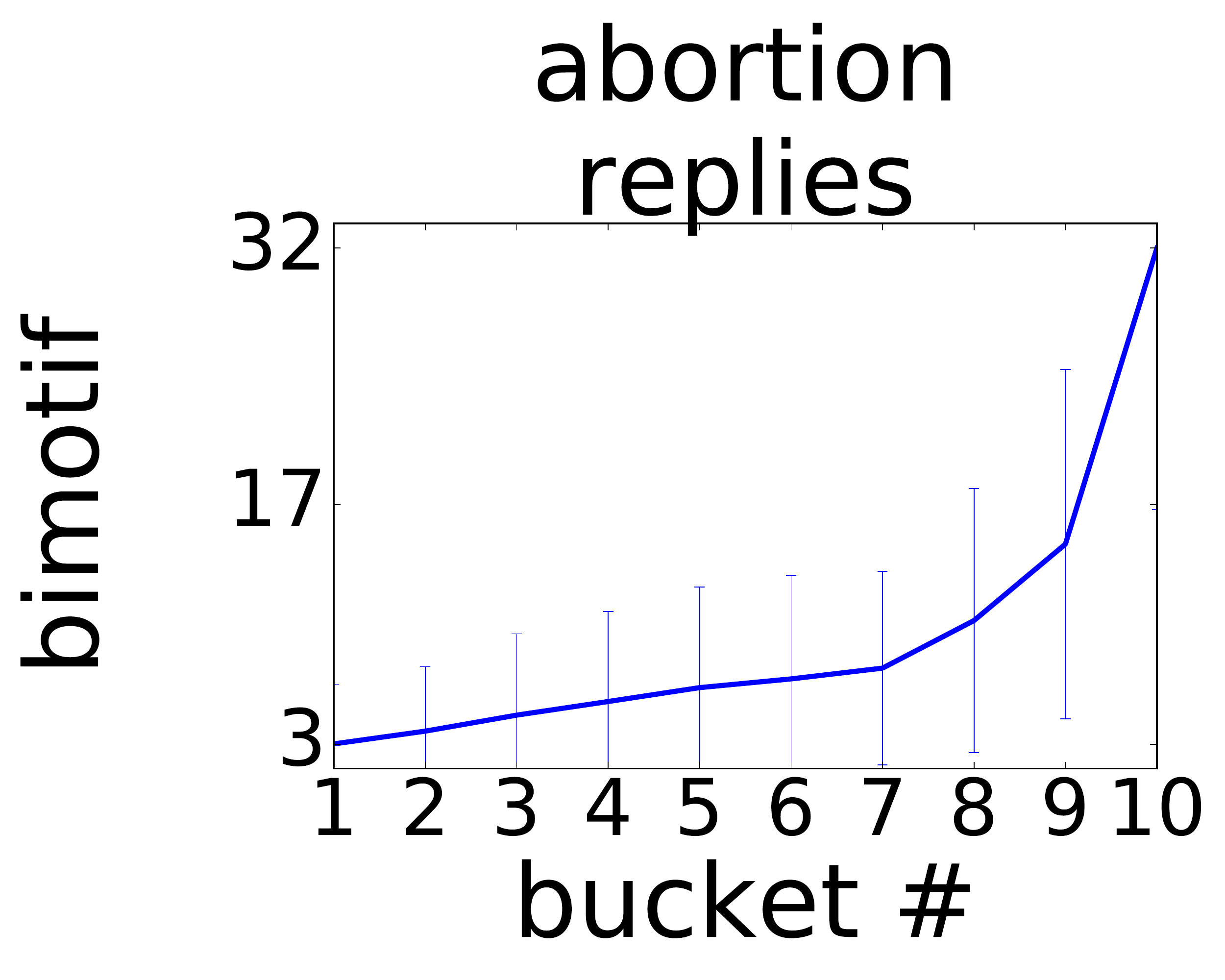}}
\end{minipage}%
\begin{minipage}{.25\linewidth}
\centering
\subfloat{\label{}\includegraphics[width=\textwidth]{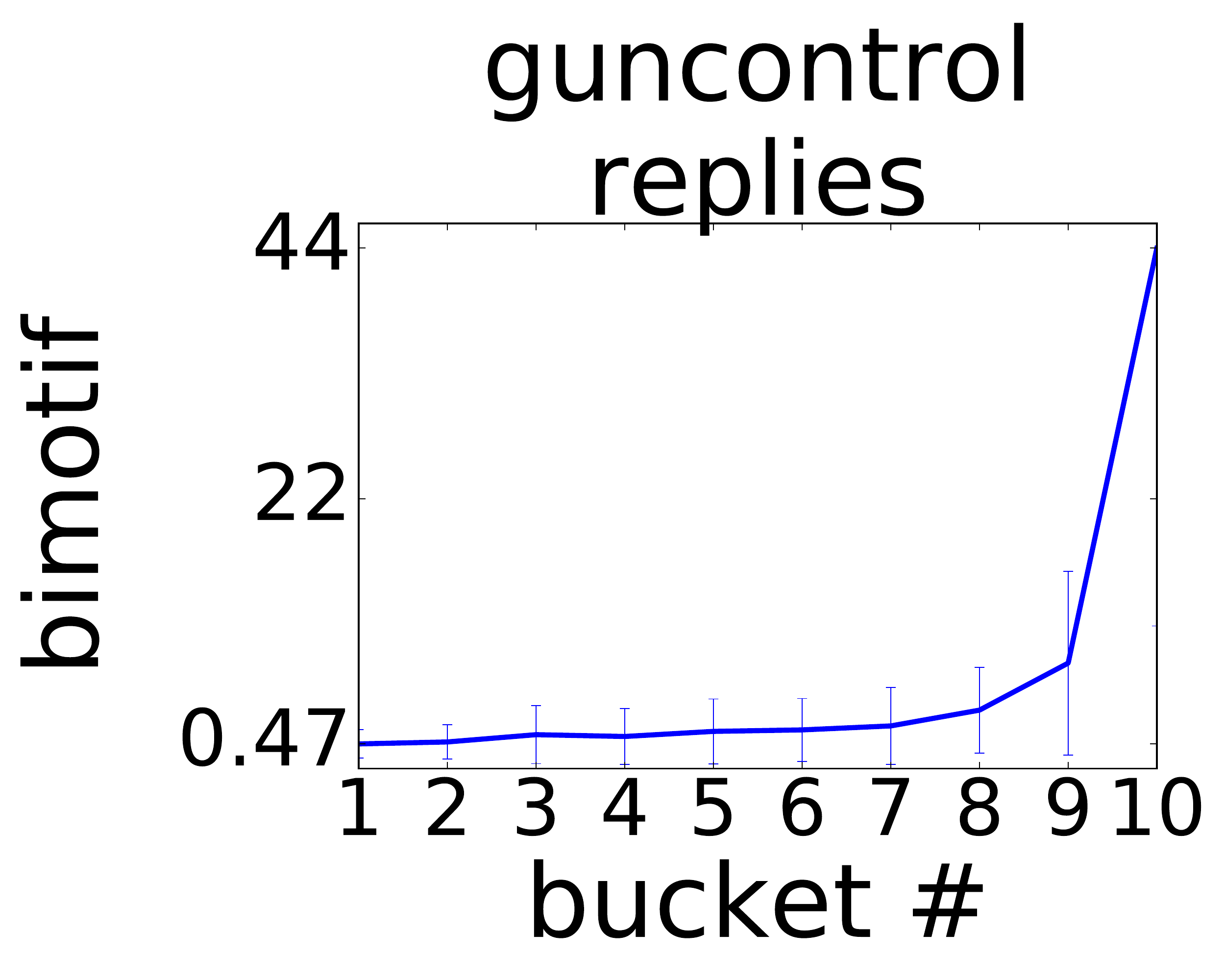}}
\end{minipage}%
\begin{minipage}{.25\linewidth}
\centering
\subfloat{\label{}\includegraphics[width=\textwidth]{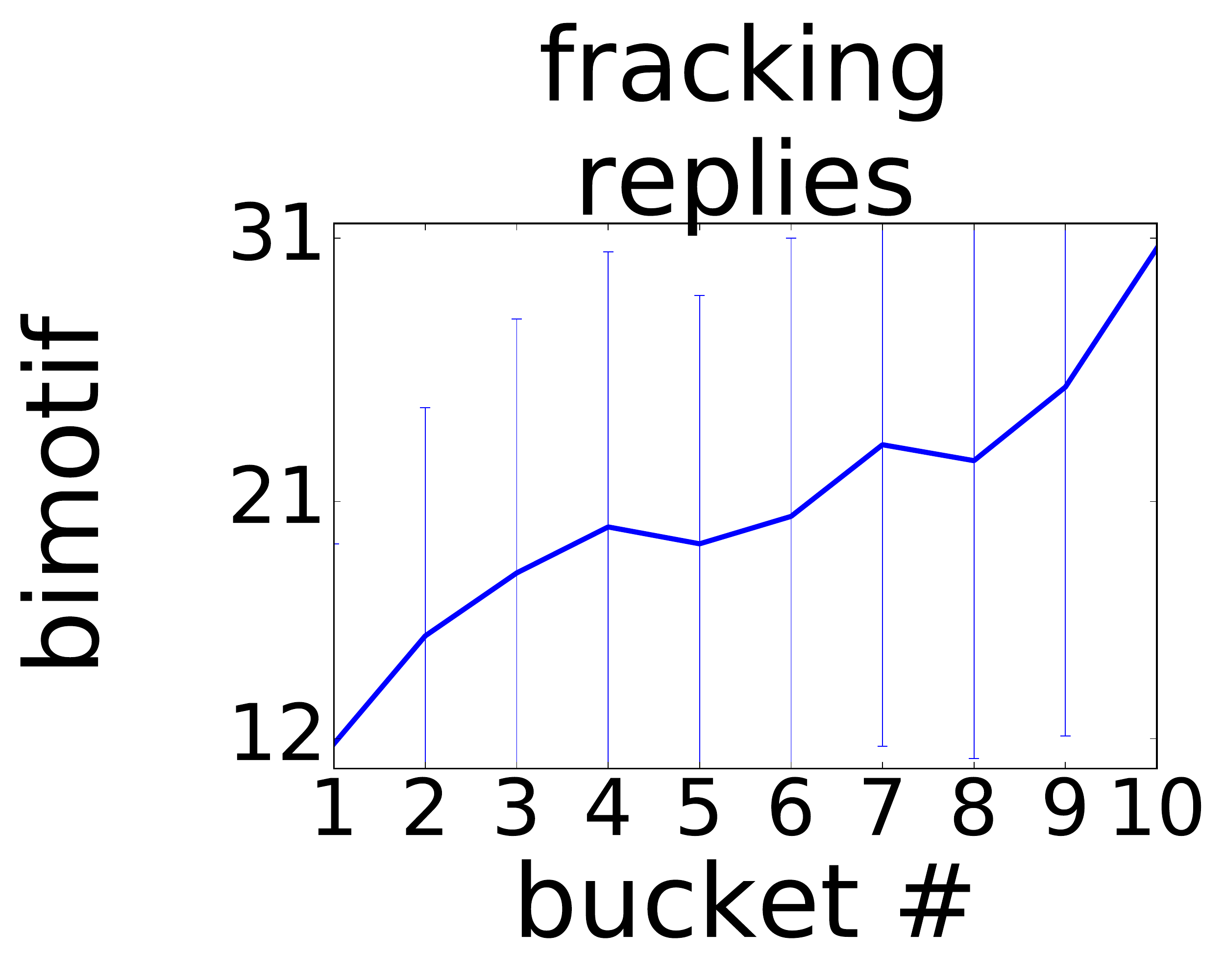}}
\end{minipage}%
\caption{Bimotifs as a function of the activity in the reply network. Users tend to reciprocate the communication more as the discussion intensifies.}
\label{fig:bi-re}
\end{figure}

In line with the above results, tie strength is correlated with retweet volume, as indicated by Figure~\ref{fig:ts-rt}.
When the discussion intensifies, users tend to endorse the opinions of their closest friends, or their trusted sources of information.
Again, this observation indicates a closing up of both sides when the debate gets heated.
Interestingly, a similar trend is present for the reply network, as shown in Figure~\ref{fig:ts-re}.
Differently from previous work, we find an increase of communication of users with their strong ties, rather than with weak ties or users of the opposing side.
We also observe an increase in back-and-forth communication, 
indicating a dialogue between users of the same side.
Figure~\ref{fig:bi-re} shows an increase in bimotifs in the reply network when the discussion intensifies.
This measure is inconclusive for the retweet network, for the reasons mentioned above.

Finally, when calculating the fractions of \emph{within-side edges} and \emph{across-side edges} for \emph{across sides edge composition}, we find that reply networks typically contain higher proportions of across-side activity compared to retweet networks, consistently with earlier work.
In fact, for retweet networks, almost all edges are classified as  \emph{within-side edges}. 
Interestingly, we also find that these proportions do not change significantly as the volume increases. The same is true for the \emph{cross-side openness} measure (not shown).

\subsection{Content}
\label{sec:content_findings}
\begin{figure}[t]
\centering
\begin{minipage}{.25\linewidth}
\centering
\subfloat{\label{}\includegraphics[width=\textwidth]{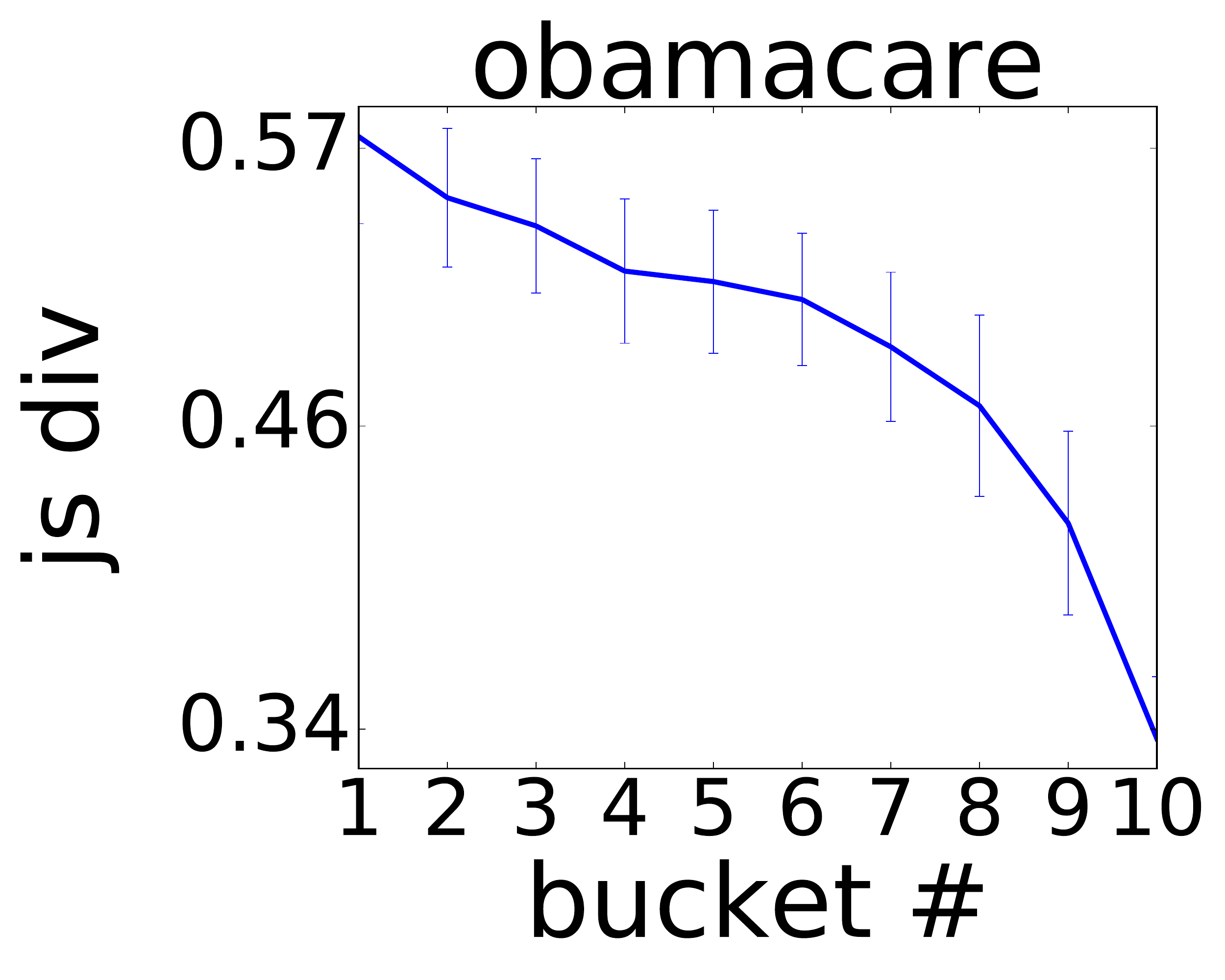}}
\end{minipage}%
\begin{minipage}{.25\linewidth}
\centering
\subfloat{\label{}\includegraphics[width=\textwidth]{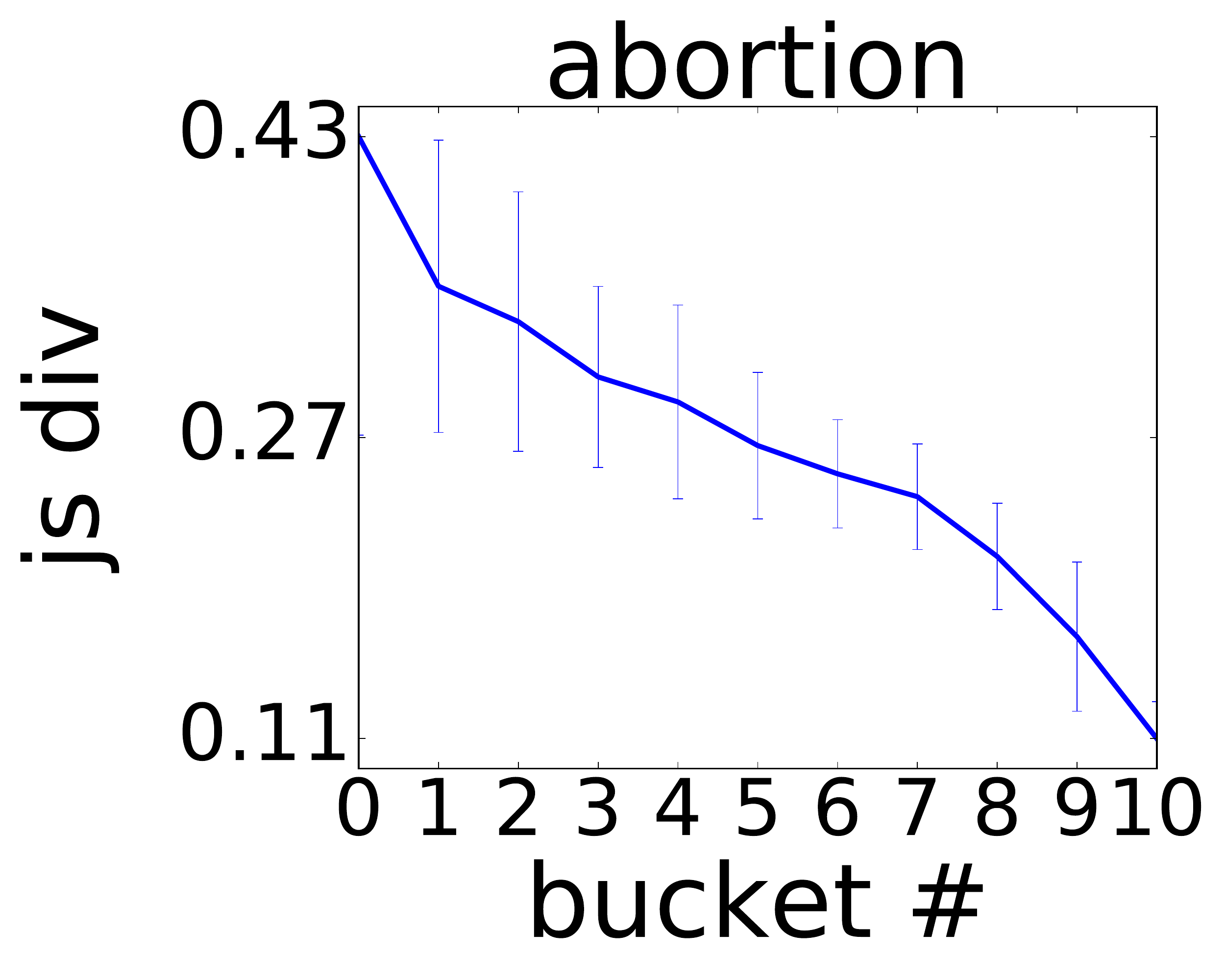}}
\end{minipage}%
\begin{minipage}{.25\linewidth}
\centering
\subfloat{\label{}\includegraphics[width=\textwidth]{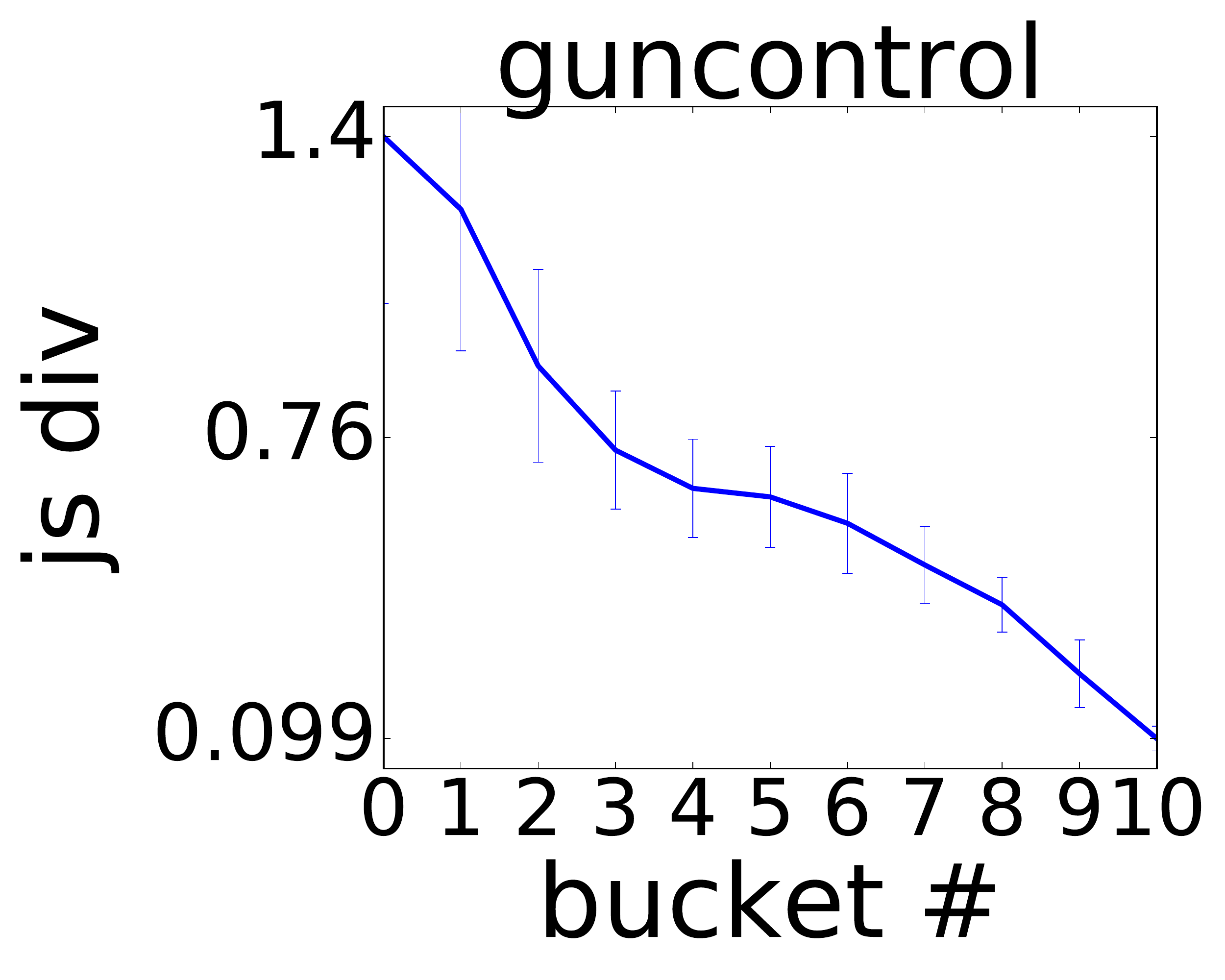}}
\end{minipage}%
\begin{minipage}{.25\linewidth}
\centering
\subfloat{\label{}\includegraphics[width=\textwidth]{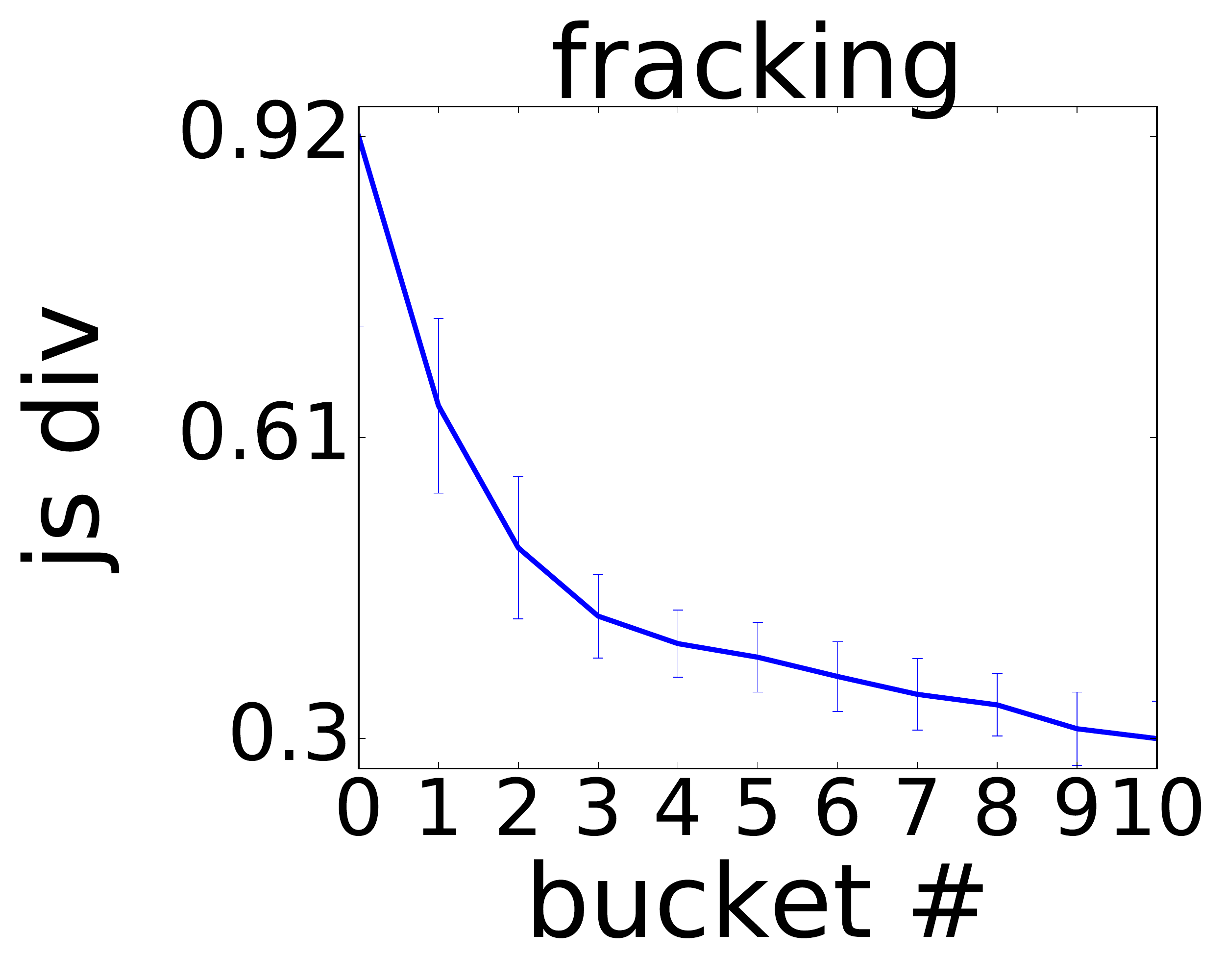}}
\end{minipage}%
\caption{Jensen-Shannon divergence of the lexicon between the two sides as a function of network activity. As the interest in the topic rises, the lexicon used by the two sides tends to converge.}
\label{fig:js}
\end{figure}

Let us now switch our attention to the content measures.
Recall that for these measures we do not distinguish between retweet and reply networks, but only between the two sides of the discussion.
The main observation is that the Jensen-Shannon divergence between the two sides decreases, as shown by Figure~\ref{fig:js}.
This decrease indicates that the lexicon of the two sides tends to converge.
The cause of this phenomenon might be the participation of casual users to the discussions, who contribute a more general lexicon to the discussion.
Alternatively, the cause might be in the event that sparks the discussion, which brings the whole network to adopt similar lexicon to speak about it, i.e., there is an event-based convergence.

\begin{figure}[tb]
\centering
\begin{minipage}{.25\linewidth}
\centering
\subfloat{\label{}\includegraphics[width=\textwidth]{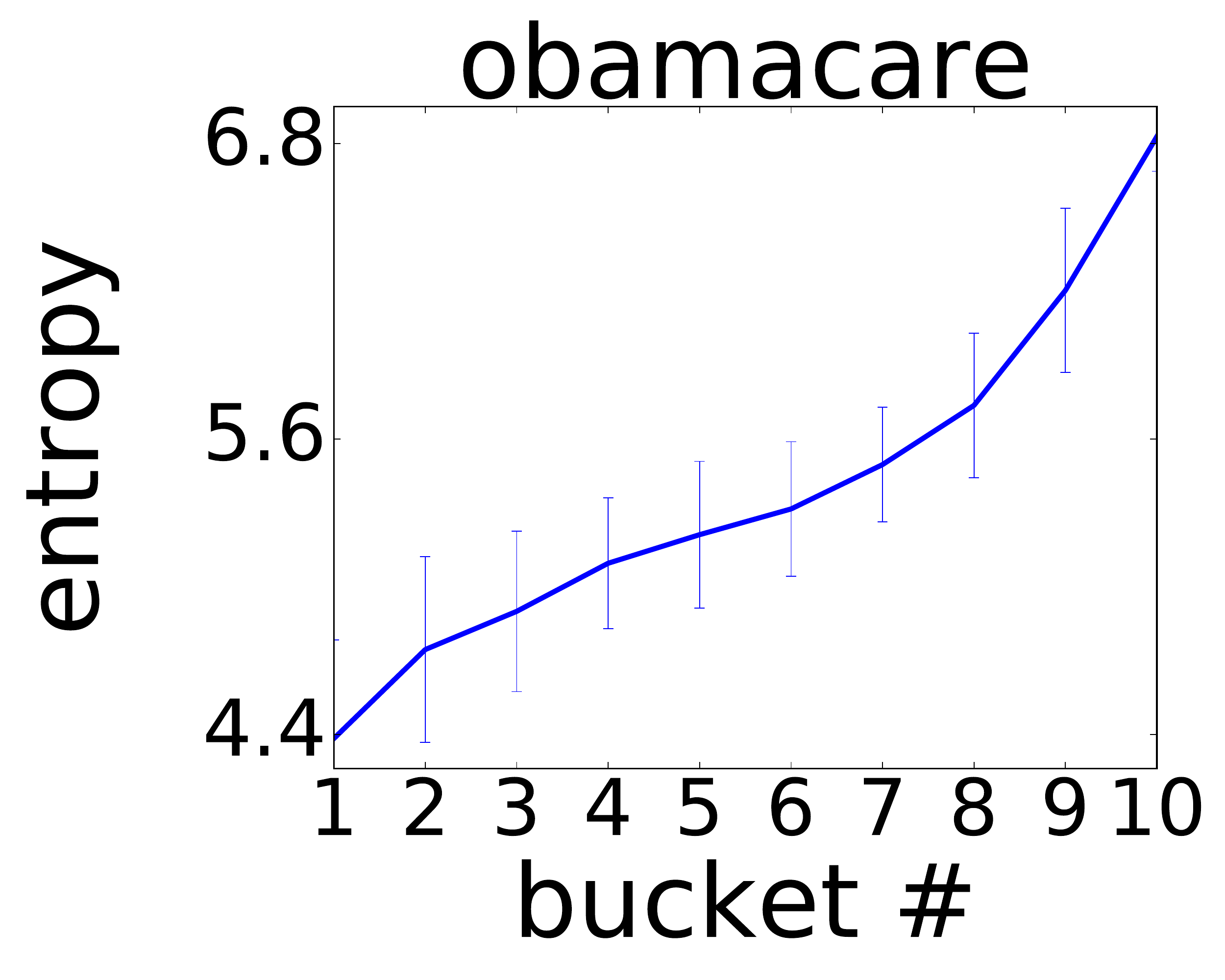}}
\end{minipage}%
\begin{minipage}{.25\linewidth}
\centering
\subfloat{\label{}\includegraphics[width=\textwidth]{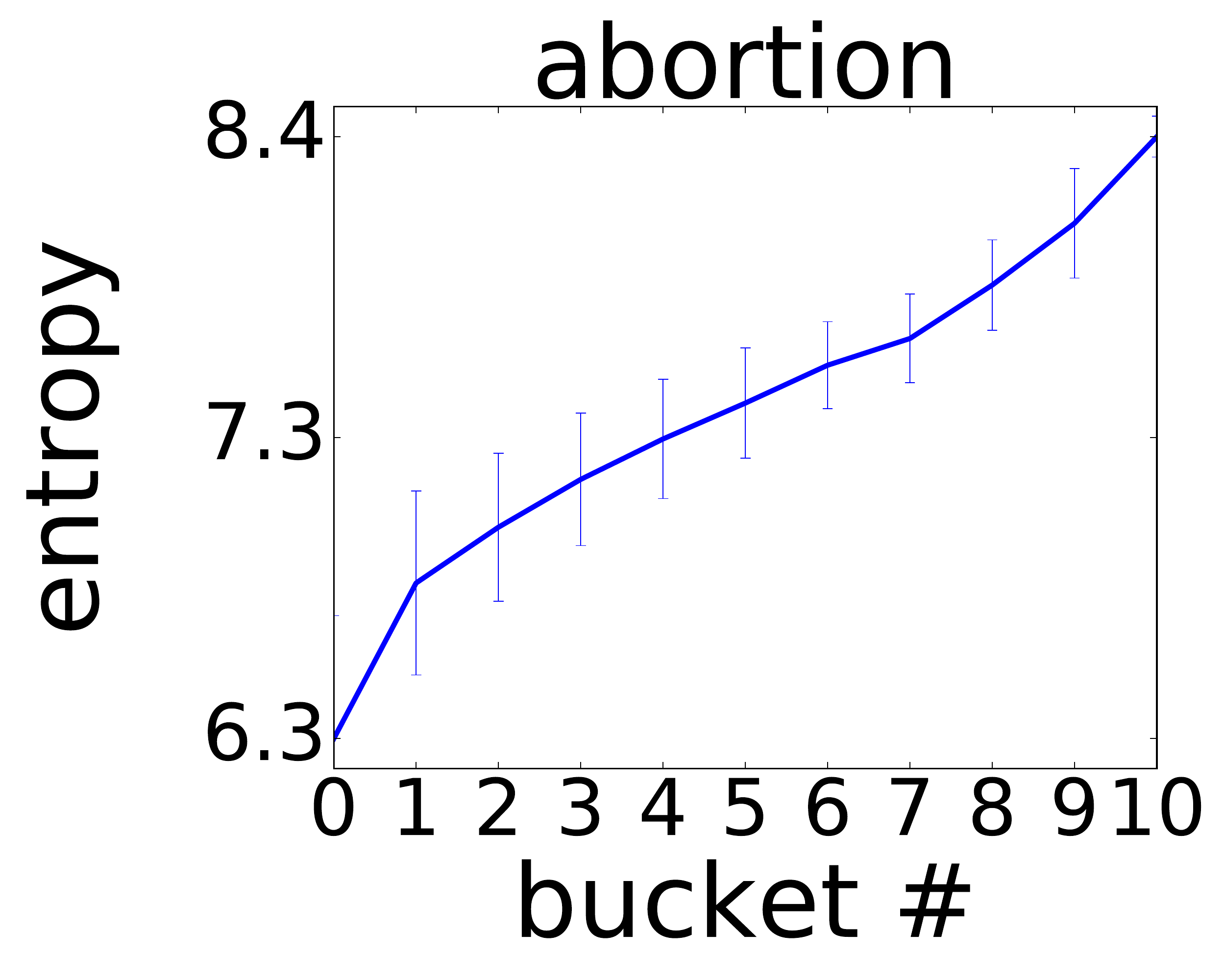}}
\end{minipage}%
\begin{minipage}{.25\linewidth}
\centering
\subfloat{\label{}\includegraphics[width=\textwidth]{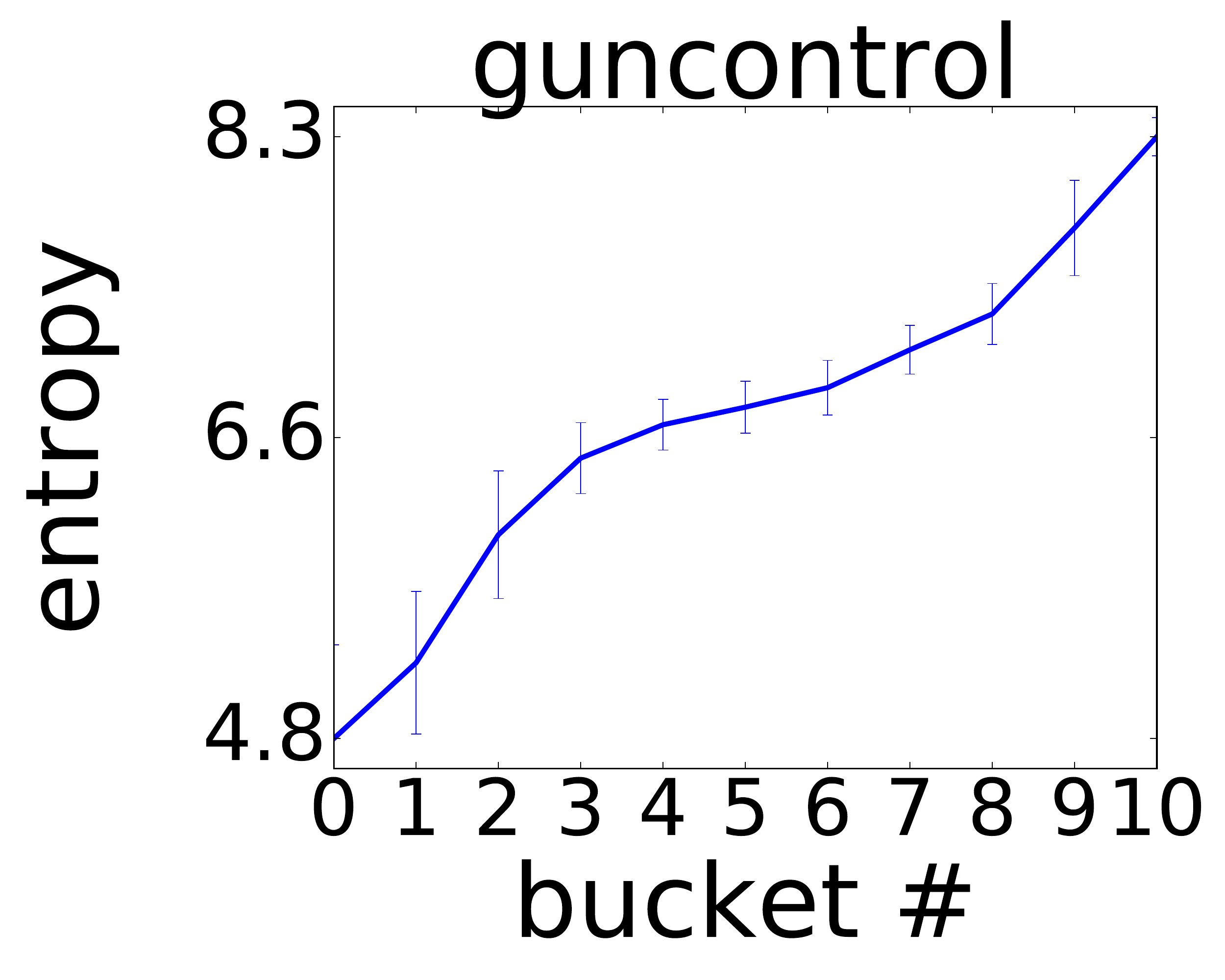}}
\end{minipage}%
\begin{minipage}{.25\linewidth}
\centering
\subfloat{\label{}\includegraphics[width=\textwidth]{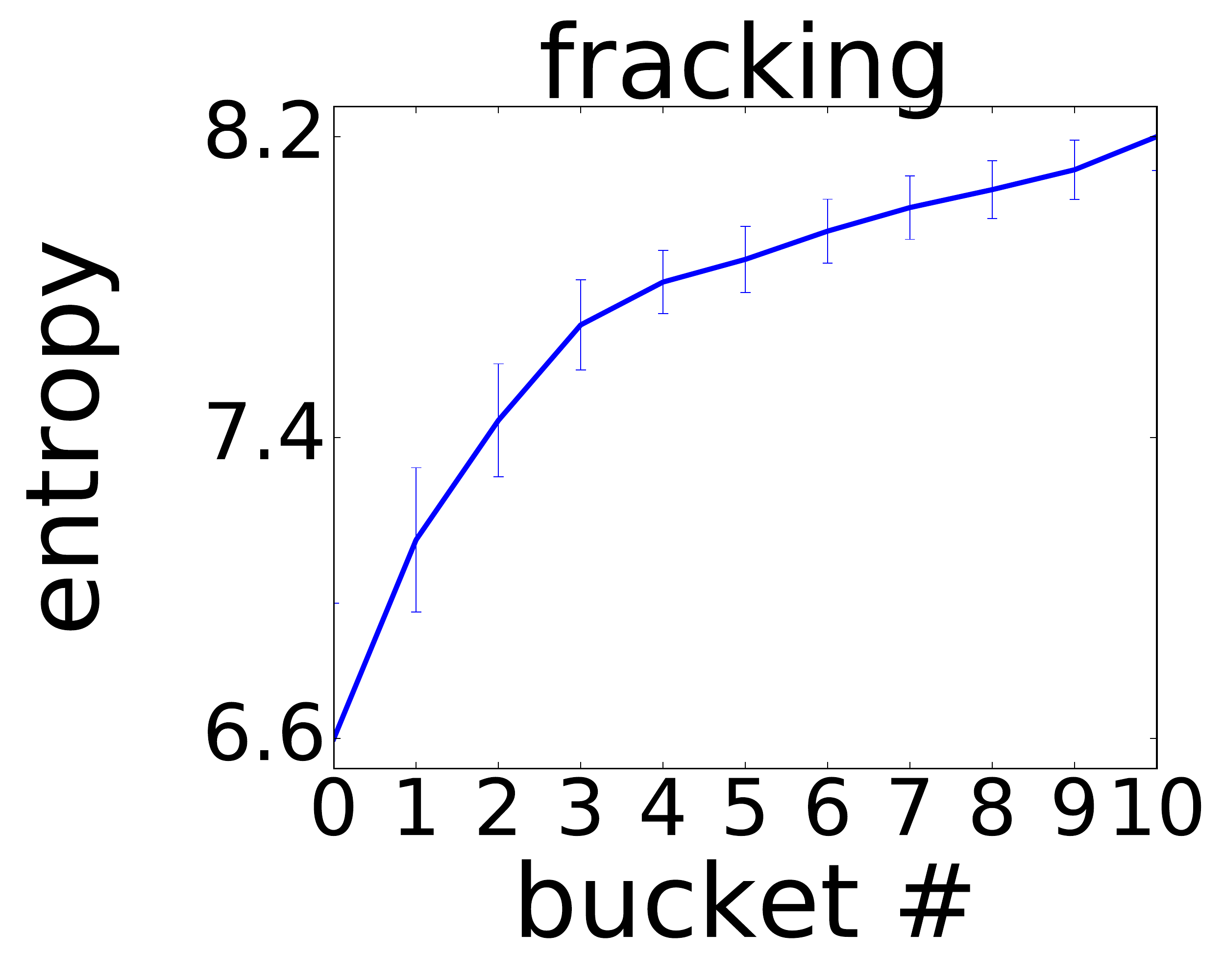}}
\end{minipage}%
\caption{Entropy of the distribution over the lexicon for one side of the discussion as a function of the activity in the network (the other side shows similar patterns). As the interest increases, the entropy increases, thus indicating the use of a wider lexicon.}
\label{fig:entropy1}
\end{figure}

To further examine the cause of the convergence of lexicon, we report the entropy of the unigram distribution.
Figure~\ref{fig:entropy1} shows that the entropy for one of the sides increases as interest increases (results for the other side show similar trends).
Thus, we find that the lexicon is more uniform and less skewed, which supports the hypothesis that a larger group of users brings a more general lexicon to the discussion, rather than the alternative hypothesis of event-based convergence.

\begin{figure}[tb]
\centering
\begin{minipage}{.25\linewidth}
\centering
\subfloat{\label{}\includegraphics[width=\textwidth]{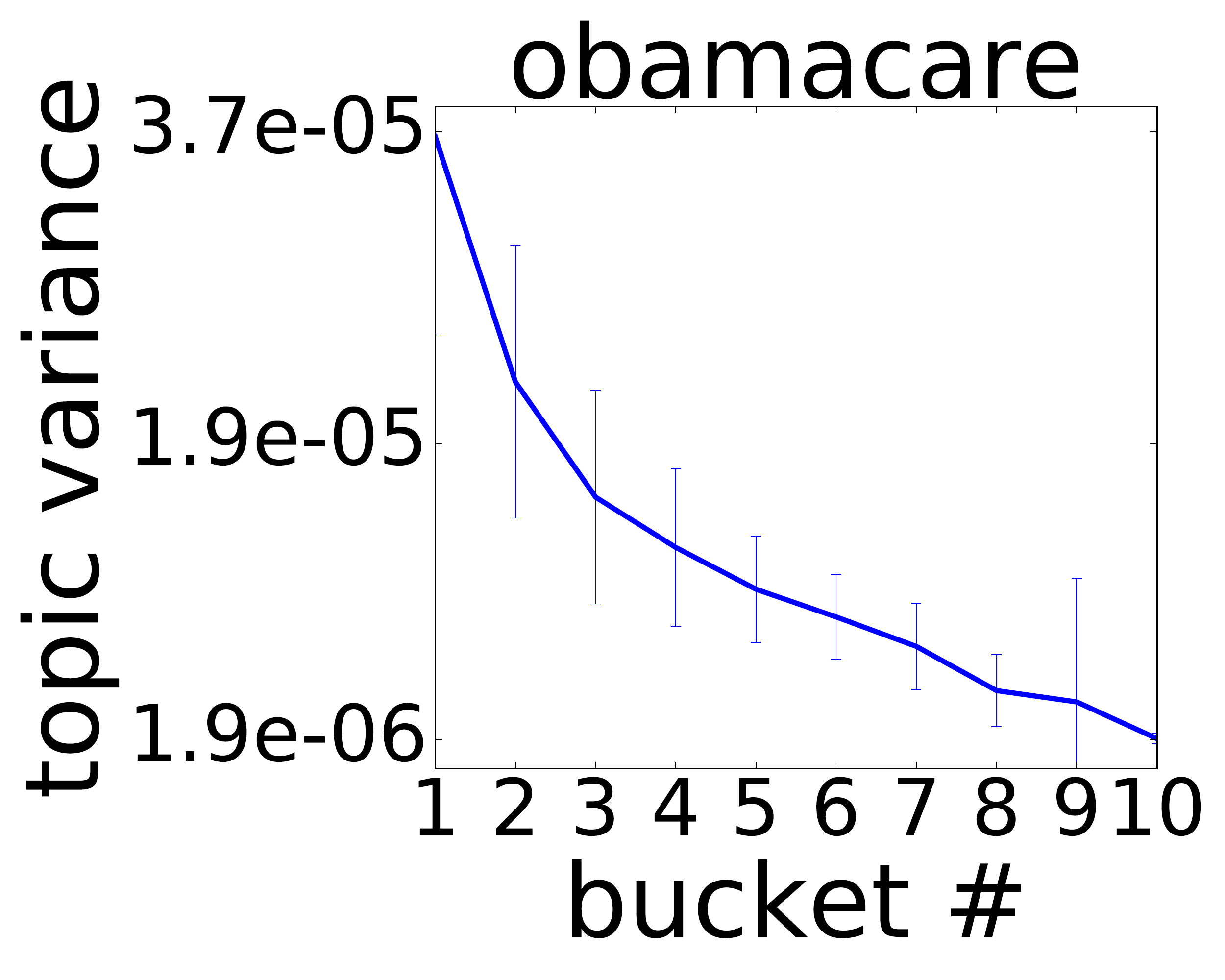}}
\end{minipage}%
\begin{minipage}{.25\linewidth}
\centering
\subfloat{\label{}\includegraphics[width=\textwidth]{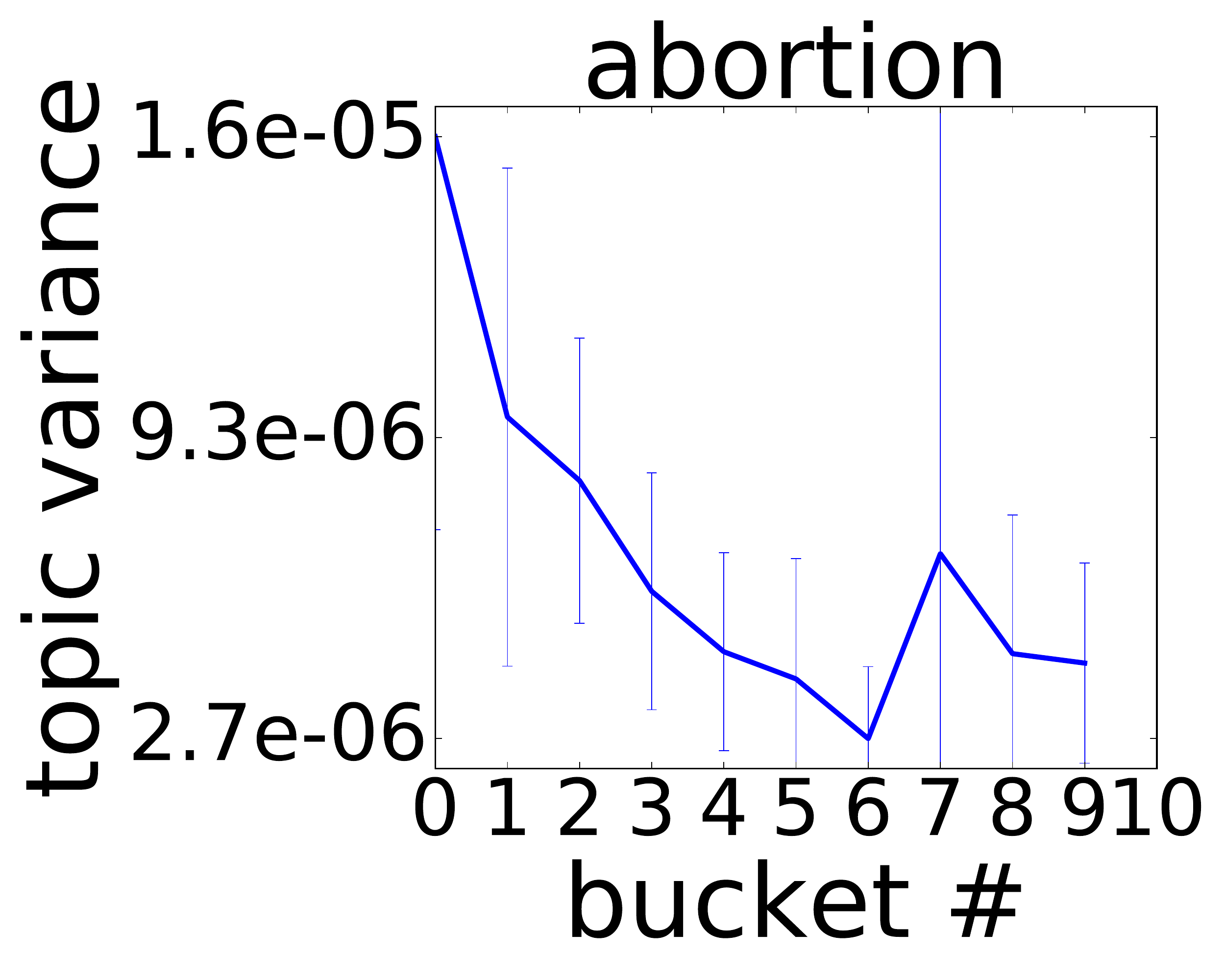}}
\end{minipage}%
\begin{minipage}{.25\linewidth}
\centering
\subfloat{\label{}\includegraphics[width=\textwidth]{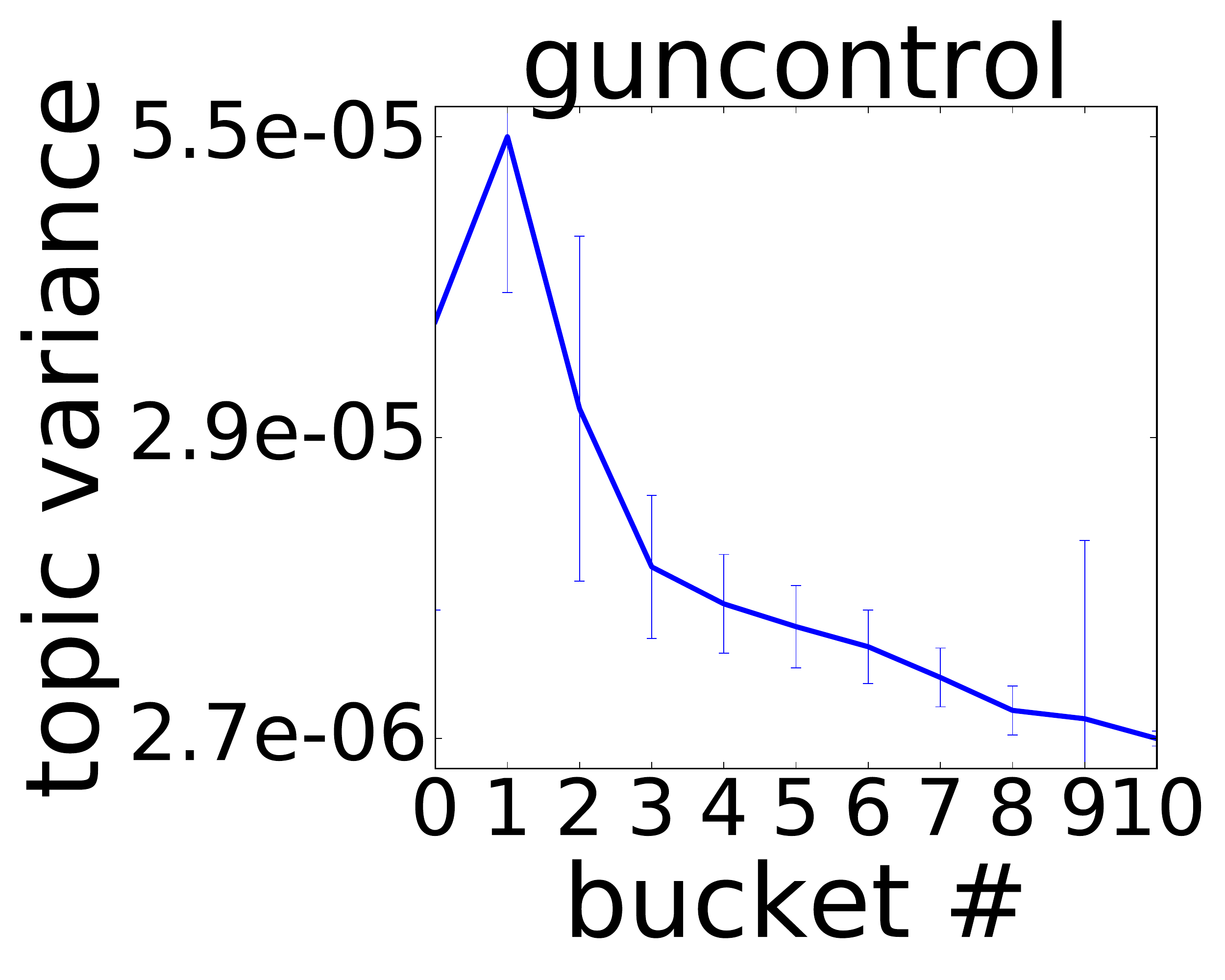}}
\end{minipage}%
\begin{minipage}{.25\linewidth}
\centering
\subfloat{\label{}\includegraphics[width=\textwidth]{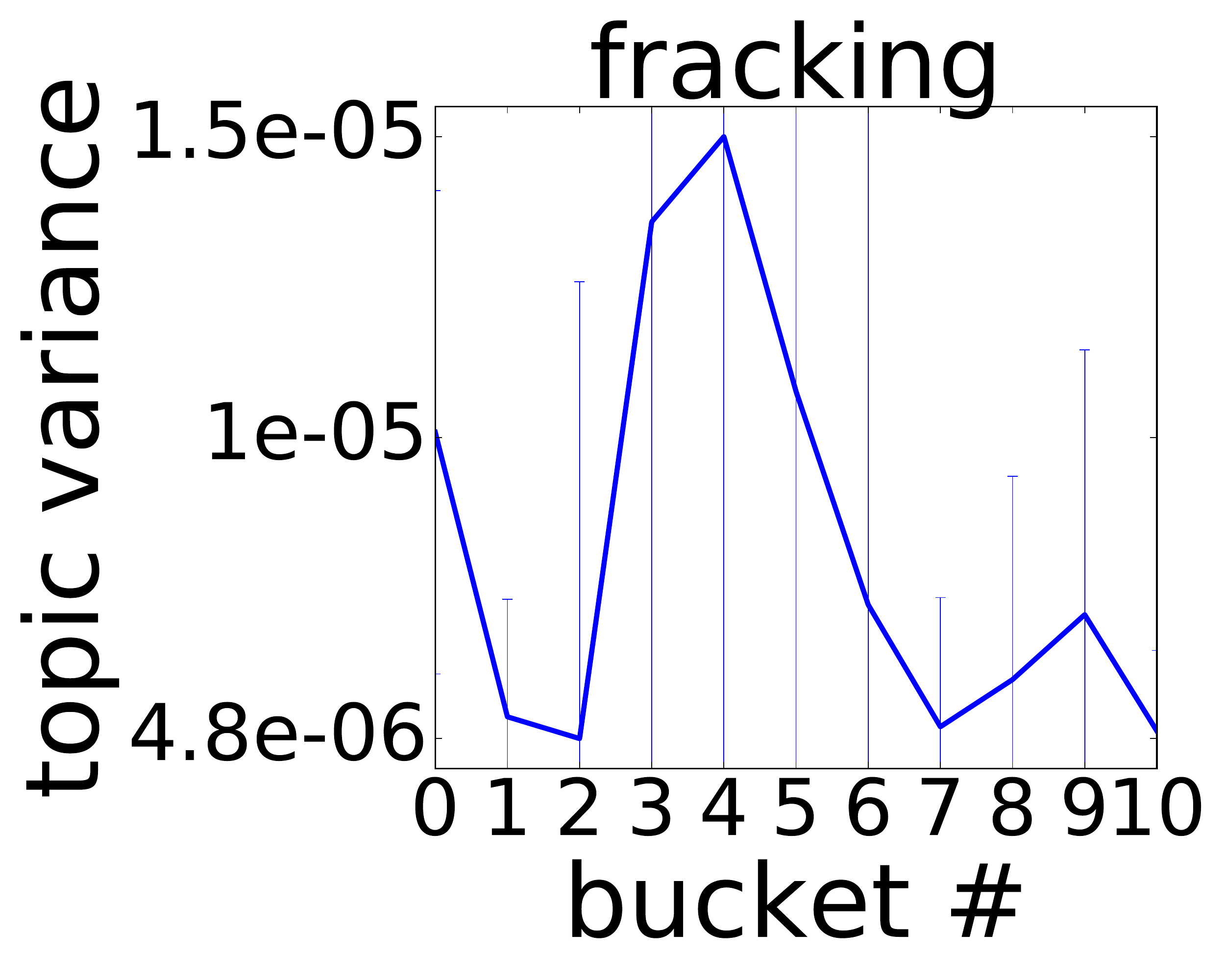}}
\end{minipage}%
\caption{Variance of the topic distribution. As the interest increases, variance decreases, indicating that a wider range of topics are being discussed.} 
\label{fig:topic_variance}
\end{figure}

To investigate \emph{what} causes the lexicon to be generalized, we compute the variance of the topic distribution for each bucket.
As we see from Figure~\ref{fig:topic_variance}, the variance decreases with increased activity, meaning that the topic distribution becomes more uniform\footnote{\small The term `fracking' is also sometimes used as an expletive, which might explain why the effects we measure are not as pronounced for this topic as the other ones. E.g. see \url{https://twitter.com/KitKat0122/status/19820978435522561}}.
This result provides evidence that users do indeed discuss a wider range of topics when there is a spike in activity.

Finally, we also examine how the sentiment and other linguistic cues change with interest.
We measure the variance in sentiment, fraction of tweets containing various LIWC categories, such as anger, sadness, positive and negative emotion, and anxiety.
Previous work shows that sentiment variance is a measure able to separate controversial from non-controversial topics~\citep{garimella2016quantifying} and linguistic patterns of communication change during shocks~\cite{romero2016social}.
However, we do not see any consistent trend. 
We hypothesise that this might be due to the noise in language (slang, sarcasm, short text, etc) on social media.

\subsection{Core}
\label{sec:core_findings}

\begin{figure}[tb]
\centering
\begin{minipage}{.25\linewidth}
\centering
\subfloat{\label{}\includegraphics[width=\textwidth]{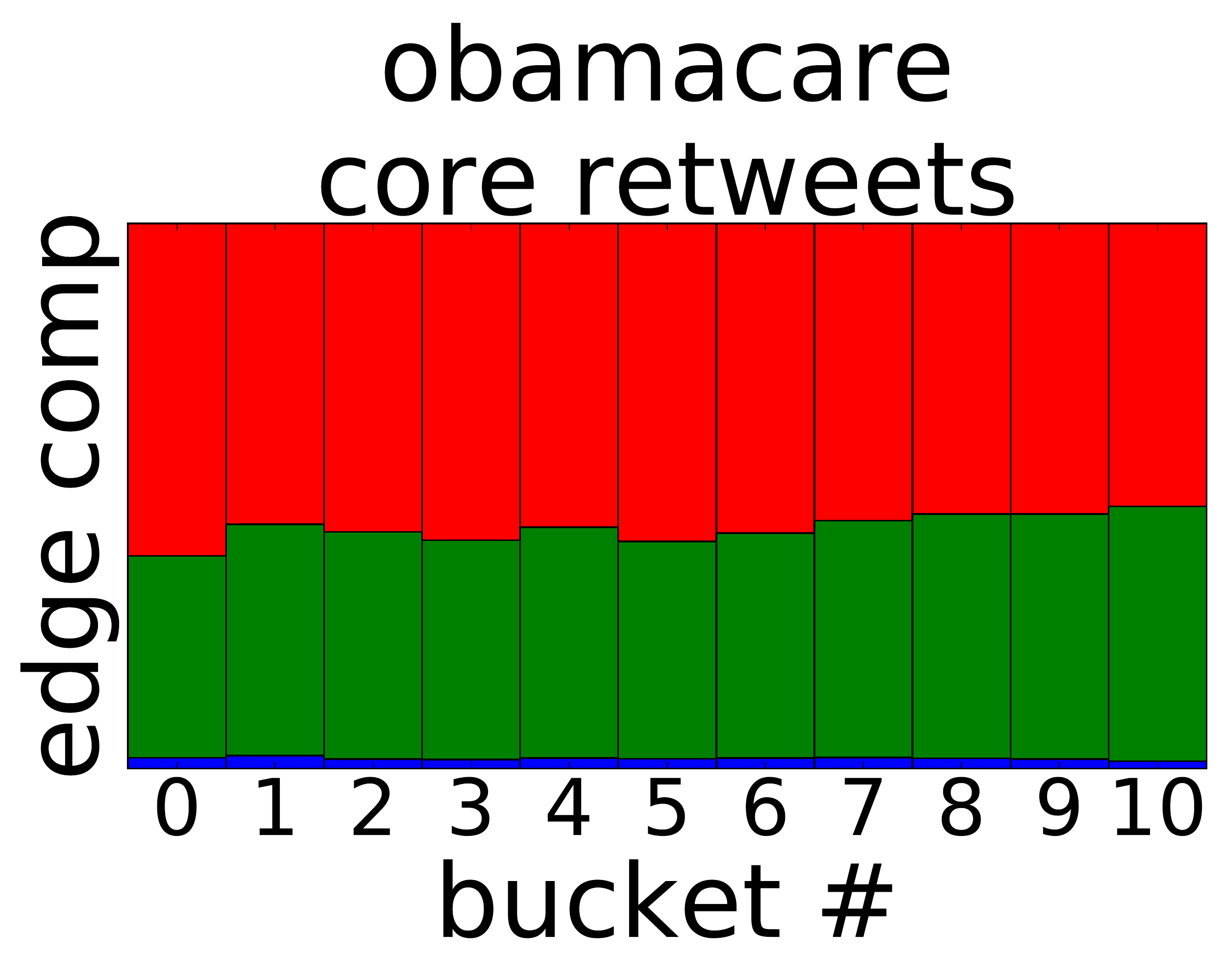}}
\end{minipage}%
\begin{minipage}{.25\linewidth}
\centering
\subfloat{\label{}\includegraphics[width=\textwidth]{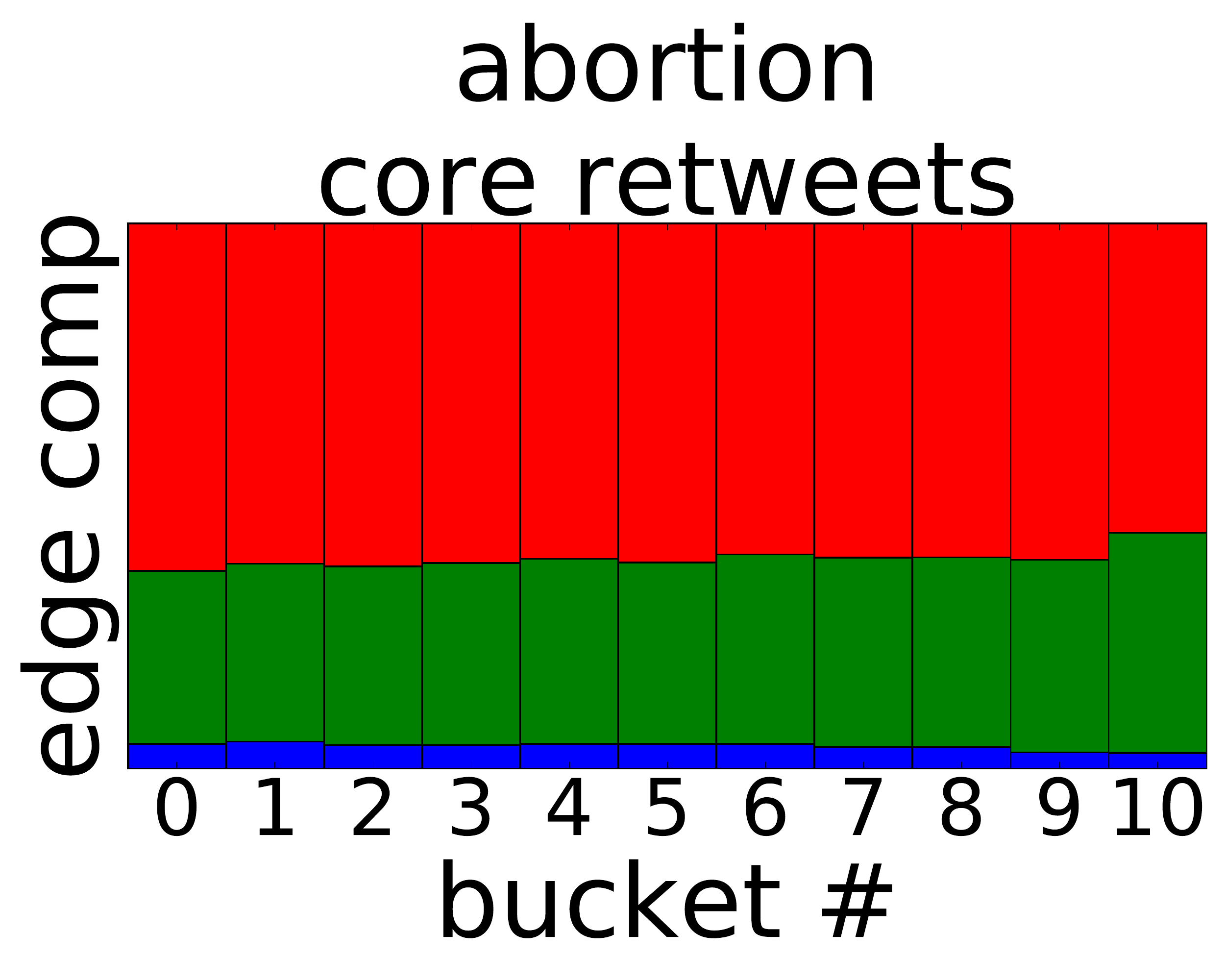}}
\end{minipage}%
\begin{minipage}{.25\linewidth}
\centering
\subfloat{\label{}\includegraphics[width=\textwidth]{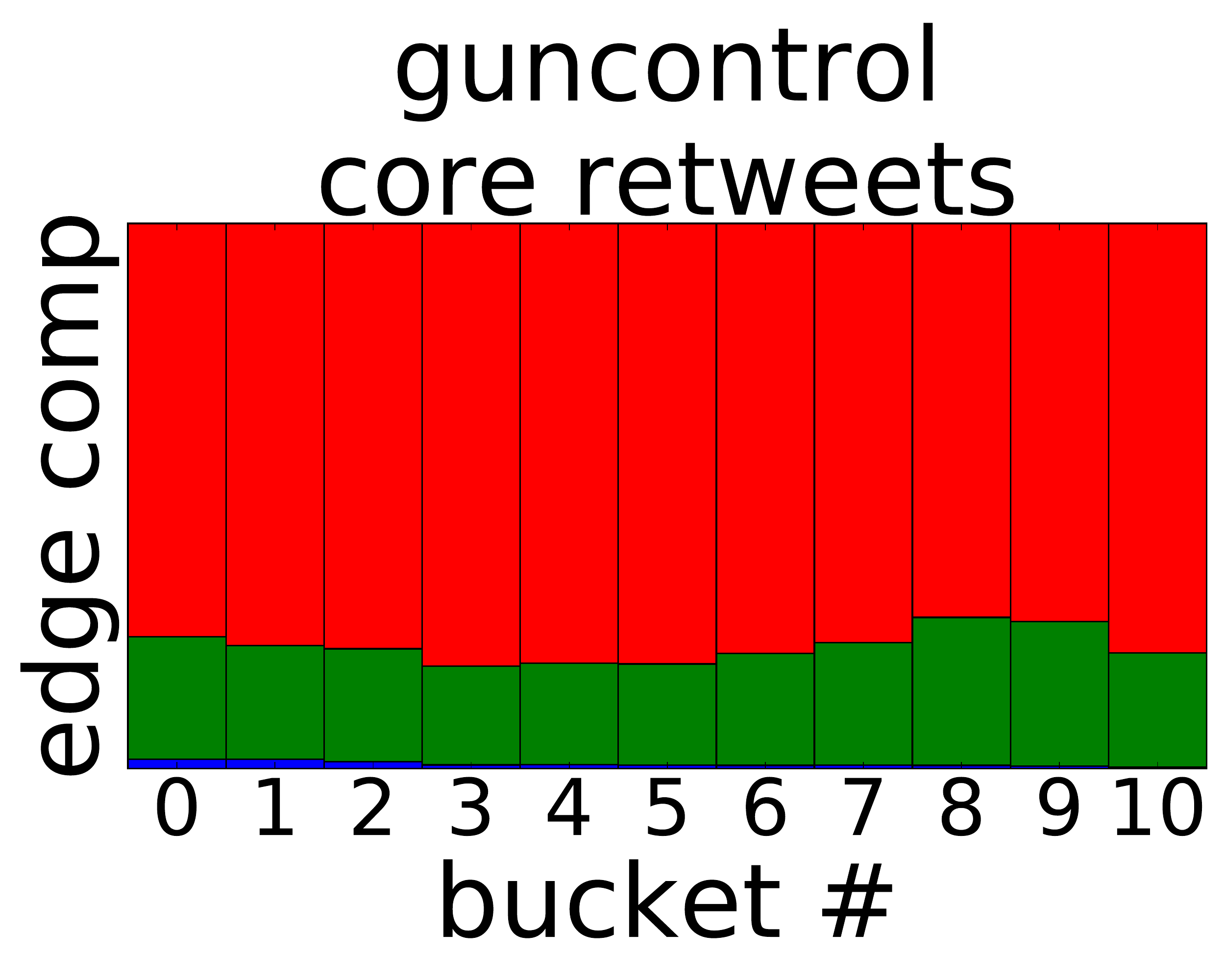}}
\end{minipage}%
\begin{minipage}{.25\linewidth}
\centering
\subfloat{\label{}\includegraphics[width=\textwidth]{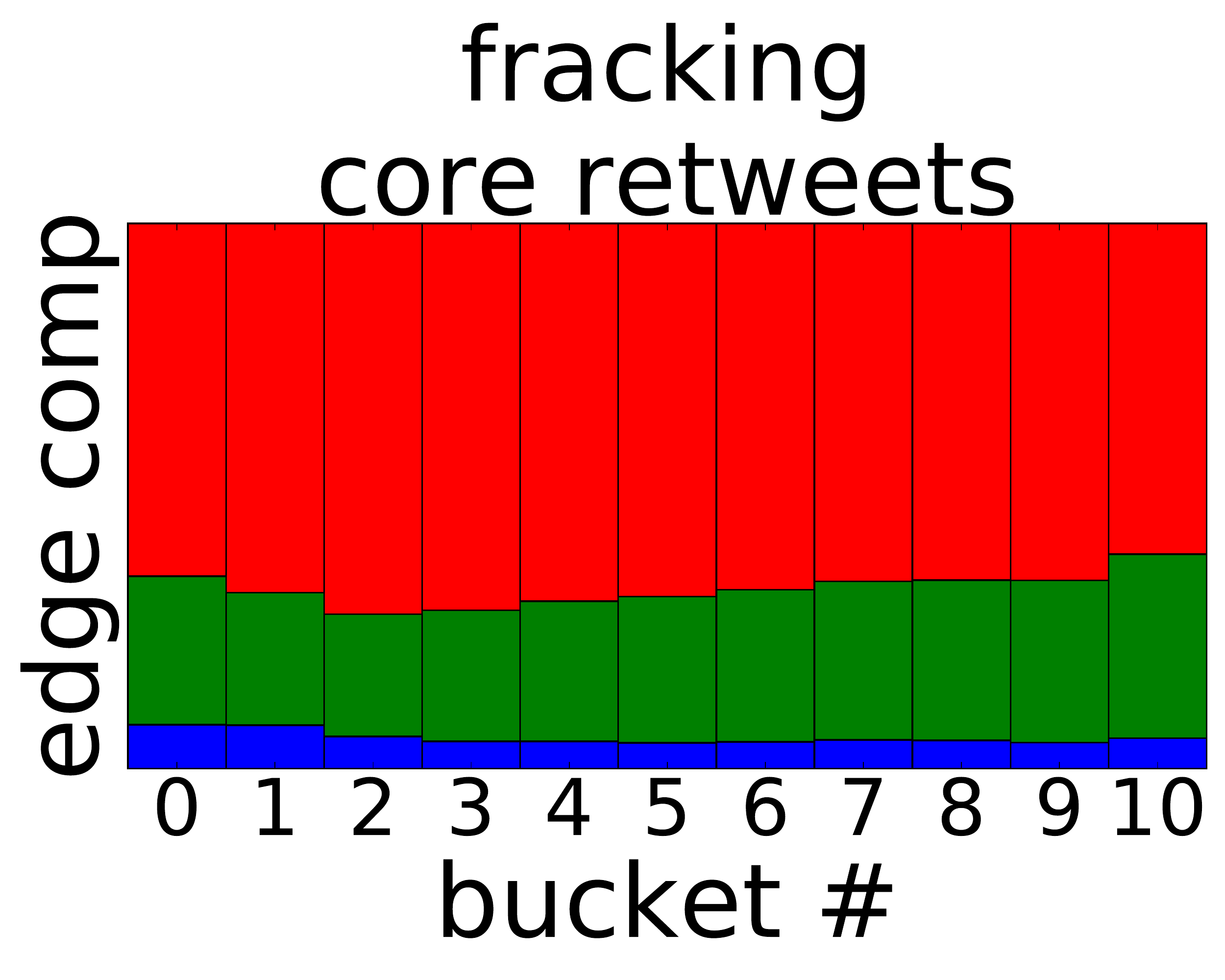}}
\end{minipage}%
\caption{Edge composition as a function of network activity in the retweet network. As the interest increases, there are no major changes in the fractions of core-core (blue), core-periphery (green), and periphery-periphery (red) edges.}
\label{fig:retweet_triplets}
\end{figure}

\begin{figure}[tb]
\centering
\begin{minipage}{.25\linewidth}
\centering
\subfloat{\label{}\includegraphics[width=\textwidth]{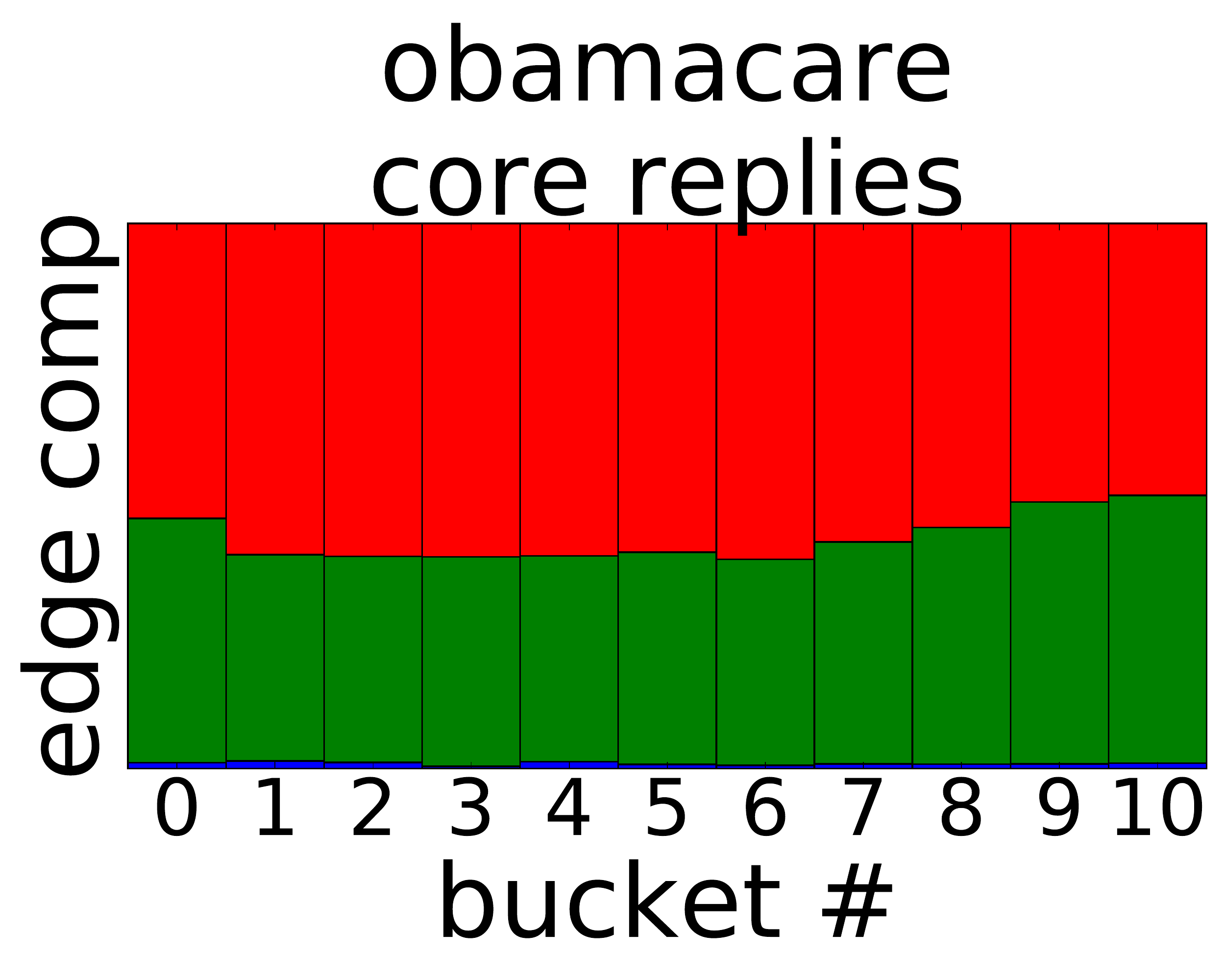}}
\end{minipage}%
\begin{minipage}{.25\linewidth}
\centering
\subfloat{\label{}\includegraphics[width=\textwidth]{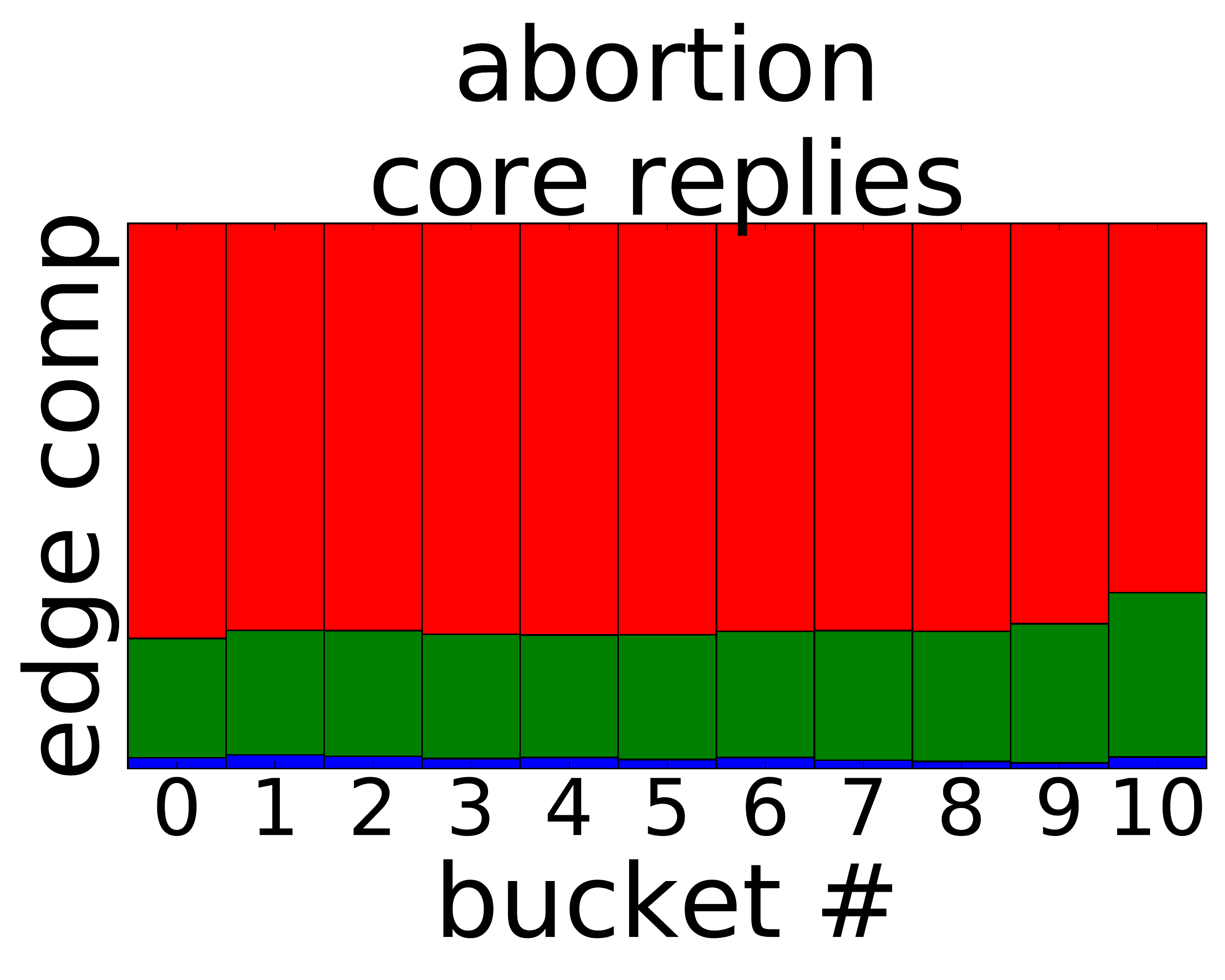}}
\end{minipage}%
\begin{minipage}{.25\linewidth}
\centering
\subfloat{\label{}\includegraphics[width=\textwidth]{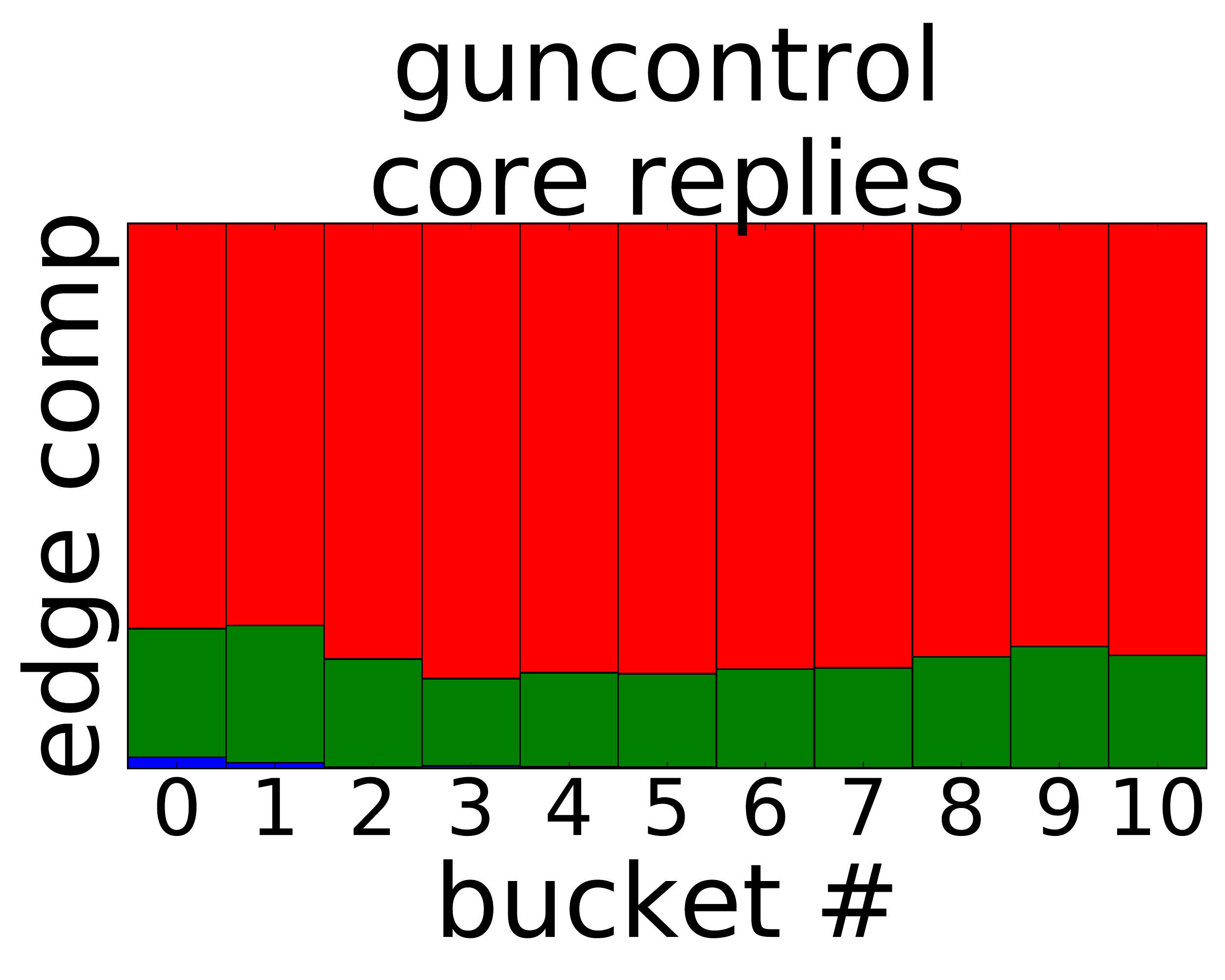}}
\end{minipage}%
\begin{minipage}{.25\linewidth}
\centering
\subfloat{\label{}\includegraphics[width=\textwidth]{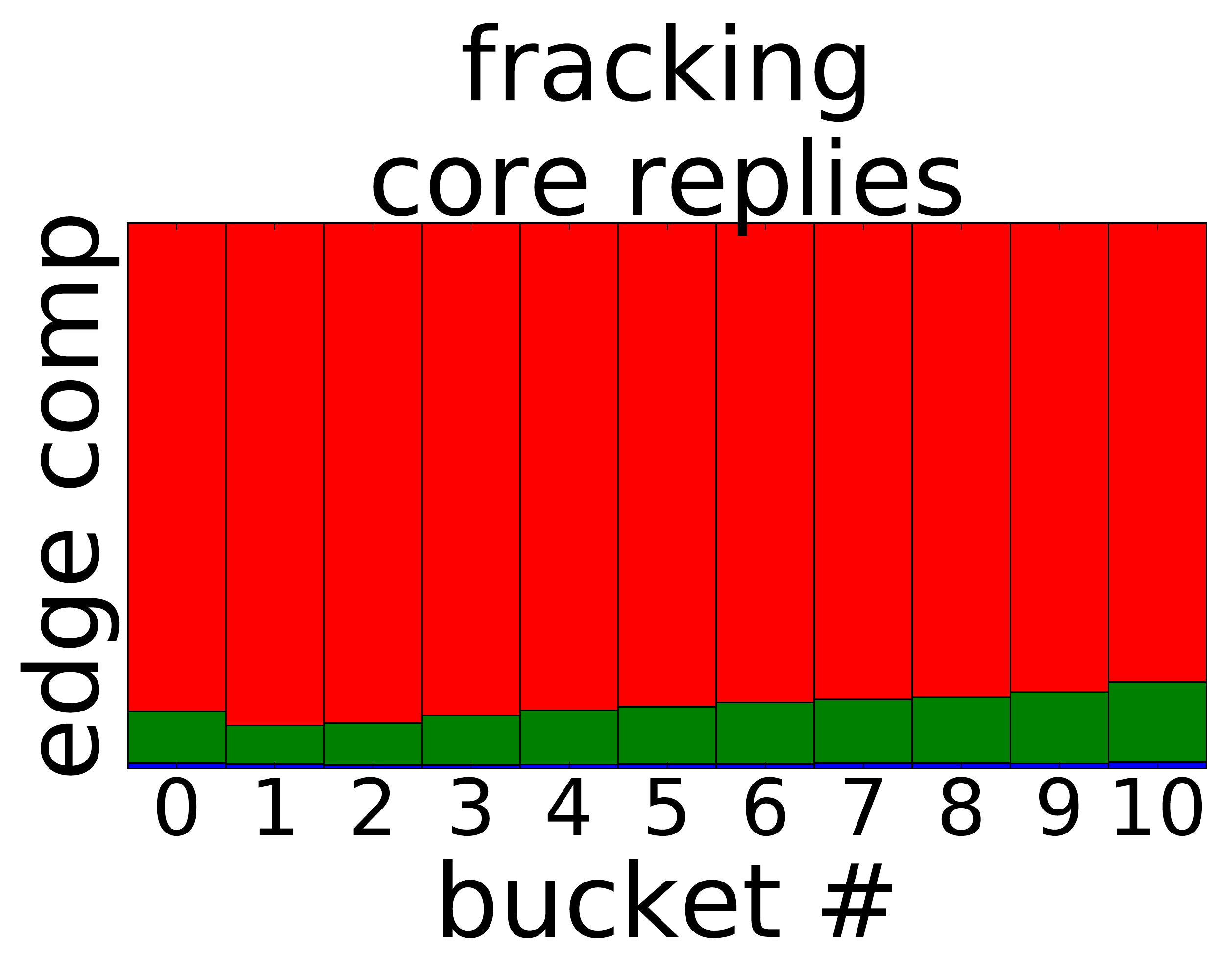}}
\end{minipage}%
\caption{Edge composition as a function of network activity in the reply network. As the interest increases, there are no major changes in the fractions of core-core (blue), core-periphery (green), and periphery-periphery (red) edges.}
\label{fig:reply_triplets}
\end{figure}

Looking at the fractions of the different types of edges (core--core, core--periphery, and periphery--periphery) across the volume buckets in Figures~\ref{fig:retweet_triplets} and~\ref{fig:reply_triplets}, we see that the composition of edges does not change significantly with increase collective attention.
This result suggests that the discussion grows in a self-similar way.

A disproportionately large fraction of edges link the periphery to the core, when taking into account the core size, as seen in Figure~\ref{fig:retweet_triplets}.
During a spike in interest, most casual users, who seldom participate in the discussion, endorse opinions from the core of the side they belong to (red bars).
For replies, we see a similar trend with respect to activity volume in Figure~\ref{fig:reply_triplets}.
In general, the core is less prevalent in the discussion, as shown by the lower fraction of core-periphery edges (green bars).

However, when looking at the \emph{core--periphery openness} (Figures~\ref{fig:cpo-rt} and~\ref{fig:cpo-re}), we see that the \emph{normalized} number of edges between core and periphery increases, i.e., the number of edges between core and periphery increases compared to the expected number based on a random-graph null model.
To interpret this result, note that when the network grows, given that the periphery is much larger than the core, most edges for the null model are among periphery nodes.
Therefore, the interaction networks show a clear hierarchical structure when growing.



\afterpage{
\begin{figure}[tb]
\centering
\begin{minipage}{.25\linewidth}
\centering
\subfloat{\label{}\includegraphics[width=\textwidth]{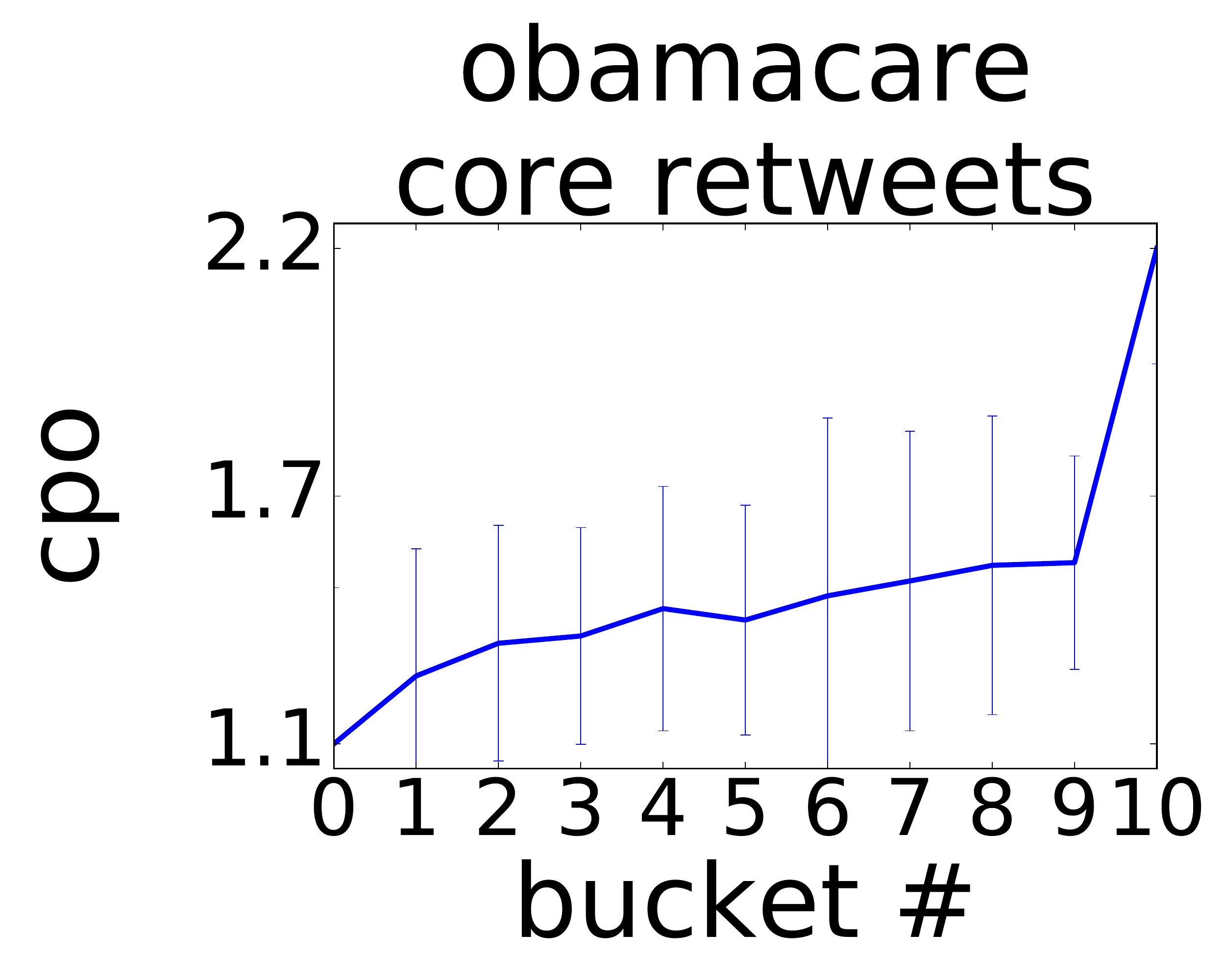}}
\end{minipage}%
\begin{minipage}{.25\linewidth}
\centering
\subfloat{\label{}\includegraphics[width=\textwidth]{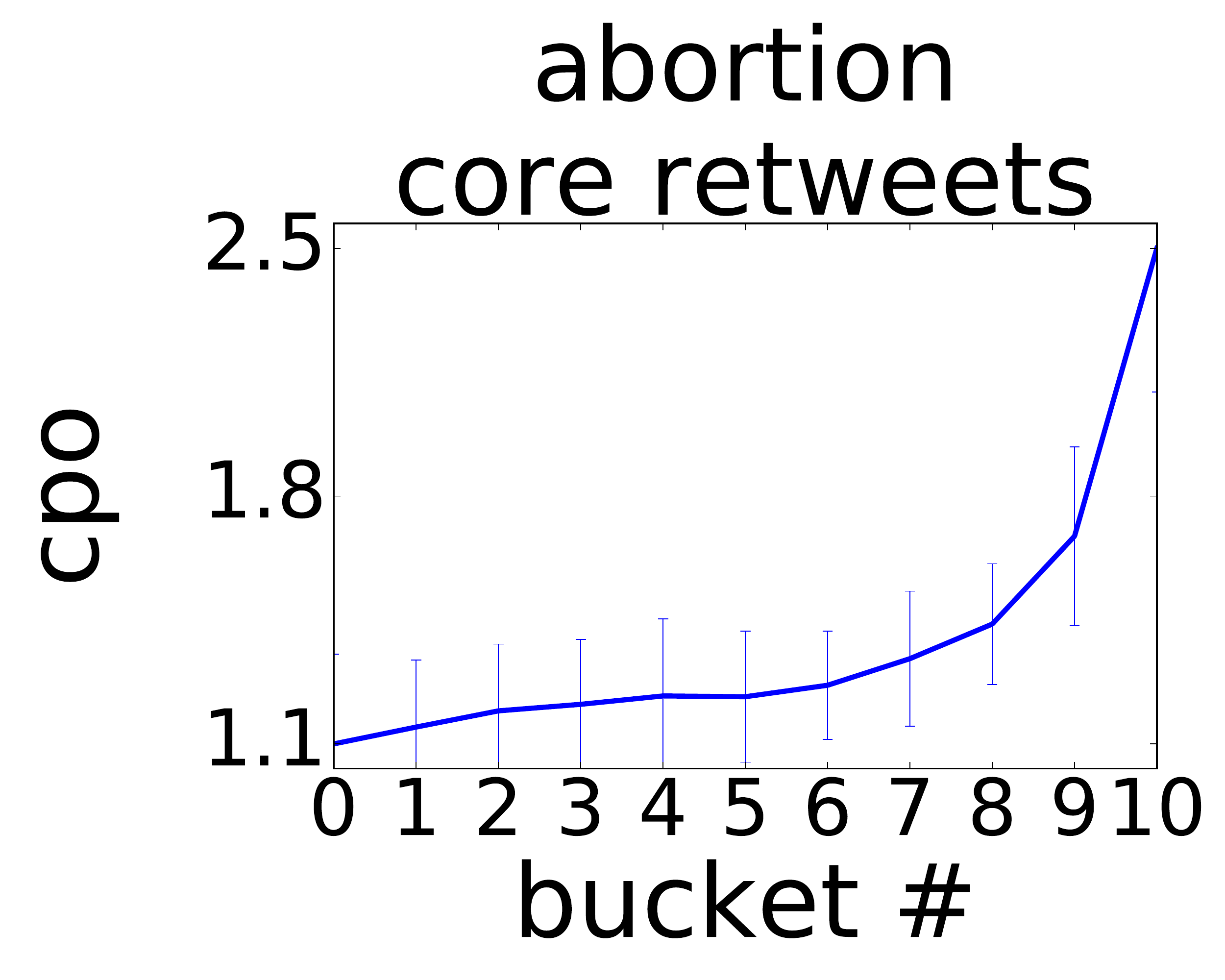}}
\end{minipage}%
\begin{minipage}{.25\linewidth}
\centering
\subfloat{\label{}\includegraphics[width=\textwidth]{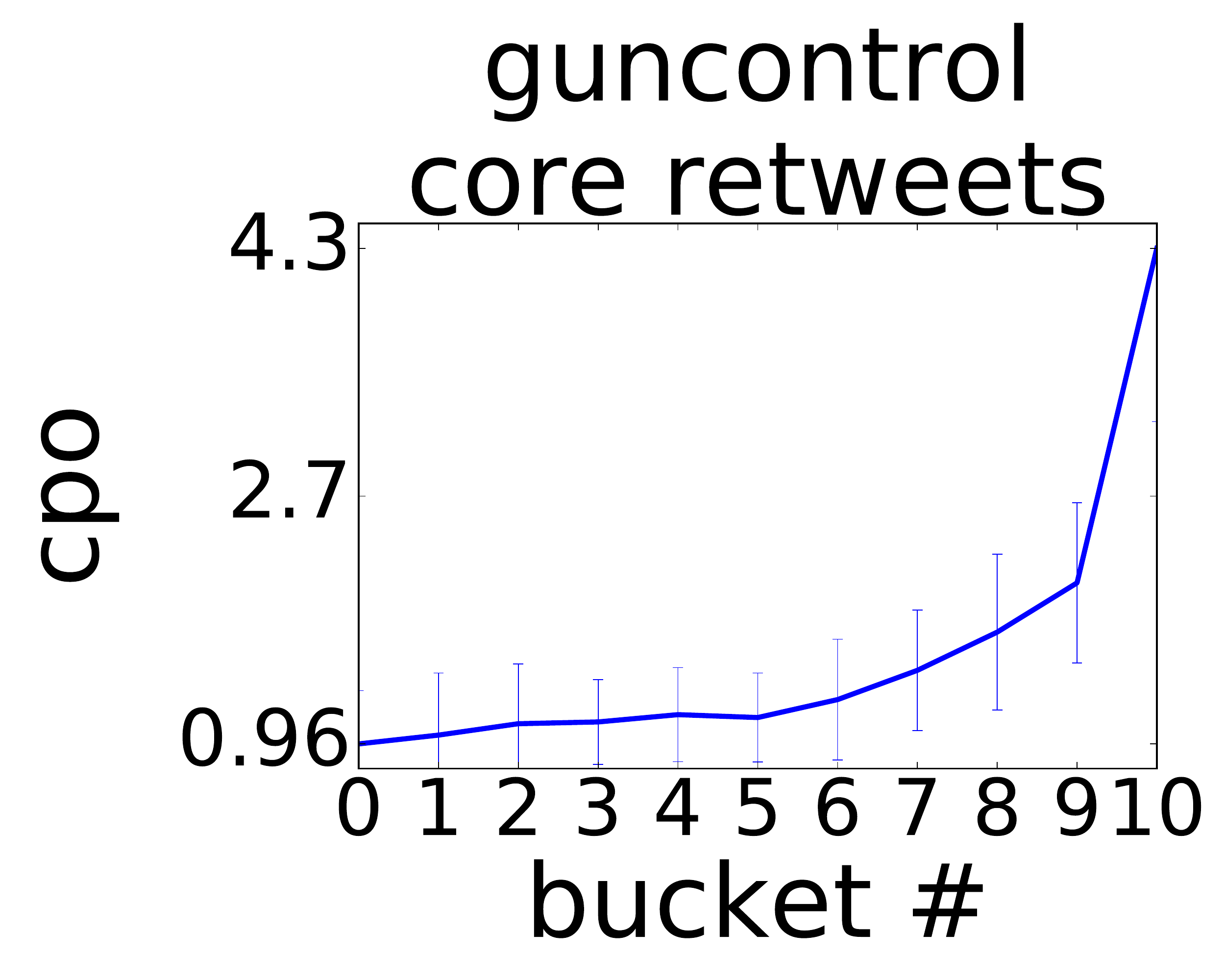}}
\end{minipage}%
\begin{minipage}{.25\linewidth}
\centering
\subfloat{\label{}\includegraphics[width=\textwidth]{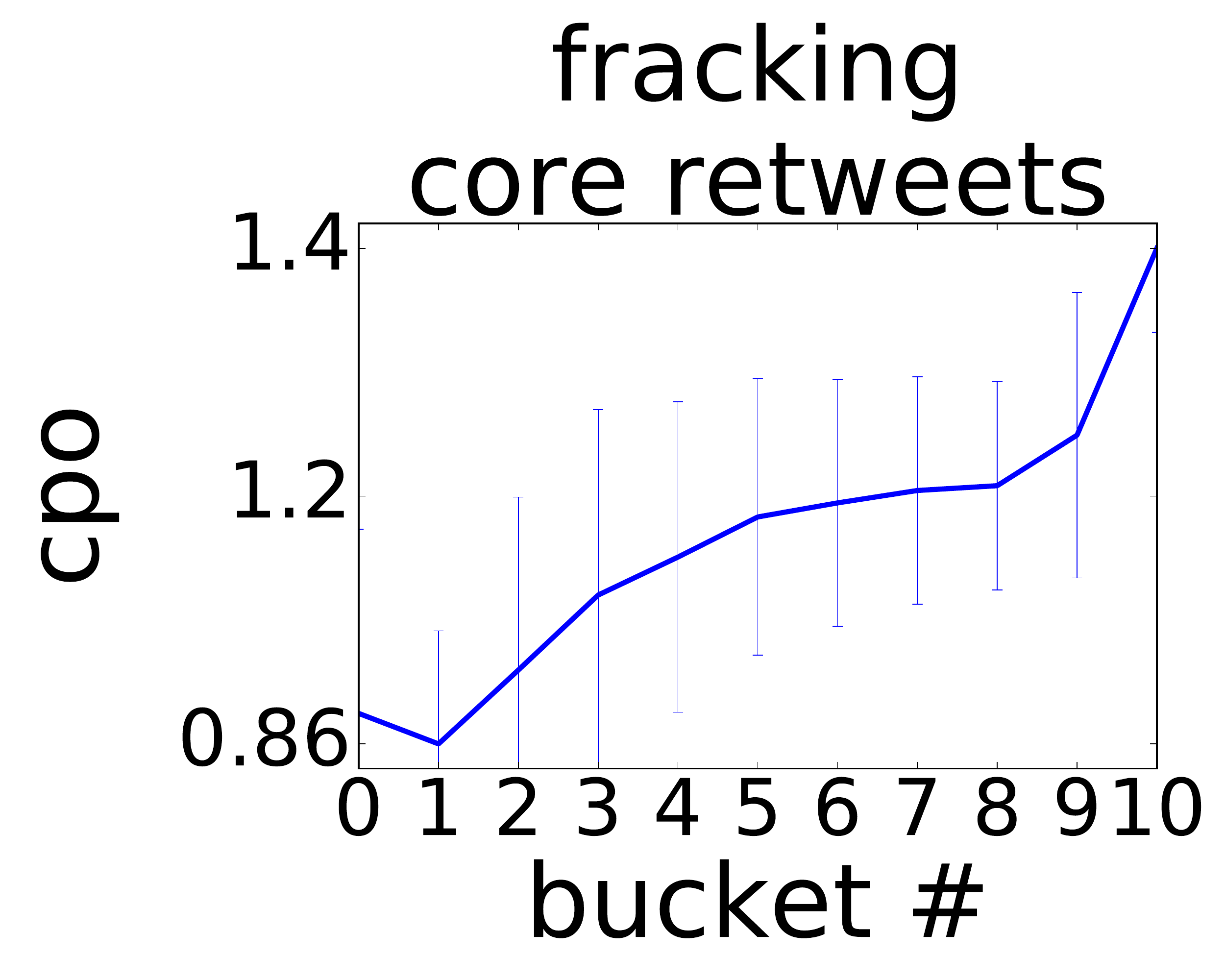}}
\end{minipage}%
\caption{Core--periphery openness as a function of activity in the retweet network. As the interest increases, the number of core-periphery edges, normalized by the expected number of edges in a random network, increases. This suggests a propensity of periphery nodes to connect with the core nodes when interest increases.}
\label{fig:cpo-rt}
\end{figure}

\begin{figure}[tb]
\centering
\begin{minipage}{.25\linewidth}
\centering
\subfloat{\label{}\includegraphics[width=\textwidth]{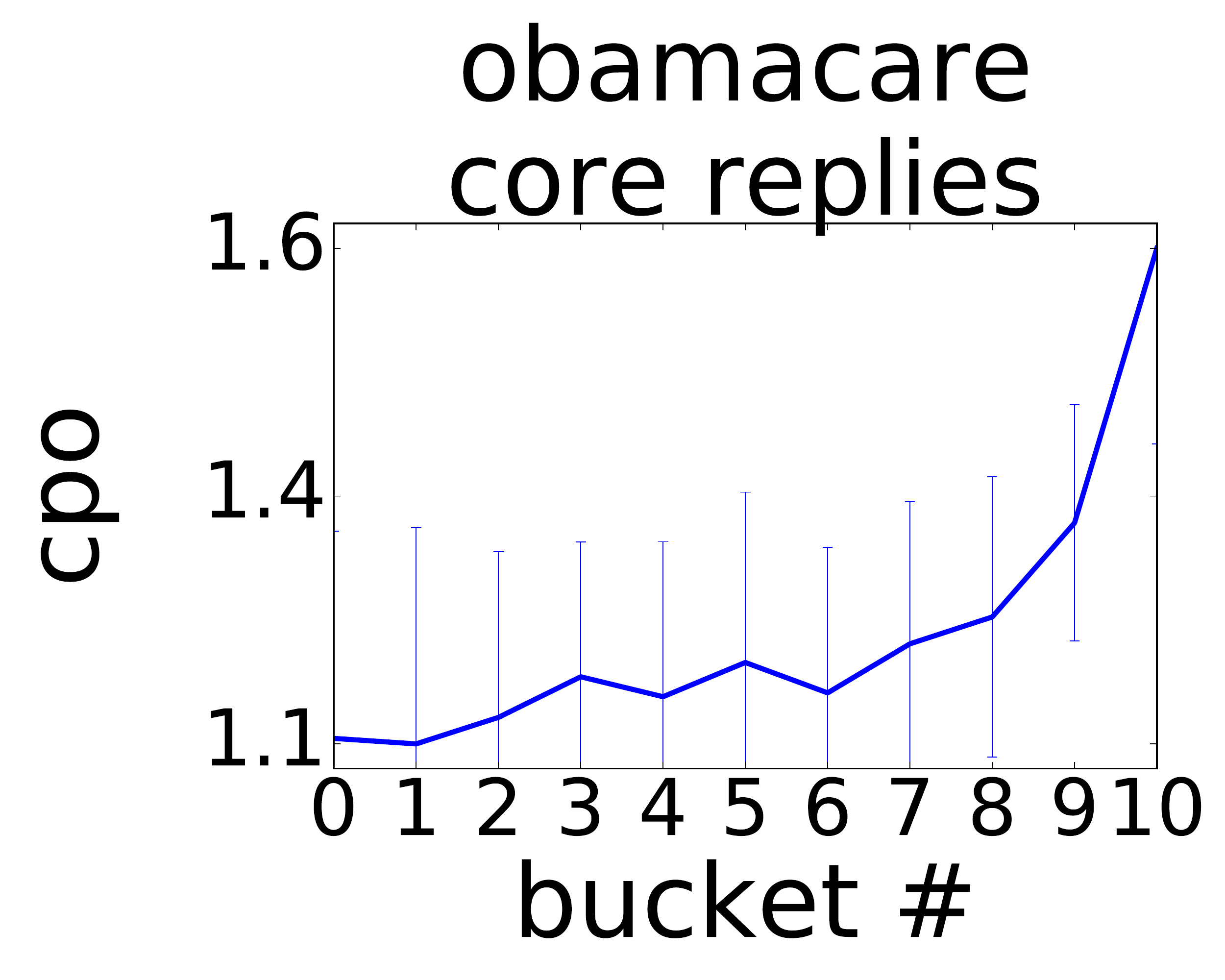}}
\end{minipage}%
\begin{minipage}{.25\linewidth}
\centering
\subfloat{\label{}\includegraphics[width=\textwidth]{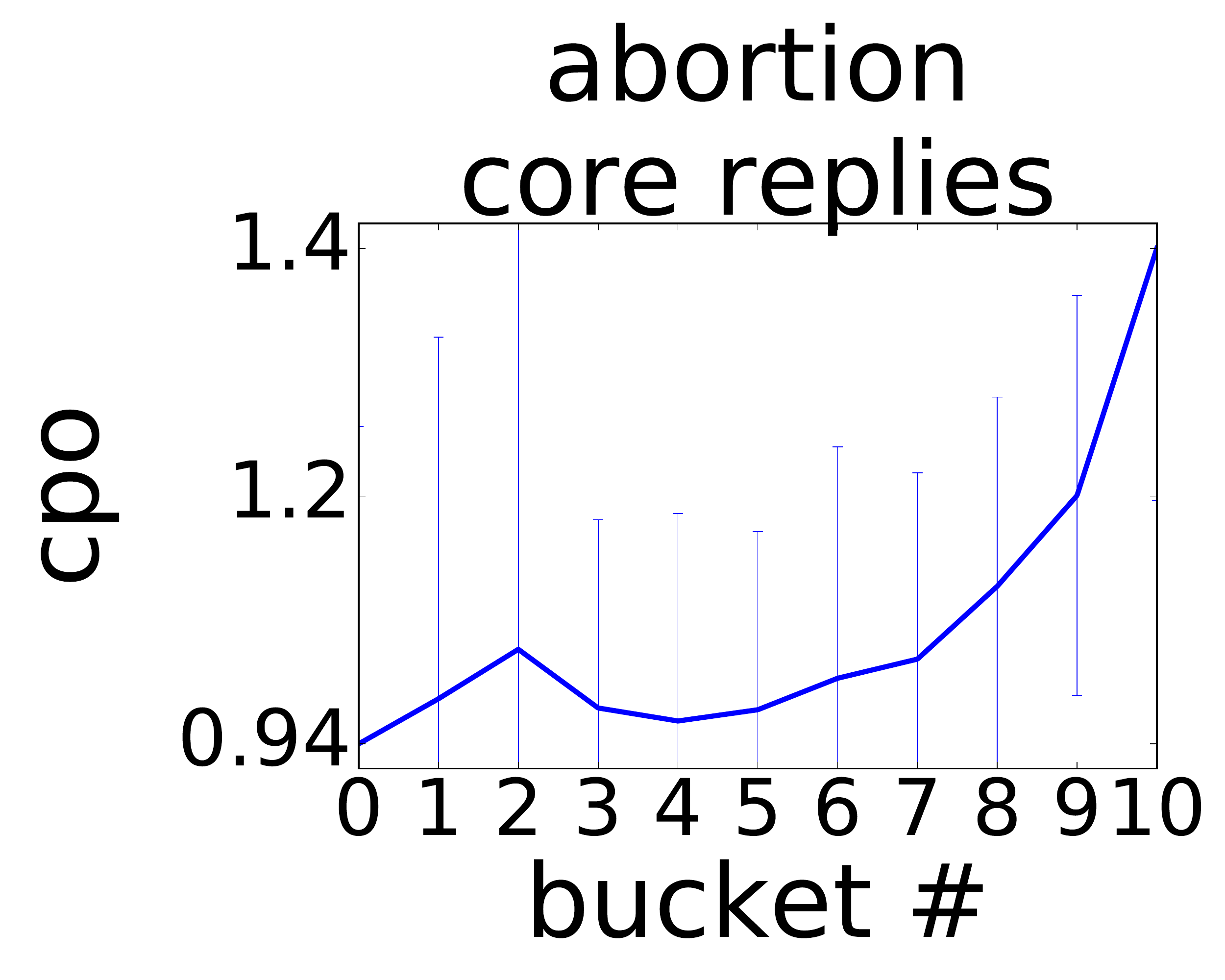}}
\end{minipage}%
\begin{minipage}{.25\linewidth}
\centering
\subfloat{\label{}\includegraphics[width=\textwidth]{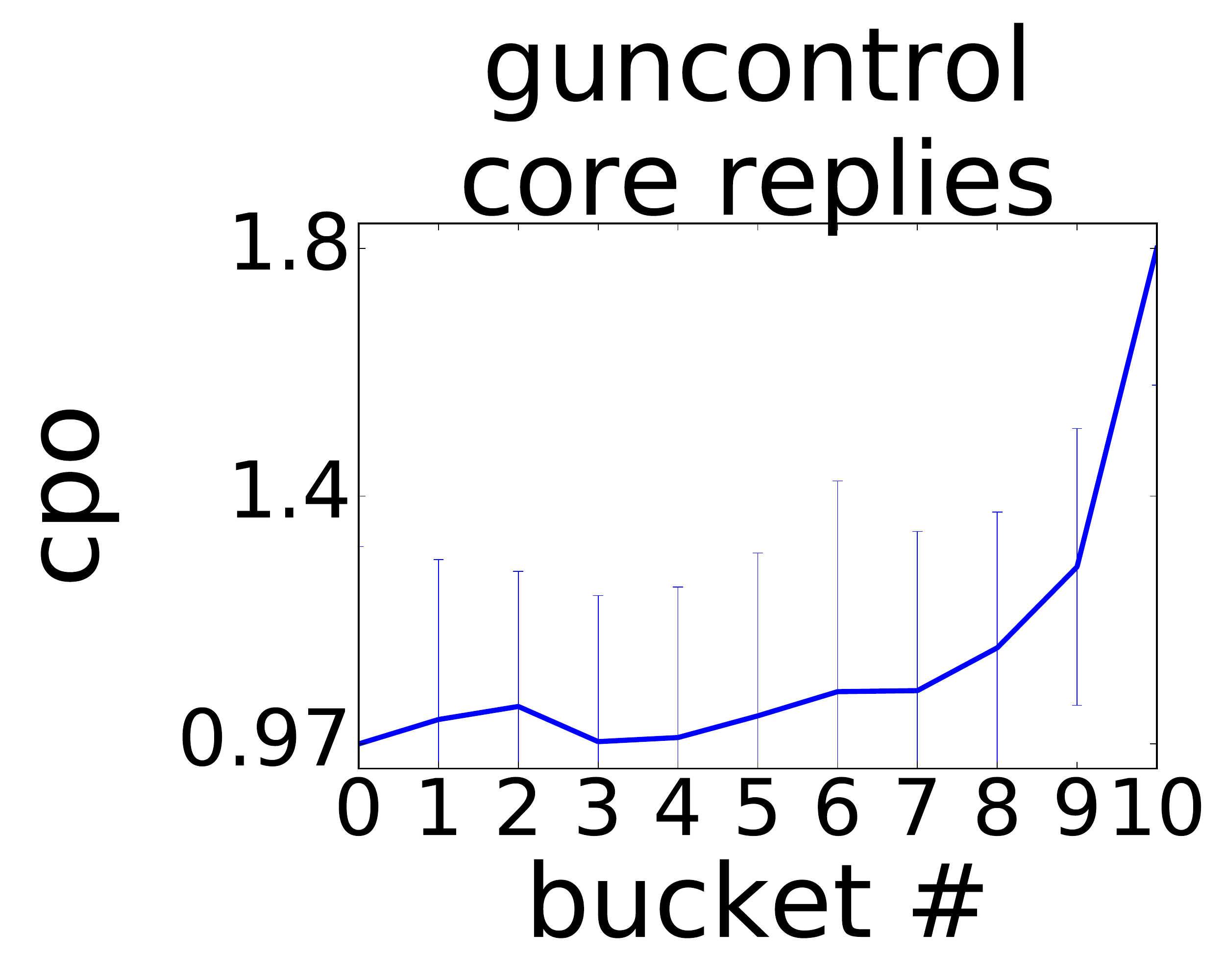}}
\end{minipage}%
\begin{minipage}{.25\linewidth}
\centering
\subfloat{\label{}\includegraphics[width=\textwidth]{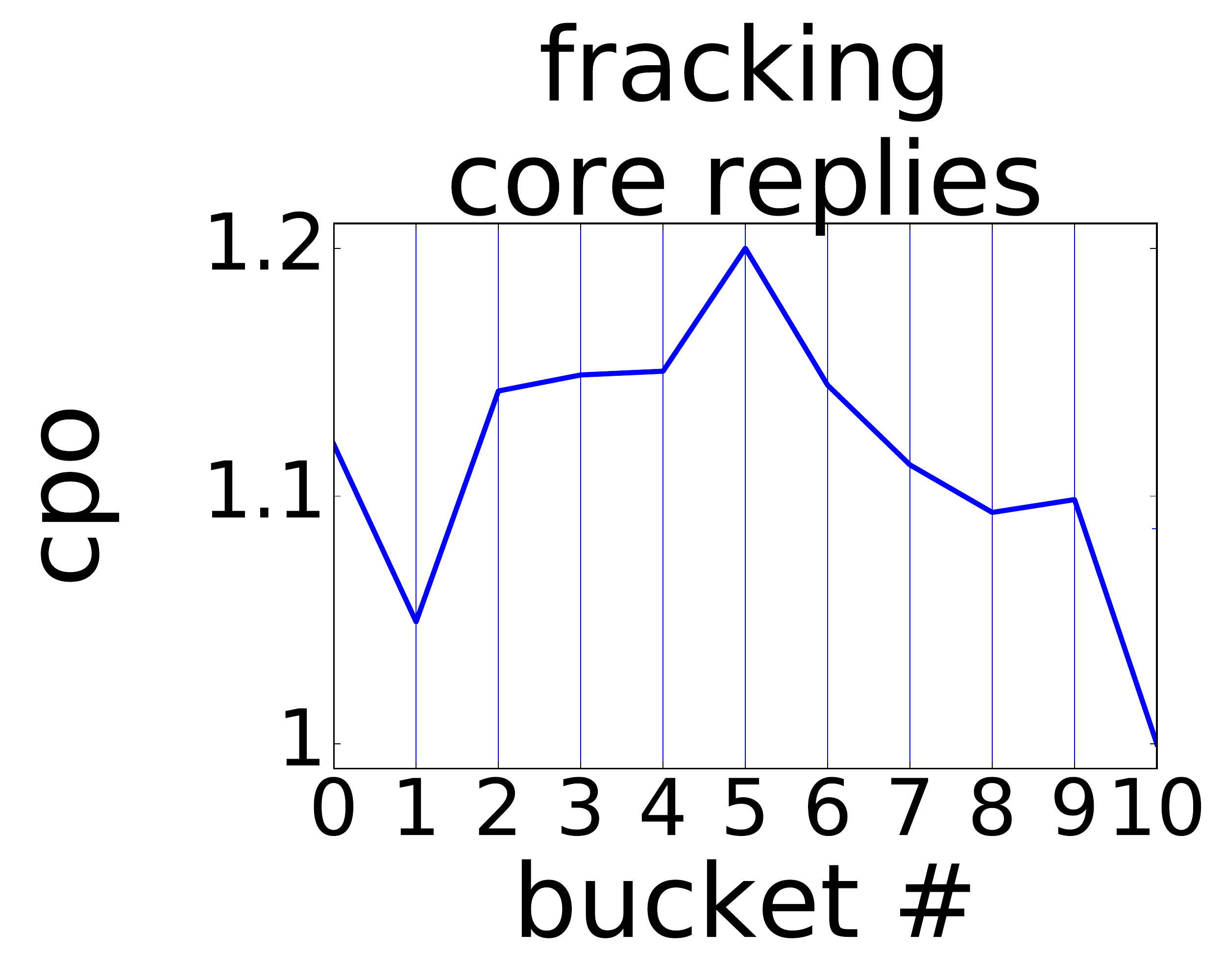}}
\end{minipage}%
\caption{Core-periphery openness as a function of activity in the reply network. As the interest increases, the number of core-periphery edges, normalized by the expected number of edges in a random network, increases for most topics. This suggests a propensity of periphery nodes to connect with the core nodes when interest increases.}
\label{fig:cpo-re}
\end{figure}
}

\subsection{Local analysis}
\label{sec:local_findings}
So far, we have analyzed global trends across the time series.
We now focus on local trends, to drill down on what happens around the spikes, 
and look at local variations of the measures just before and after the spike.
We mark a day in the time series as a spike if the volume of active users is at least two standard deviations above the mean.
Table~\ref{tab:time} shows the Pearson correlation between various measures and network activity, one week before and after the spike.
The trends observed globally still hold.
There is a positive correlation of RWC with activity, which adds more evidence to our finding that polarization increases during spikes.
The trends for bimotif, tie strength, and content divergence also persist, and are much stronger locally.

In addition to the previous measures, we also analyze other content features, such as the fraction of retweets, replies, mentions, and URLs around the spike.
Interestingly, we find strong positive correlation of retweets, mentions, and URLs with volume, which indicates that discussion and endorsement increase during a spike.
This finding is consistent with the ones by~\citet{smith2013role}, who find that users tend to add URLs to their tweets when discussing controversial topics.
Note that these additional content measures are only indicative for the local analysis, and do not produce consistent results at the global level.

\begin{table}[tb]
\centering
\small
\caption{Pearson correlation of various measures with volume one week before, during and after a spike in interest. All values except those marked with an asterisk (*) are significant at $p < 0.05$.}
\label{tab:time}
\begin{tabular}{lcccc}
\toprule
Measure            & Obamacare & Gun Control & Abortion & Fracking \\
\midrule
RWC          & 0.20      & 0.21       & 0.19     & 0.23     \\
Openness     & -0.09*    & 0.81       & 0.23     & 0.08     \\
Bimotif      & 0.27      & 0.36       & 0.33     & 0.23     \\
Tie Strength & 0.96      & 0.98       & 0.95     & 0.86   \\
JSD          & -0.66     & -0.86      & -0.63    & -0.46    \\
Entropy      & 0.42      & 0.46       & 0.67     & 0.26     \\
Frac. RT     & 0.15*     & 0.6        & 0.59     & 0.56     \\
Frac. Men.    & 0.20      & 0.71       & 0.54     & 0.51     \\
Frac. URL    & 0.32     & 0.36      & 0.39    & 0.40    \\
\bottomrule
\end{tabular}
\end{table}

\subsection{Evolution over time}
\label{sec:time_findings}

\begin{figure}
\centering
\includegraphics[width=\linewidth]{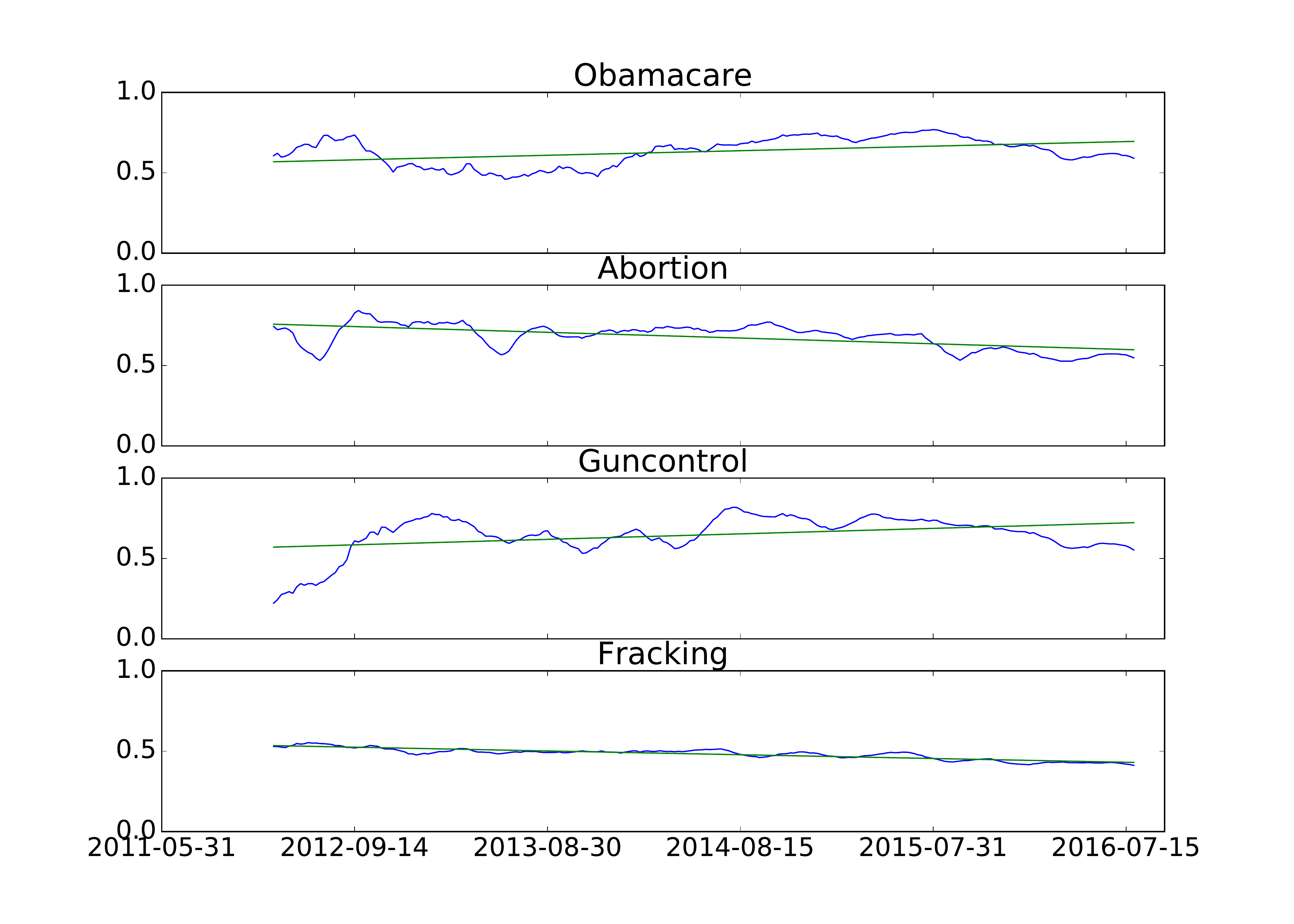}
\caption{Long-term trends of RWC (controversy) score in our dataset. No consistent trend can be observed, which contradicts the narrative that social media is making our society more divided.}
\label{fig:time-trends}
\end{figure}

Let us now focus on how the measures change throughout time.
The longitudinal span of the dataset of five years allows us to track the long-term evolution of discussion on controversial topics.
A common point of view holds that social media is aggravating the polarization of society and exacerbating the divisions in it~\citep{benkler2006wealth}.
At the same time, the political debate (in U.S.) itself has become more polarized in recent years~\citep{andris2015rise}.
However, we do not find conclusive evidence for this argument with our analysis on this dataset.

Figure~\ref{fig:time-trends} shows the long-term trends of the RWC measure for the four topics.
The trend is downwards for `abortion' and `fracking', while it is upwards for `obamacare' and `gun control'.
One could argue that the latter topics are more politically linked to the current administration in U.S., and for this reason have received increasing attention with the elections approaching.
However, the only safe conclusion that can be drawn from this dataset is that there is no clear signal.
The figure suggests that social media, and in particular Twitter, are better suited at capturing the `twitch' response of the public to events and news.
In addition, while our dataset spans a quite long time span for typical social media studies, it is still much shorter than other ones used typically in social science (coming from, e.g., census, polls, congress votes).
This limit is intrinsic of the tool, given that social media have risen in popularity only relatively recently (e.g., Twitter is 10 years old).

\subsection{Non-controversial topics}

For comparison, we perform measurements over a set of non-controversial topics, defined by the hashtags {\it \#ff}, standing for `Follow Friday', used every Friday by users to recommend interesting accounts to follow; {\it \#nba} and {\it \#nfl}, used to discuss sports games; {\it \#sxsw}, used to comment on the {\it South-by-South-West} conference; {\it \#tbt}, standing for `Throwback Thursday', used every Thursday by users to share memories (news, pictures, stories) from the past.

\begin{figure}[tb]
\centering
\begin{minipage}{.3\linewidth}
\centering
\subfloat{\label{}\includegraphics[width=\textwidth]{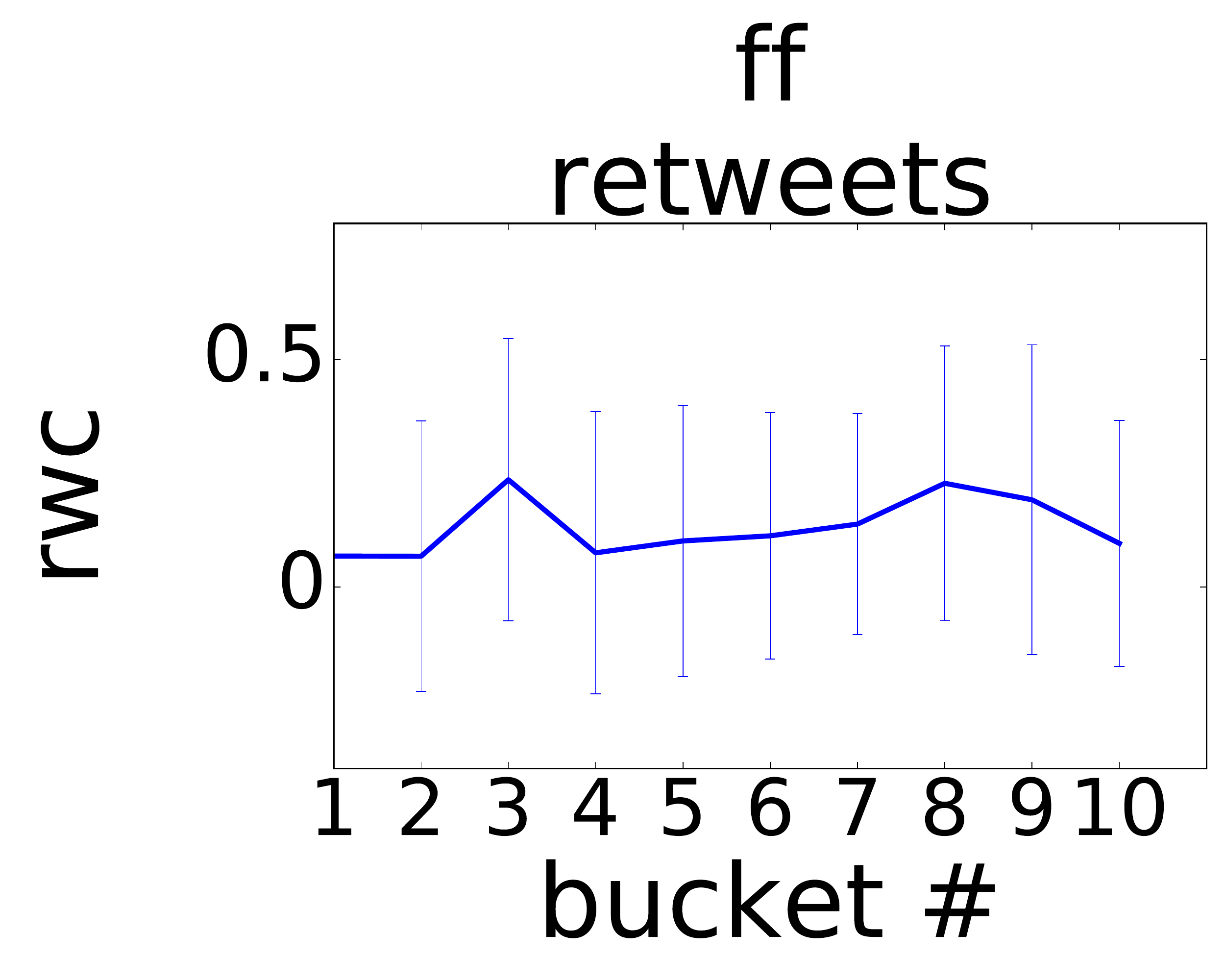}}
\end{minipage}%
\begin{minipage}{.3\linewidth}
\centering
\subfloat{\label{}\includegraphics[width=\textwidth]{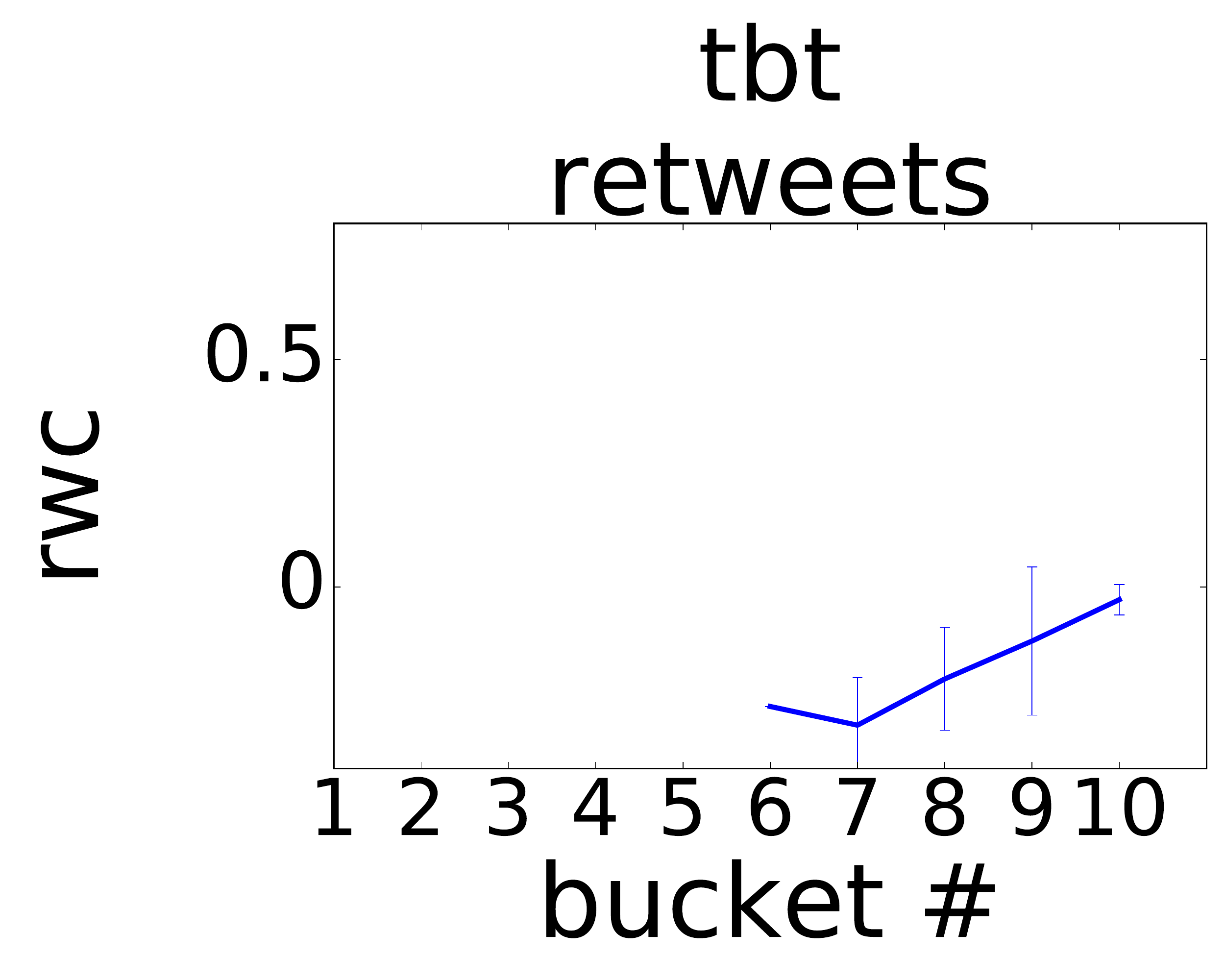}}
\end{minipage}%
\begin{minipage}{.3\linewidth}
\centering
\subfloat{\label{}\includegraphics[width=\textwidth]{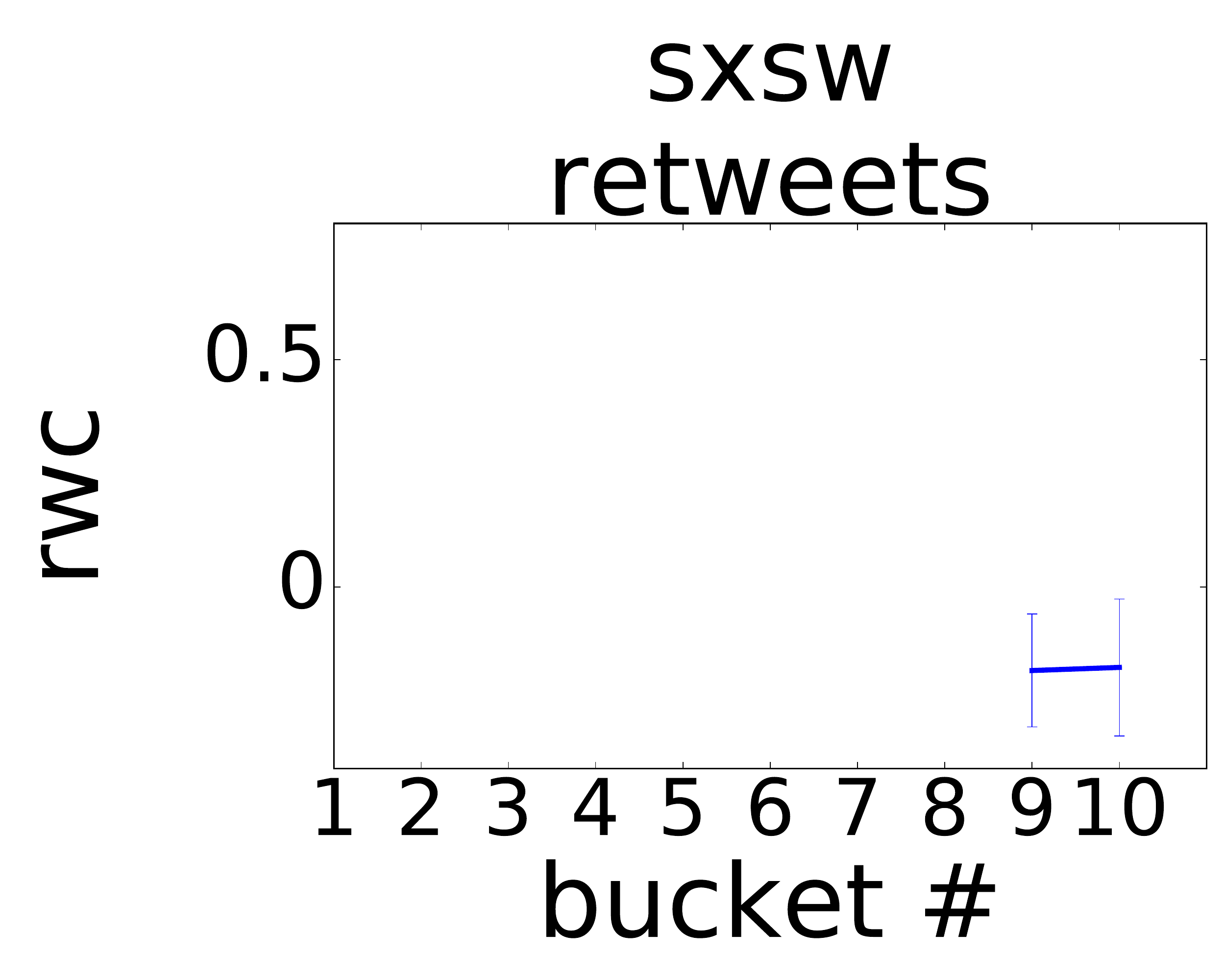}}
\end{minipage}%

\begin{minipage}{.3\linewidth}
\centering
\subfloat{\label{}\includegraphics[width=\textwidth]{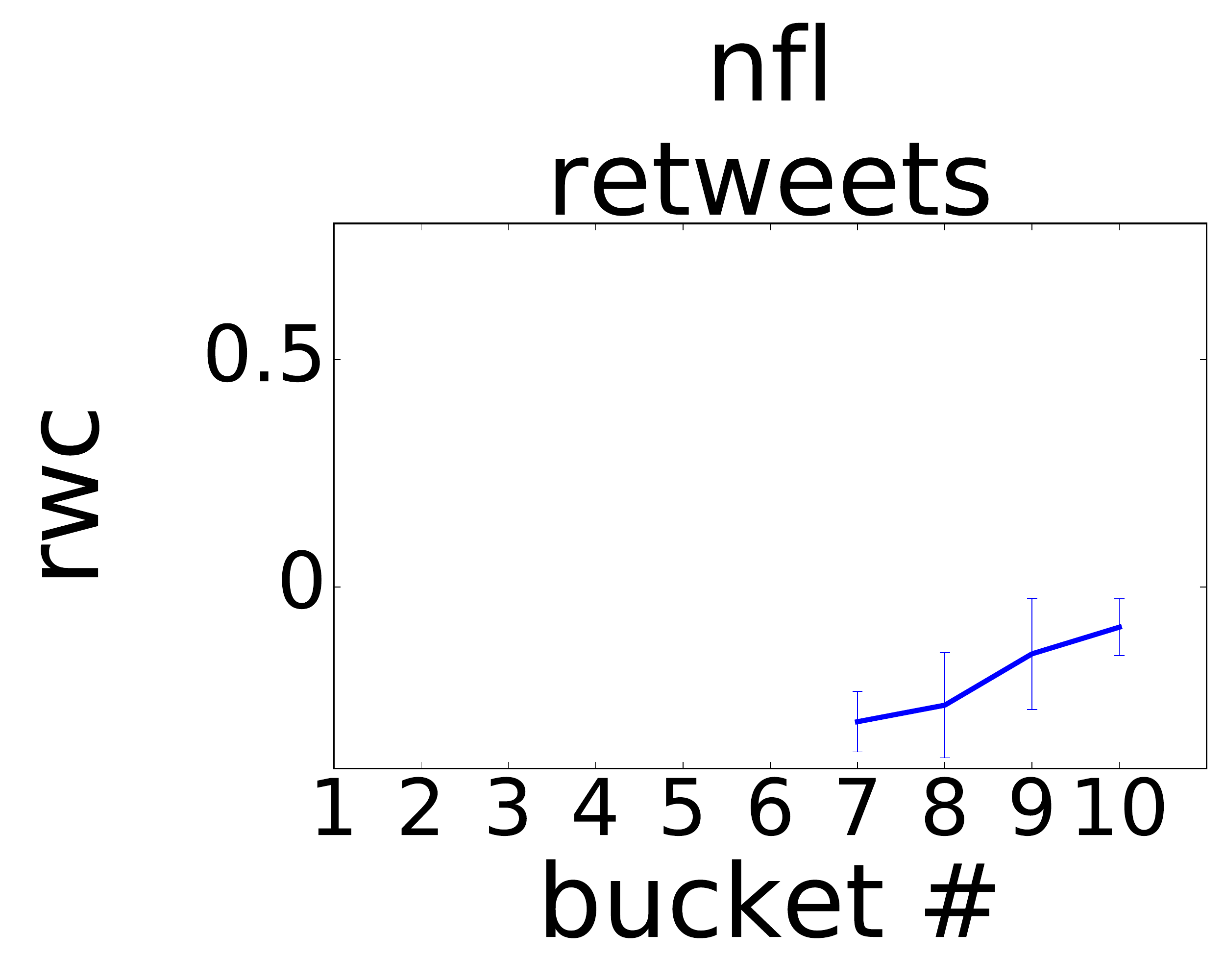}} 
\end{minipage}%
\begin{minipage}{.3\linewidth}
\centering
\subfloat{\label{}\includegraphics[width=\textwidth]{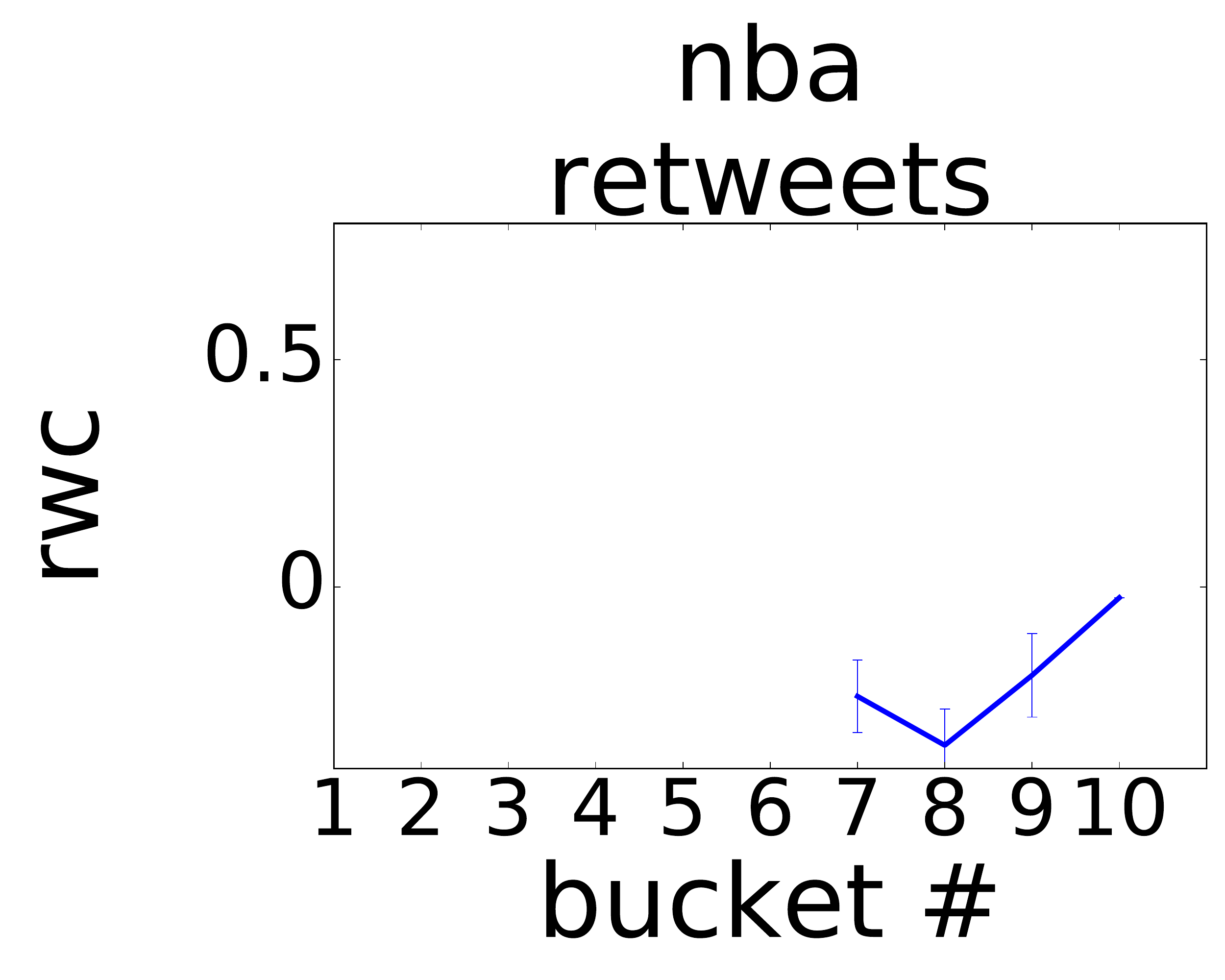}}
\end{minipage}%
\caption{Non-controversial topics: RWC score as a function of the activity in the retweet network.}
\label{fig:noncontrrwc}
\end{figure}

We find that several structural measures, namely \emph{clustering coefficient}, \emph{tie strength}, and \emph{bimotif}, behave similarly to the controversial topics, in that they obtain increased values for increased volume of activity.
This result is in accordance with the ones by \citet{romero2016social}.
Conversely, the values of the \rwc measure typically remain in ranges that indicate low presence of controversy, even as the volume of activity spikes (Figure~\ref{fig:noncontrrwc}).
Additionally, with the definition of `core' introduced above, we could only identify a negligibly small core for these topics (i.e., found very few users who were consistently active on these topics).

\begin{figure}[t]
\centering
\begin{minipage}{.33\linewidth}
\centering
\subfloat{\label{}\includegraphics[width=\textwidth]{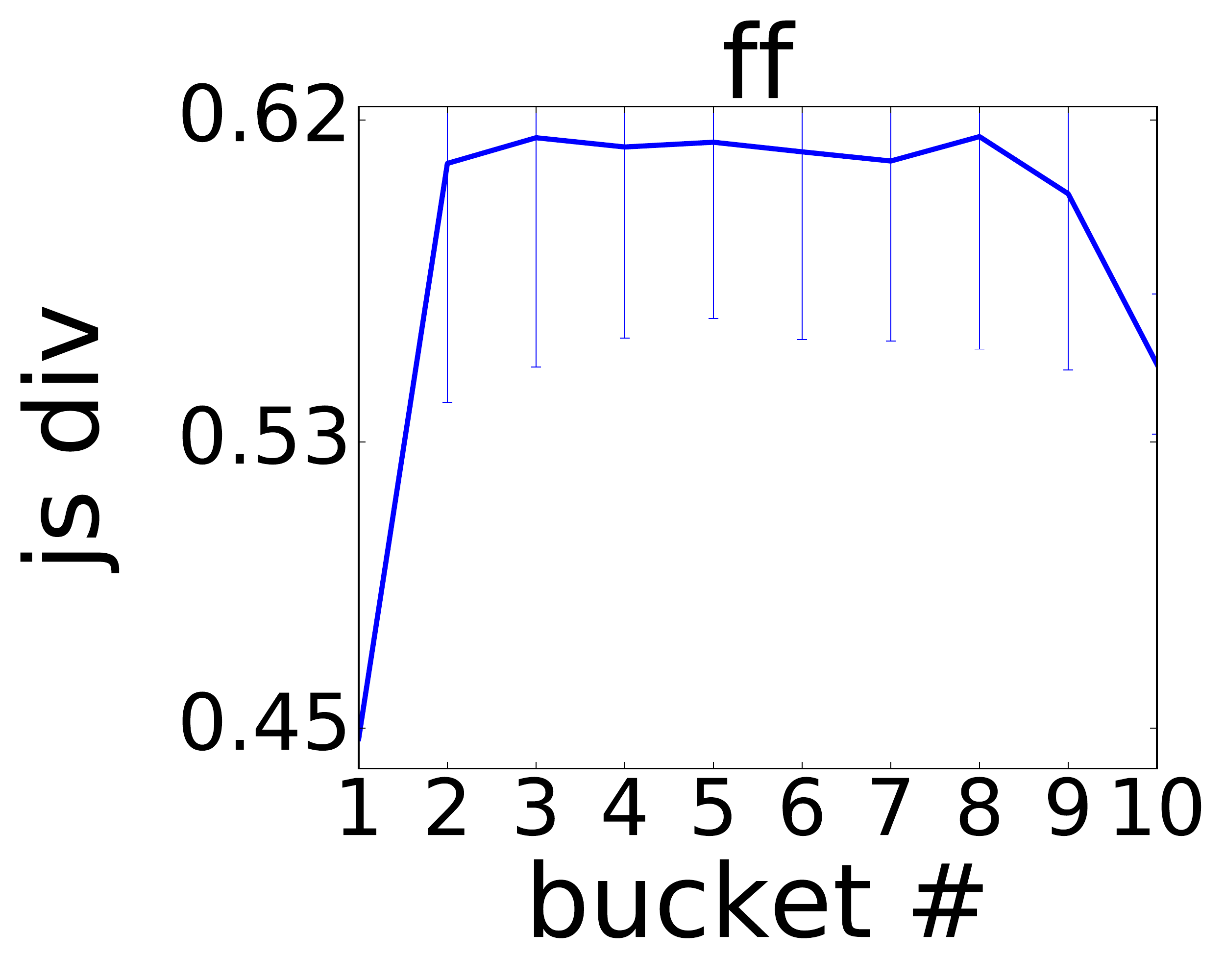}}
\end{minipage}%
\begin{minipage}{.33\linewidth}
\centering
\subfloat{\label{}\includegraphics[width=\textwidth]{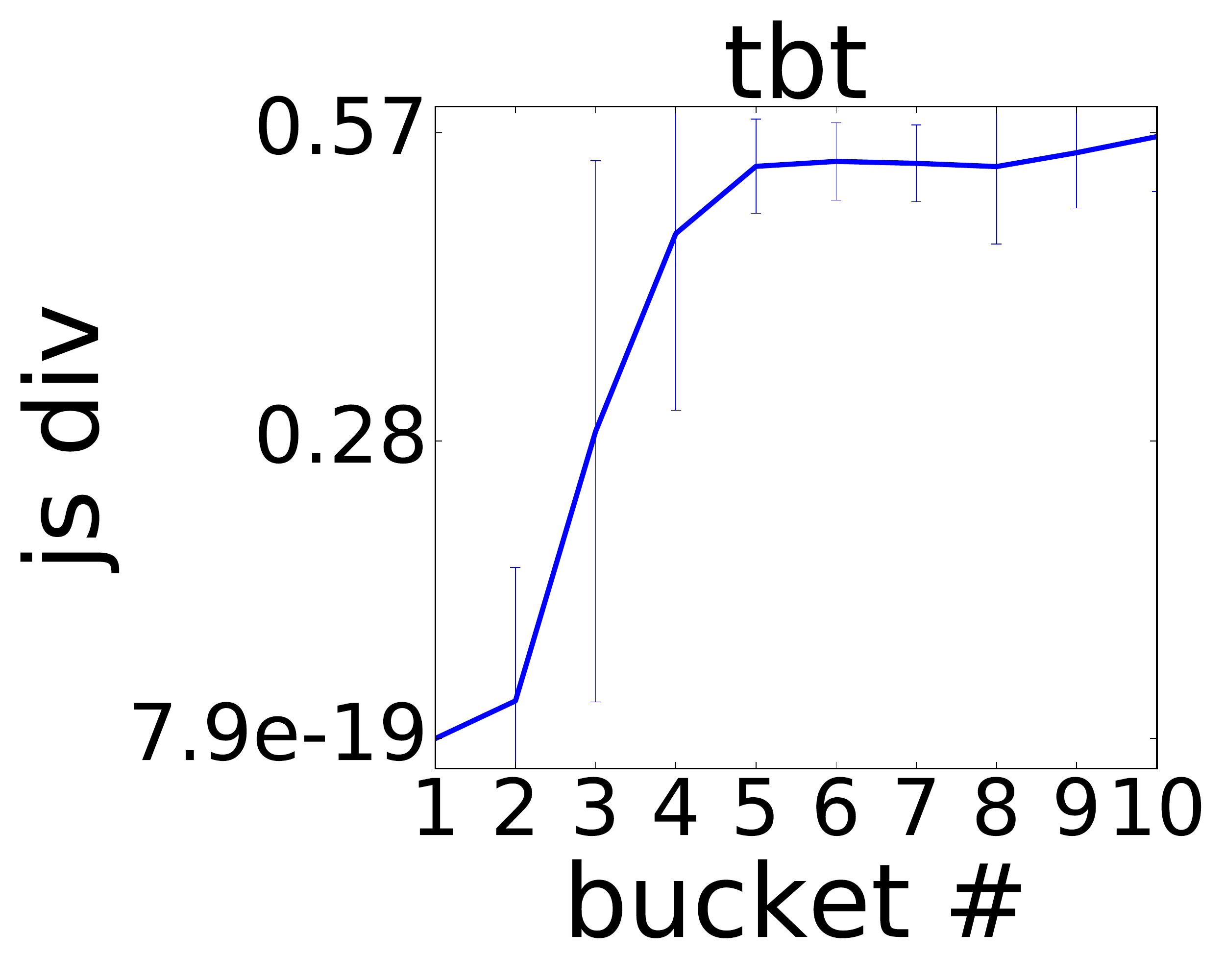}}
\end{minipage}%
\begin{minipage}{.33\linewidth}
\centering
\subfloat{\label{}\includegraphics[width=\textwidth]{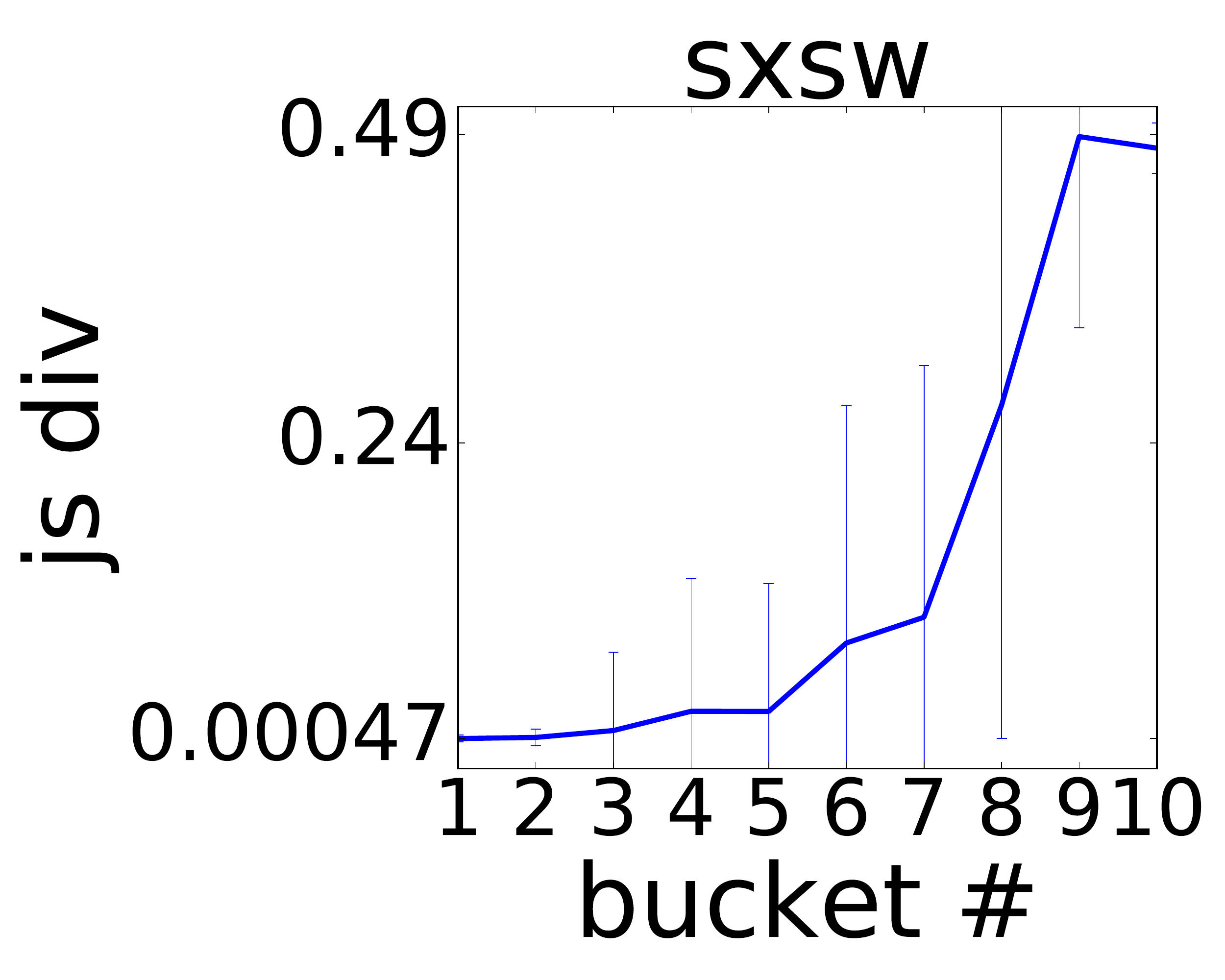}}
\end{minipage}%

\begin{minipage}{.33\linewidth}
\centering
\subfloat{\label{}\includegraphics[width=\textwidth]{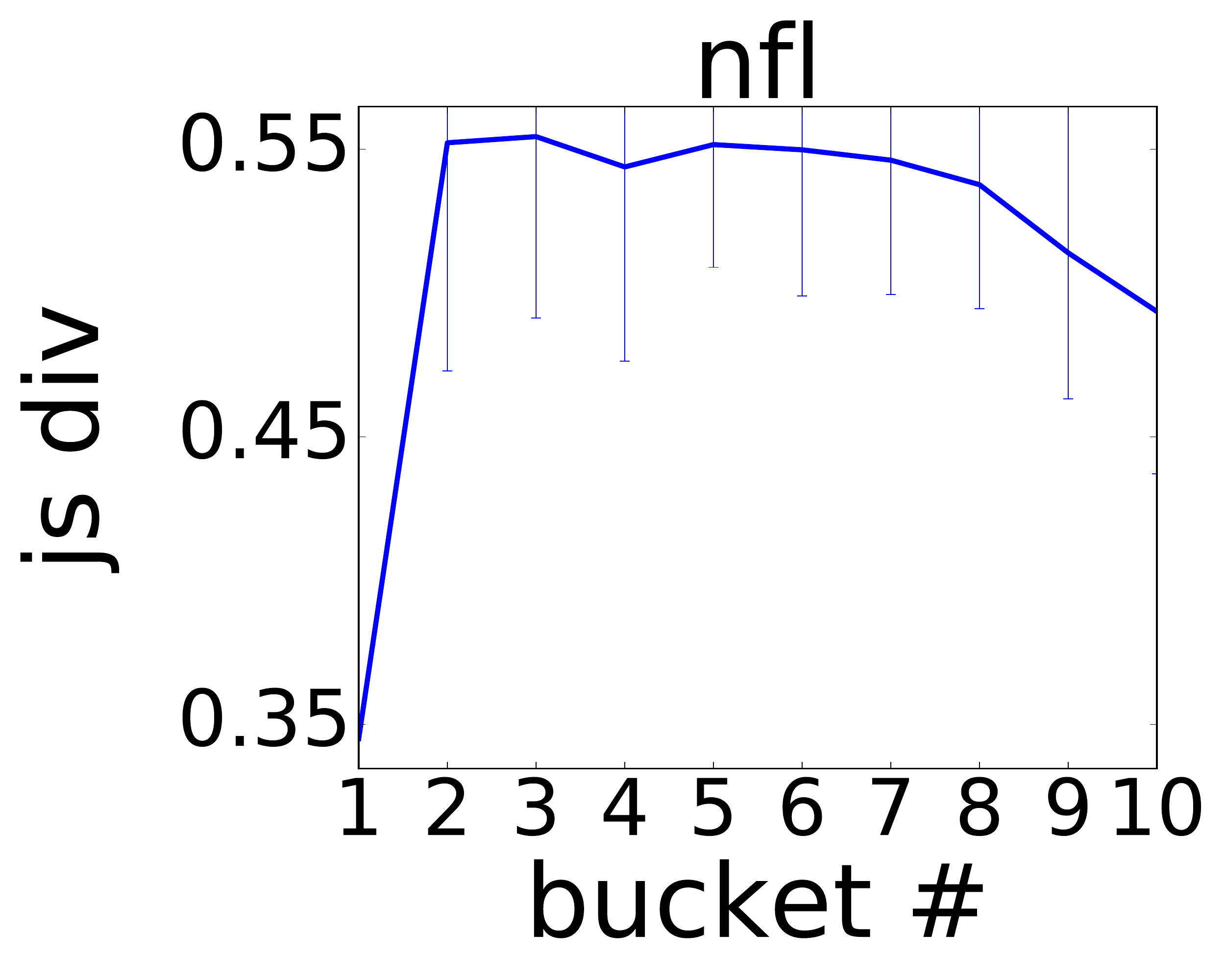}}
\end{minipage}%
\centering
\begin{minipage}{.33\linewidth}
\centering
\subfloat{\label{}\includegraphics[width=\textwidth]{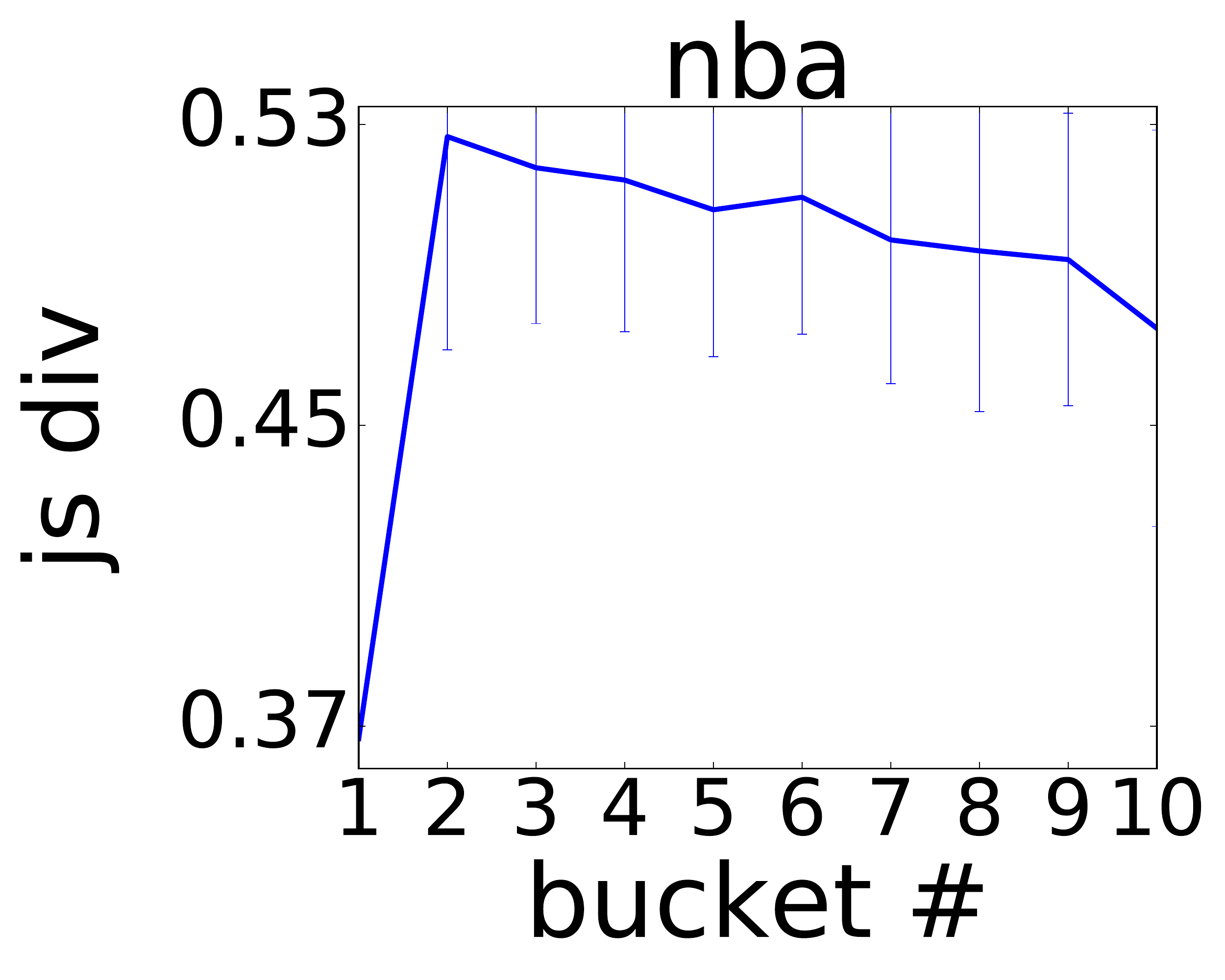}}
\end{minipage}%
\caption{Non-controversial topics: Jensen-Shannon divergence of the lexicon between the two sides as a function of network activity. As the interest in the topic rises, the lexicon used by the two sides tends to converge.}
\label{fig:noncontrjsdiv}
\end{figure}

\begin{figure}[tb]
\centering
\begin{minipage}{.33\linewidth}
\centering
\subfloat{\label{}\includegraphics[width=\textwidth]{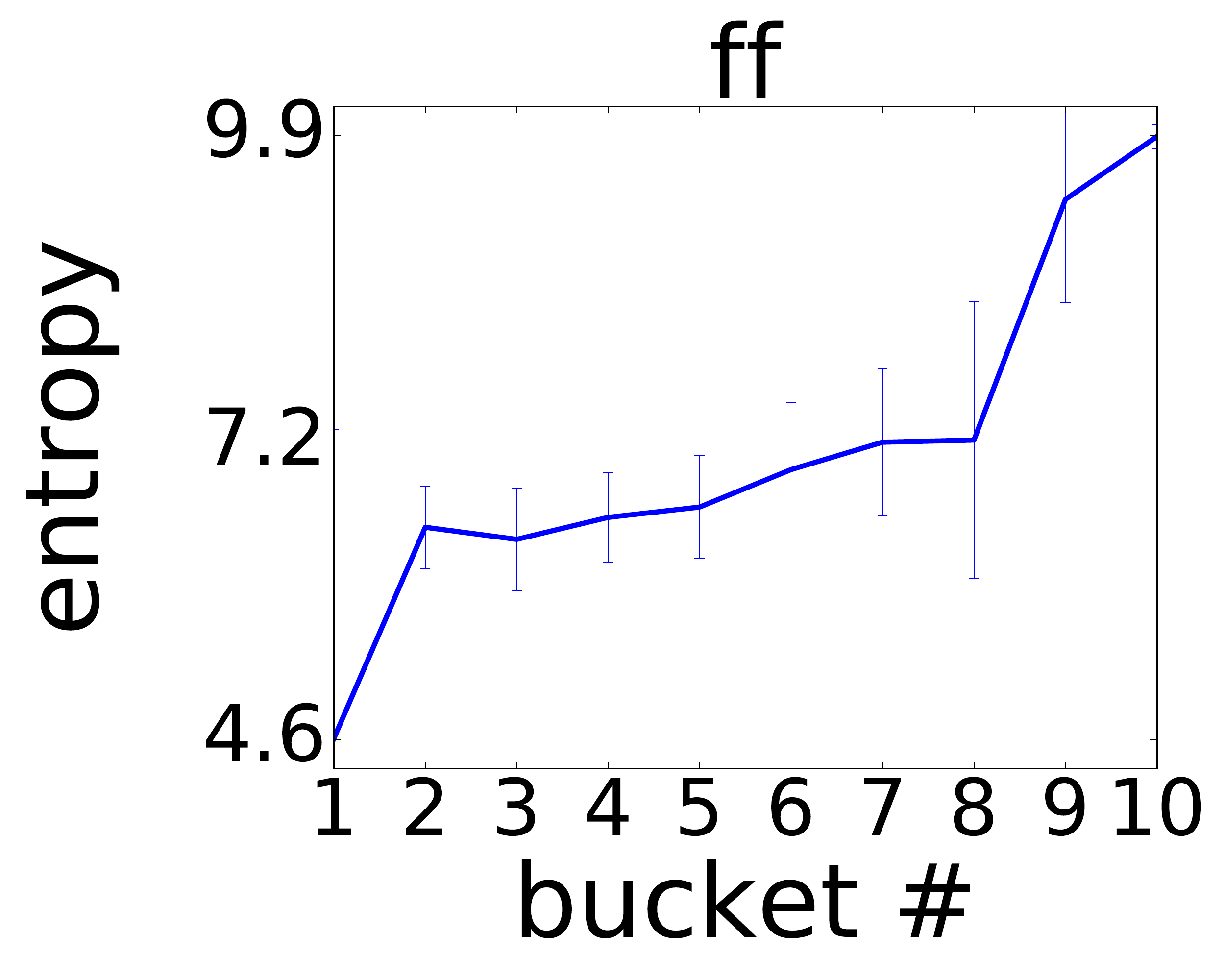}}
\end{minipage}%
\begin{minipage}{.33\linewidth}
\centering
\subfloat{\label{}\includegraphics[width=\textwidth]{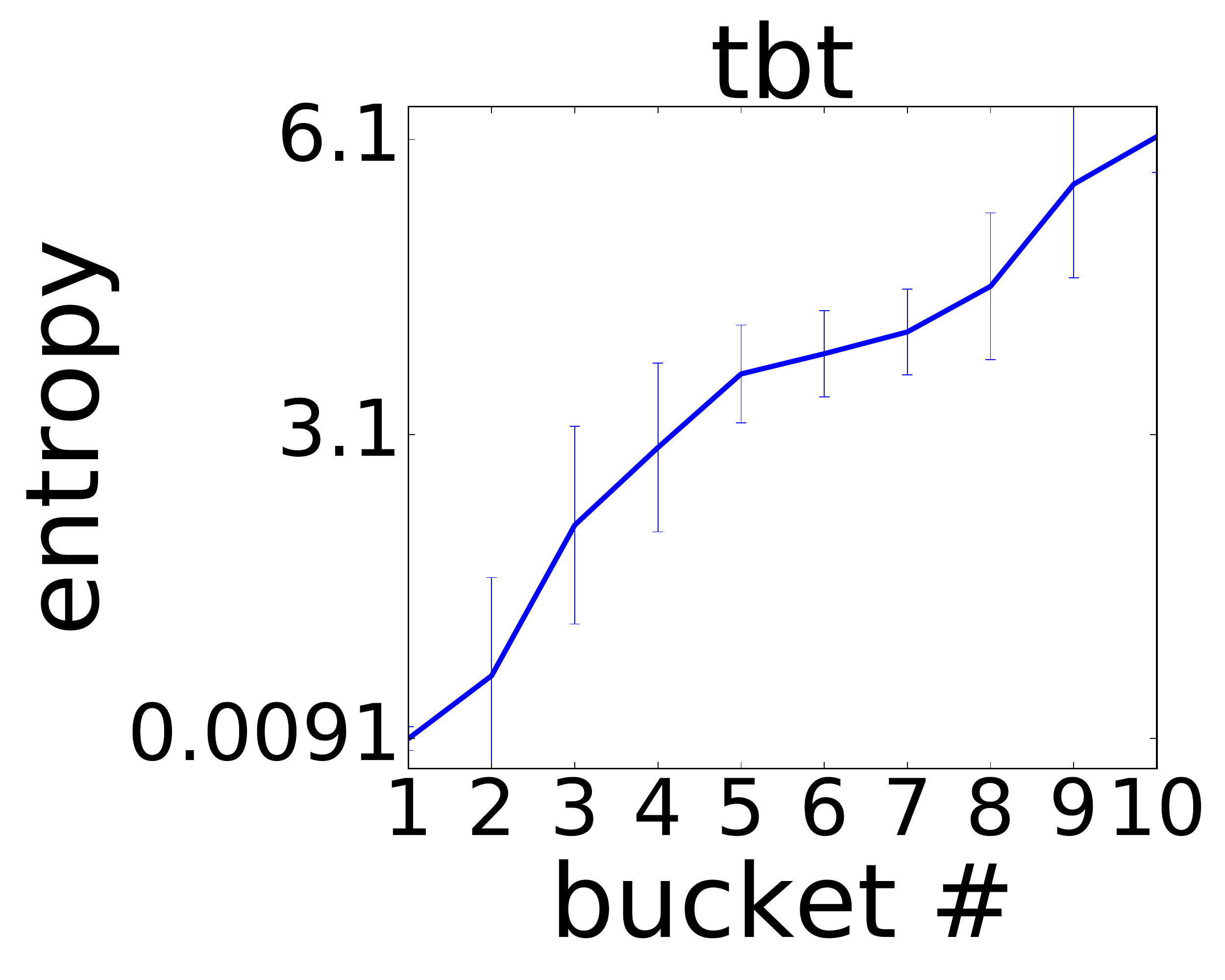}}
\end{minipage}%
\begin{minipage}{.33\linewidth}
\centering
\subfloat{\label{}\includegraphics[width=\textwidth]{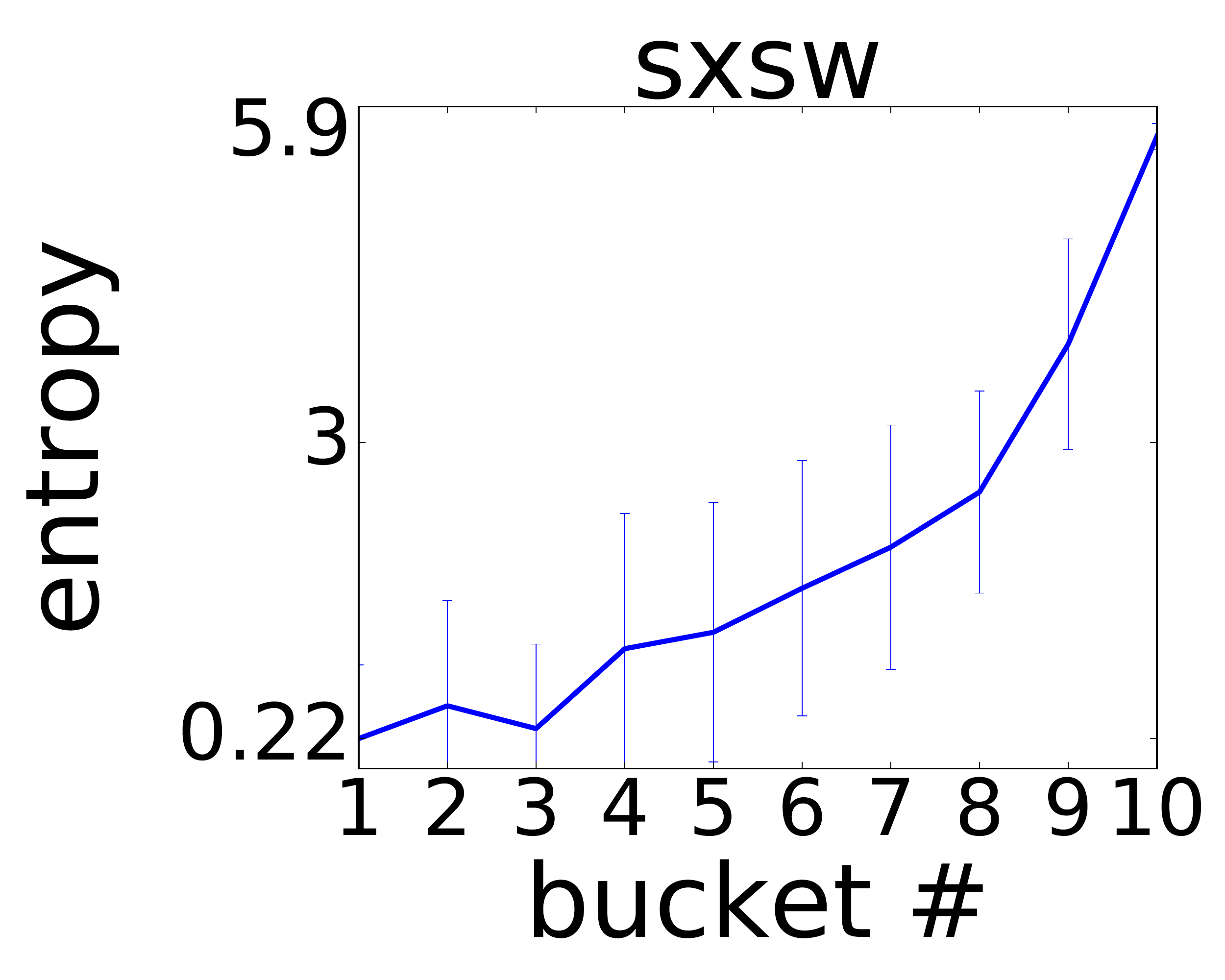}}
\end{minipage}%

\begin{minipage}{.33\linewidth}
\centering
\subfloat{\label{}\includegraphics[width=\textwidth]{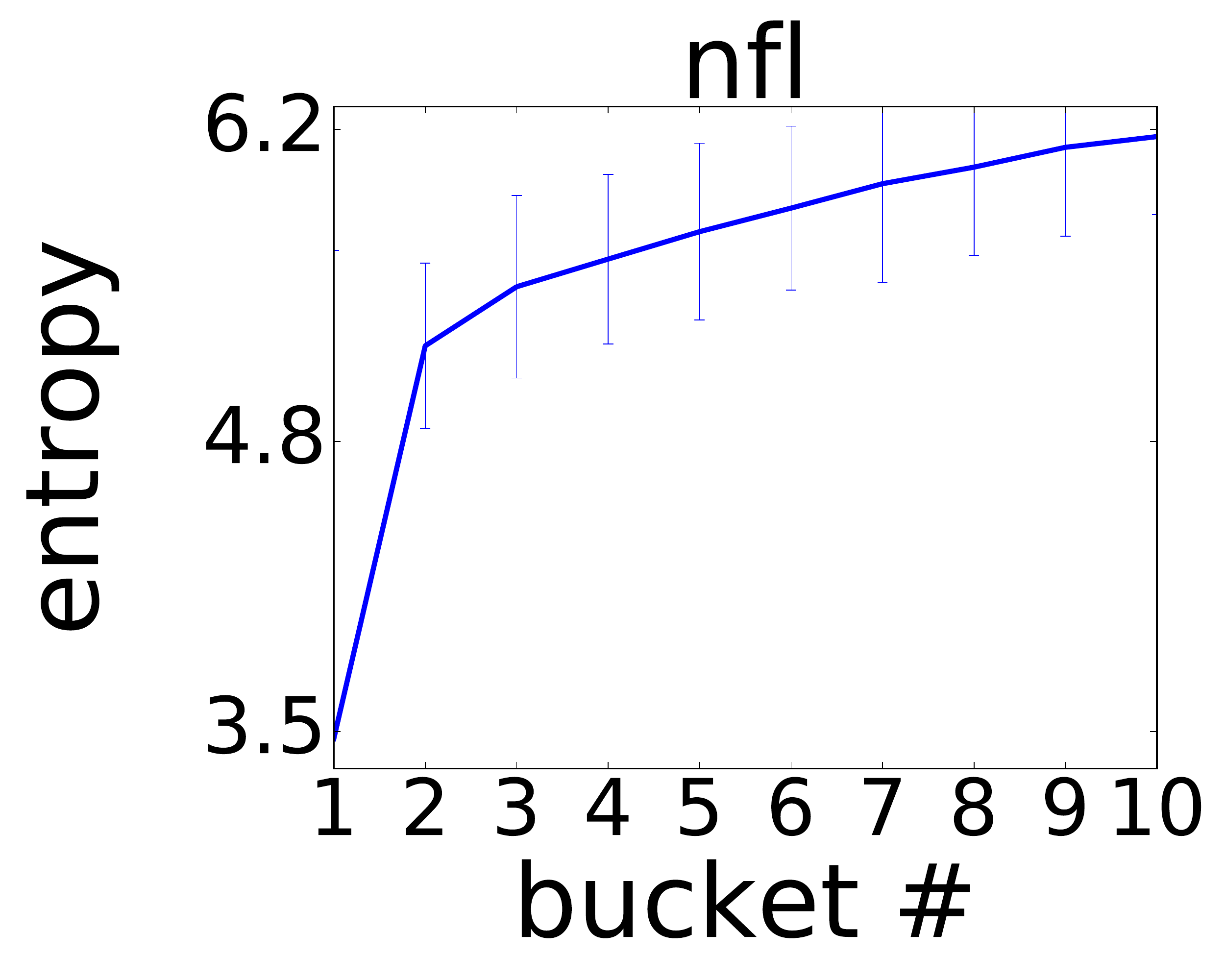}}
\end{minipage}%
\centering
\begin{minipage}{.33\linewidth}
\centering
\subfloat{\label{}\includegraphics[width=\textwidth]{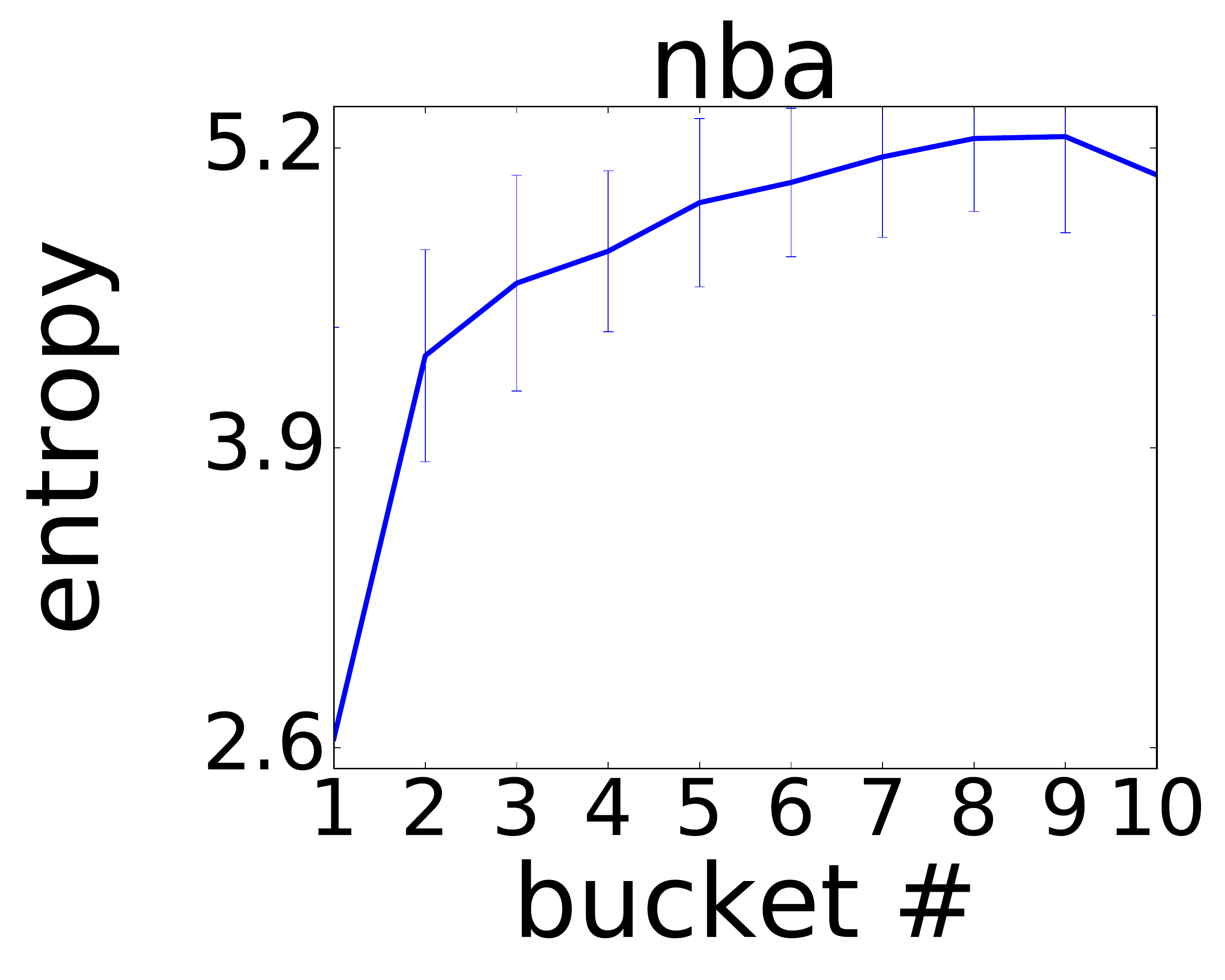}}
\end{minipage}%
\caption{Non-controversial topics: Entropy of the distribution over the lexicon for one side of the discussion as a function of the activity in the network (the other side shows similar patterns).
}
\label{fig:noncontrentropy}
\end{figure}

\begin{figure}[tb]
\centering
\begin{minipage}{.33\linewidth}
\centering
\subfloat{\label{}\includegraphics[width=\textwidth]{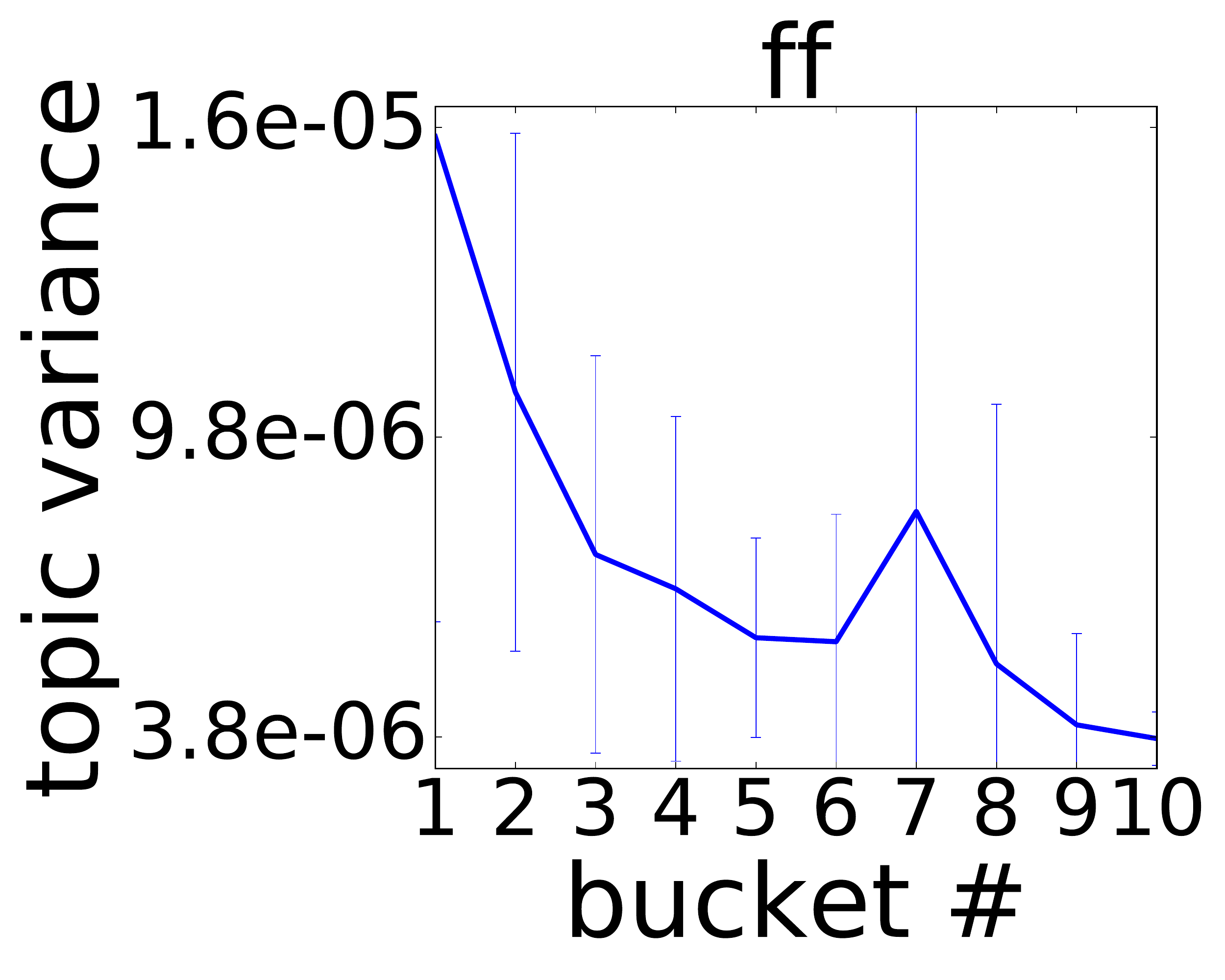}}
\end{minipage}%
\begin{minipage}{.33\linewidth}
\centering
\subfloat{\label{}\includegraphics[width=\textwidth]{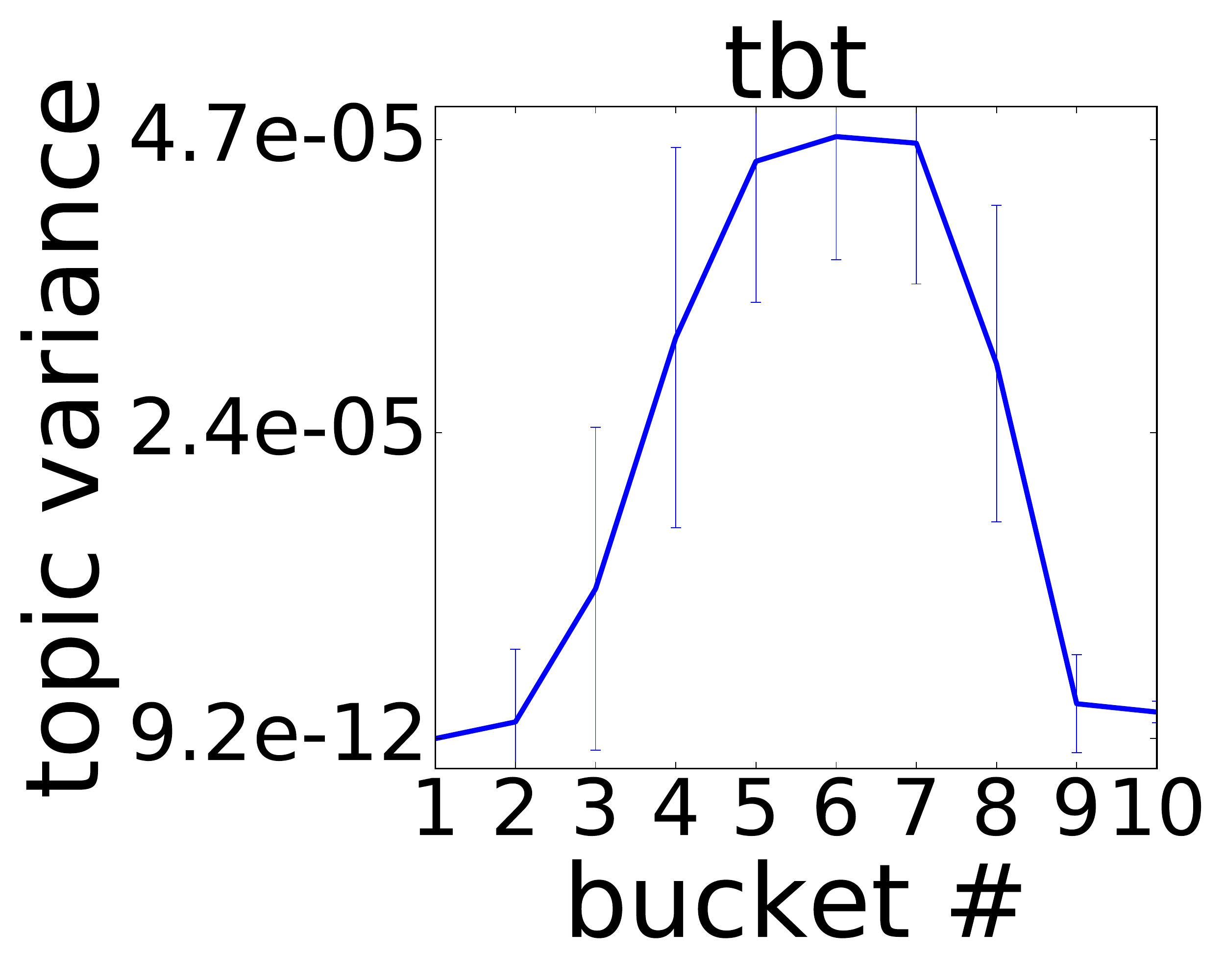}}
\end{minipage}%
\begin{minipage}{.33\linewidth}
\centering
\subfloat{\label{}\includegraphics[width=\textwidth]{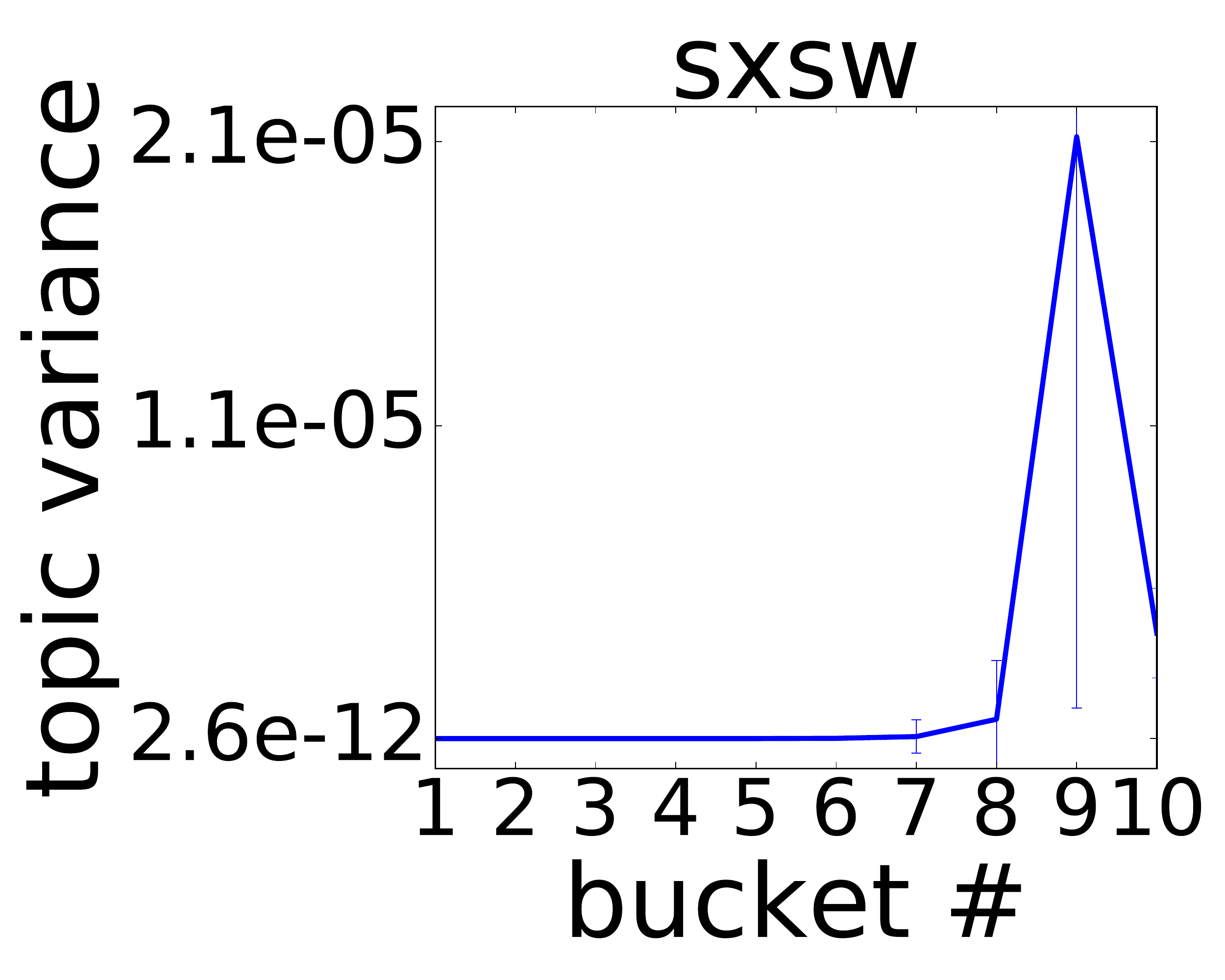}}
\end{minipage}%

\begin{minipage}{.33\linewidth}
\centering
\subfloat{\label{}\includegraphics[width=\textwidth]{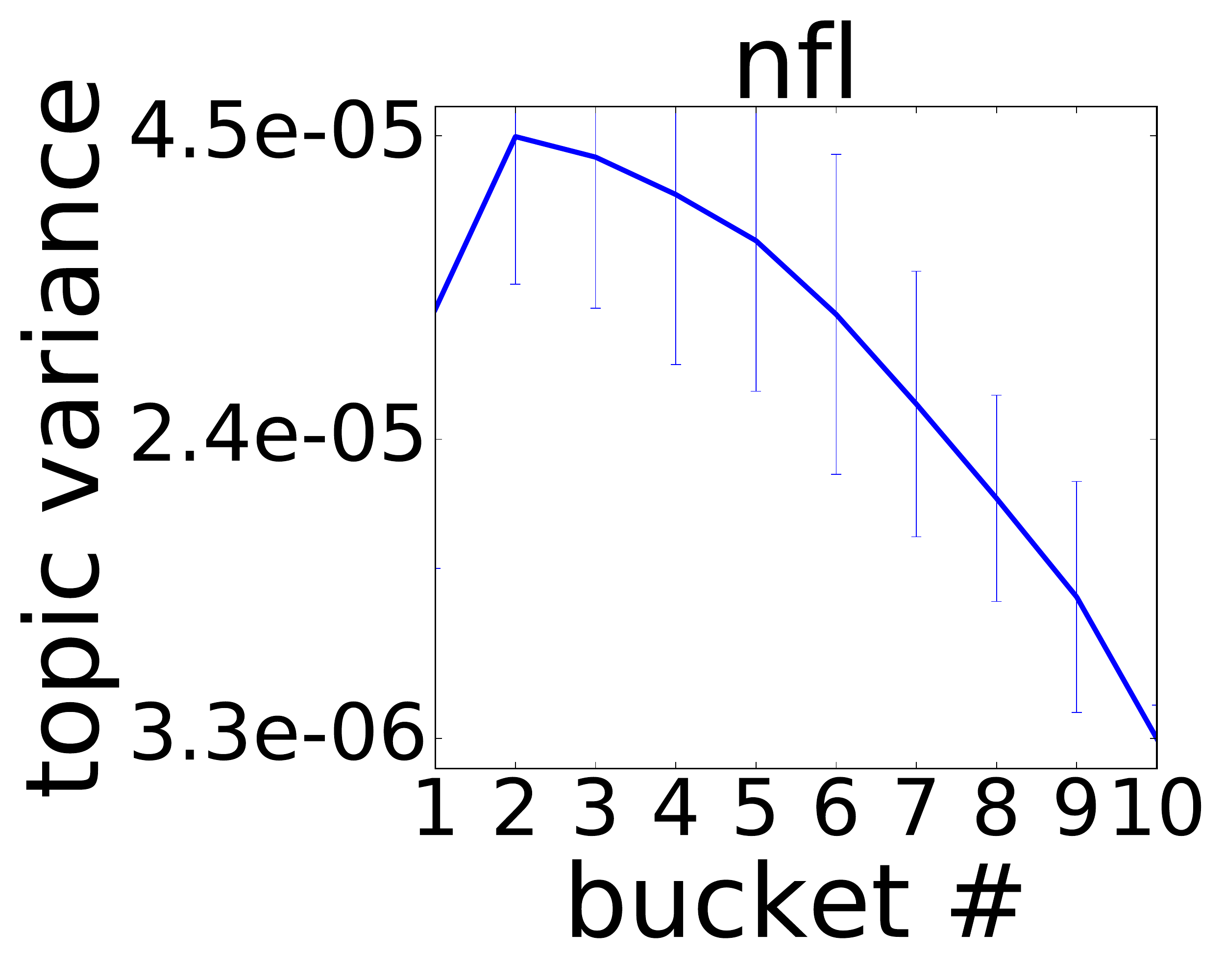}}
\end{minipage}%
\centering
\begin{minipage}{.33\linewidth}
\centering
\subfloat{\label{}\includegraphics[width=\textwidth]{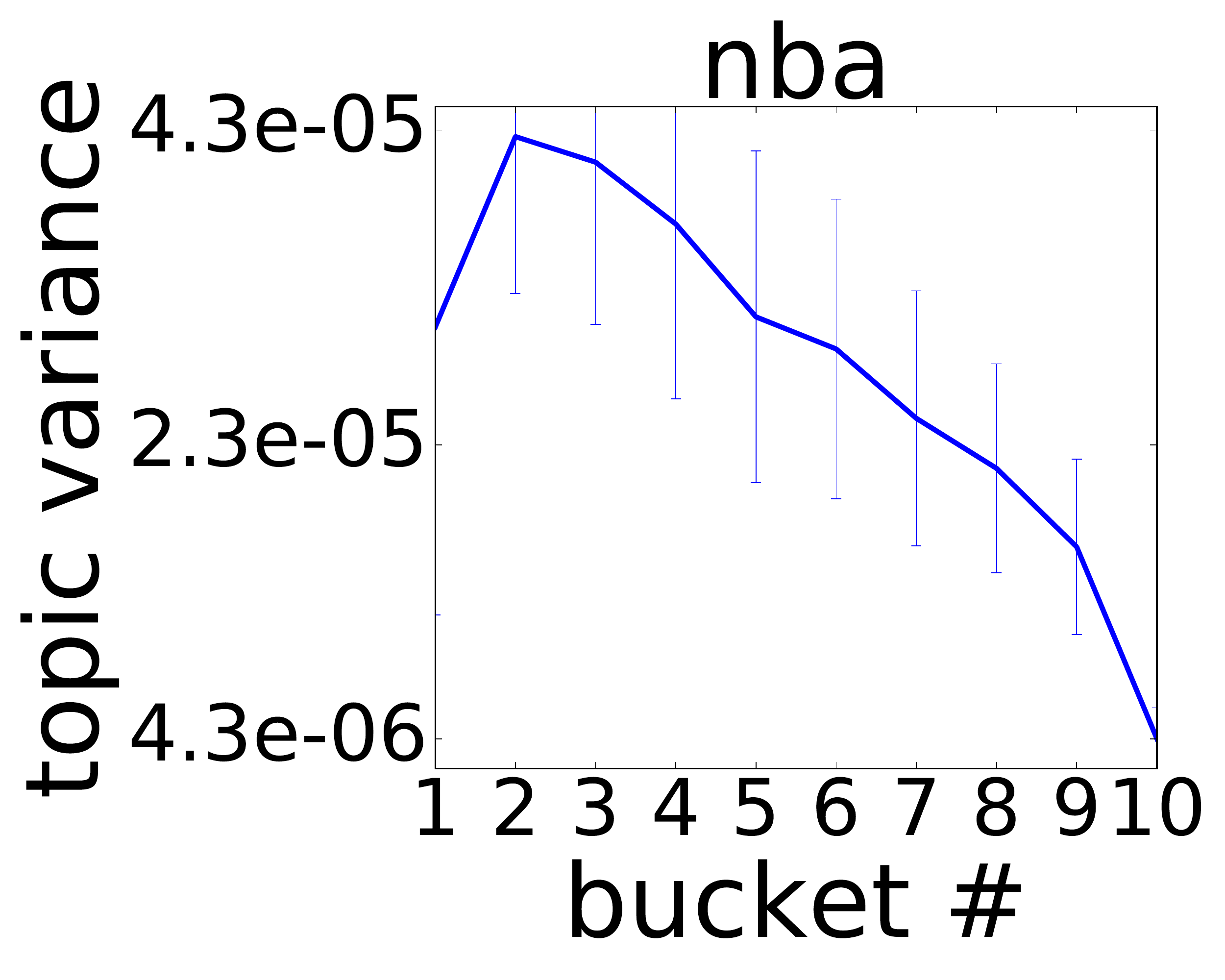}}
\end{minipage}%
\caption{Non-controversial topics: Variance of the topic distribution. As the interest increases, variance decreases, indicating that a wider range of topics are being discussed.} 
\label{fig:noncontrtopicvariance}
\end{figure}

Finally, in terms of content measures we find that, as for the controversial topics, the entropy of the lexicon increases with volume (Figure~\ref{fig:noncontrentropy}).
Topic variance also decreases with volume in most cases, meaning that a wider range of topics are discussed (Figure~\ref{fig:noncontrtopicvariance}).
On the contrary, the Jensen-Shannon divergence stays at relatively constant values across volume levels (Figure~\ref{fig:noncontrjsdiv}).
It thus behaves differently compared to controversial topics (Figure~\ref{fig:js}).
This result is to be expected, as the two `sides' identified by METIS on the networks of non-controversial topics are not as well defined as they are in the case of controversial topics.

\section{Conclusion}
\label{sec:conclusions}
The evolution of networks is a well-studied phenomenon in social sciences, 
physics, and computer science.
However, the evolution of \emph{interaction} networks 
has received substantially less attention so far.
In particular, interaction networks related to discussions of controversial topics, 
which are important from a sociological point of view, have not been analyzed before.
This study is a first step towards understanding this important social phenomenon.

We analyzed four highly controversial topics of discussion on Twitter for a period 
of five years.
By examining the endorsement and communication networks of users involved in these discussions, we found that spikes in interest correspond to an increase in the controversy of the discussion.
This result is supported by a wide array of network analysis measures, and is consistent across topics.
We also found that interest spikes correspond to a convergence of the lexicon used by the opposite sides of a controversy, and a more uniform lexicon overall.
The code and datasets used in the paper are available on the project website.\textsuperscript{\ref{footnote:website}}

Implications of this work relate to the understanding of 
how our society evolves via continuous debates, 
and how culture wars develop~\citep{abramowitz2005can,highton2011long,lindaman2002issue}.
It is often argued that technology, and social media in particular, is having a negative impact on our ability to relate to the unfamiliar~\citep{benkler2006wealth}, due to the ``echo chamber'' and ``filter bubble'' effects.
However, while we found instantaneous temporary increase in controversy in relation to external events, our study did not find evidence of long term increase in polarization of the discussions, neither after these events nor as a general longitudinal trend.
At the same time, 
investigating how to reduce the polarization of these discussions on controversial topics 
is a research-worthy problem~\citep{garimella2016connecting,garimella2017mary},
and taking into account the dynamics of the process is a promising direction to explore.

Our observations pave the way to the development of models of evolution for controversial interaction networks, similarly to how studies about measuring the Web and social media were the stepping stone to developing models for them.
A logical next step for this line of work is to investigate 
how to use early signals from social media network structure and content to predict the impact of an event.
Equally of interest is whether the observations made in this study translate to other social media beside Twitter, for instance, Facebook or Reddit.
Finally, while we did not find any consistent long-term trend in the polarization of the discussions, it is worth continuing this line of investigation, as the effects of increased polarization might not be easily discoverable from social-media analysis alone.

\spara{Acknowledgements.}
This work has been supported by the Academy of Finland project ``Nestor'' (286211) and the EC H2020 RIA project ``SoBigData'' (654024).

\bibliographystyle{ACM-Reference-Format}
\bibliography{biblio}

\end{document}